\documentclass[preprint,nopreprintline,3p,times]{elsarticle}
\usepackage{import}  % subimport
\usepackage{lineno,hyperref}
\modulolinenumbers[5]
\usepackage[Export]{adjustbox}  % improved includegraphics (ie. with frame options)
\usepackage{afterpage} % helps layouting page breaks
\usepackage[
 ruled,
 vlined,
 linesnumbered,
 algosection,  % number algorithms with leading section numbers
]{algorithm2e}
\usepackage{amsthm}  % for \theoremstyle
\usepackage{amssymb}  % special symbols like \vartriangleleft
\usepackage{booktabs} % Great tables
\usepackage[
 labelfont=bf,  % bold labels (eg. for "Figure 1")
 margin=3em,  % Reduce caption width by this margin on both sides
]{caption}  % caption styling for figures / tables
\usepackage{enumitem}
\usepackage{float}
\usepackage[T1]{fontenc} % Allow use of \textsc and \textbf at the same time.
\usepackage{hyphenat} % allows hyphenation for words with a dash (when \hyp{} is used instead of - for the dash)
\usepackage[utf8]{inputenc}
\usepackage{longtable}
\usepackage{lscape}
\usepackage[tbtags]{mathtools}
\usepackage{multicol}
\usepackage{multirow}
\usepackage{placeins} % provides Floatbarrier (floats before and after the barrier are not mixed up)
\usepackage{rotating} % provides sidewaysfigure, sidewaystable
\usepackage[
 labelformat=simple,  % no extra braces in label. We add add those always by redefining \thesubfigure
]{subcaption}  % subfigure (several subfigures withing one figure)
\usepackage{supertabular}
\usepackage{tabularx}
\usepackage[dvipsnames]{xcolor}
\usepackage[
 capitalise,
 nameinlink,
 noabbrev]{cleveref}  % cref

\newcommand{\bkt}[1]{\bktR{#1}}
\newcommand{\Bkt}[1]{\BktR{#1}}

\newcommand{\DBkt}[1]{\Bkt{\Bkt{#1}}}

\newcommand{\bktR}[1]{\setWithBracketsBase{(}{#1}{)}}
\newcommand{\BktR}[1]{\SetWithBracketsBase{(}{#1}{)}}

\newcommand{\bktC}[1]{\setWithBrackets{\{}{#1}{\}}}
\newcommand{\BktC}[1]{\SetWithBrackets{\{}{#1}{\}}}

\newcommand{\bktU}[1]{\setWithBrackets{\lceil}{#1}{\rceil}}
\newcommand{\BktU}[1]{\SetWithBrackets{\lceil}{#1}{\rceil}}

\newcommand{\BktD}[1]{\SetWithBrackets{\lfloor}{#1}{\rfloor}}

\newcommand{\fixedbkt}[1]{\csname #1\endcsname}

\newcommand{\bktRfixed}[2]{
\setWithBracketsBase{\fixedbkt{#1}(}{#2}{\fixedbkt{#1})}}

\newcommand{\bktCfixed}[2]{
\setWithBracketsBase{\fixedbkt{#1}\{}{#2}{\fixedbkt{#1}\}}}

\newcommand{\bktUfixed}[2]{
\setWithBracketsBase{\fixedbkt{#1}\lceil}{#2}{\fixedbkt{#1}\rceil}}

\newcommand{\innerBracketSpace}{2mu}

\newcommand{\setWithBracketsBase}[3]{#1{#2}#3}
\newcommand{\SetWithBracketsBase}[3]{\setWithBracketsBase{\left#1}{#2}{\right#3}}
\DeclareRobustCommand{\setWithBrackets}[3]{%
 \ifInnermost{\setWithBrackets}{#2}{%
  \setWithBracketsBase{#1}{\mspace{\innerBracketSpace}#2\mspace{\innerBracketSpace}}{#3}%
 }{%
  \setWithBracketsBase{#1}{#2}{#3}%
 }%
}
\DeclareRobustCommand{\SetWithBrackets}[3]{%
 \setWithBrackets{\left#1}{#2}{\right#3}%
}

\makeatletter

\newcommand{\ifInnermost}[4]{%
  \let\ifInnermost@oldBraces#1%
  \gdef\ifInnermost@foundBraces{0}
  \renewcommand{#1}[3]{%
    \gdef\ifInnermost@foundBraces{1}%
    \renewcommand{#1}[3]{}%
  }%
  \sbox0{$#2$}%
  \let#1\ifInnermost@oldBraces%
  \ifnum\numexpr\ifInnermost@foundBraces\relax=0%
    #3%
  \else%
    #4%
  \fi%
}

\makeatother

\definecolor{magenta}{RGB}{255,0,255}
\definecolor{gray83}{RGB}{212,212,212}
\definecolor{gray67}{RGB}{171,171,171}
\definecolor{gray50}{RGB}{128,128,128}
\definecolor{gray33}{RGB}{86,86,86}
\definecolor{gray17}{RGB}{43,43,43}

\definecolor{reddish_brown}{RGB}{149,  82,  81}
\definecolor{brown}{RGB}{199, 141, 107}
\definecolor{purple}{RGB}{154,  51, 143}
\definecolor{violet}{RGB}{225,  15, 125}
\definecolor{pink}{RGB}{255,  62, 181}
\definecolor{flieder}{RGB}{236, 166, 192}
\definecolor{red}{RGB}{217,  38,  41}
\definecolor{mid_red}{rgb}{1,0,0}
\definecolor{orange}{RGB}{232, 121,  40}
\definecolor{mid_orange}{RGB}{250, 121,  40}
\definecolor{yellow}{RGB}{244, 236,  39}
\definecolor{mid_yellow}{rgb}{1,1,0}
\definecolor{yellowish_green}{RGB}{151, 215,   0}
\definecolor{light_green}{RGB}{111, 184,  76}
\definecolor{mid_green}{rgb}{0, 0.8, 0.3}
\definecolor{dark_green}{RGB}{ 13, 149,  75}
\definecolor{turkoise}{RGB}{125, 207, 182}
\definecolor{default_blue}{RGB}{  0, 181, 226}
\definecolor{light_blue}{RGB}{ 36, 156, 216}
\definecolor{dark_blue}{RGB}{ 29,  78, 137}
\definecolor{dark_orange}{RGB}{189, 100, 33}

\definecolor{emphcolor}{rgb}{0.175,0.175,0.64}

\newtheorem{theorem}{Theorem}
\numberwithin{theorem}{section}
\newtheorem{proposition}[theorem]{Proposition}
\newtheorem{lemma}[theorem]{Lemma}
\newtheorem{corollary}[theorem]{Corollary}
\newtheorem{conjecture}[theorem]{Conjecture}
\theoremstyle{definition}
\newtheorem{definition}[theorem]{Definition}
\newtheorem{observation}[theorem]{Observation}
\newtheorem{notation}[theorem]{Notation}

\theoremstyle{remark}

\newtheorem*{claim_no_num}{Claim}
\makeatletter
\@addtoreset{claim}{theorem}
\makeatother
\newcommand{\claimqed}{\scriptsize \qed}
\newenvironment{proof_of_claim}{\par\noindent\textit{Proof of claim:}}{\claimqed}

\Crefname{observation}{Observation}{Observations}
\Crefname{conjecture}{Conjecture}{Conjectures}
\Crefname{notation}{Notation}{Notations}

\newenvironment{qedtheorem}
 {\pushQED{\qed}\theorem}
 {\popQED\endtheorem}

\newenvironment{qedproposition}
 {\pushQED{\qed}\proposition}
 {\popQED\endproposition}

\newenvironment{qedcorollary}
 {\pushQED{\qed}\corollary}
 {\popQED\endcorollary}

\newcommand{\boxedDescription}[5]{%
 {%
  \setlength{\parskip}{\topsep}% Add the same enclosing spacing as used for AMS theorems, etc.
   {%
    \setlength{\fboxsep}{5pt}% increased spacing inside frame
     {%
      \setlength{\parindent}{0pt}% disable standard paragraph indent
       \framebox[\textwidth][c]{%
        \begin{tabular}{%
          @{\hspace{\fboxsep}}
          l@{}
          @{\hspace{\fboxsep}}
          p{\textwidth-3\fboxsep-\maxof{\widthof{#1}}{\widthof{#2}}}@{\hspace{\fboxsep}}
        }%
         \multicolumn{2}{p{\textwidth-2\fboxsep}}{%
          {\Large{\textsc{#3}}}} \\ %
         \textit{#1:} & #4 \\ %
         \textit{#2:} & #5 \\ %
        \end{tabular}%
       }%
     }%
   }%
  \vspace{\topsep}
 }%
}

\newcommand{\problemBase}[3]{%
 \boxedDescription{Instance}{Task}{#1}{#2}{#3}%
}

\newcommand{\problemIndex}[4]{
 \index{#2|(}%
 \problemBase{#1}{#3}{#4}%
 \index{#2|)}%
}

\newcommand{\problem}[4]{
 \problemIndex{#1}{#1@\textsc{#1}}{#2}{#3}
}

\newcolumntype{L}{>{\raggedright\arraybackslash}X}
\newcolumntype{R}{>{\raggedleft\arraybackslash}X}

\newcount\n
\def\tablebody{}
\makeatletter
\loop\ifnum\n<100
        \advance\n by1
        \protected@edef\tablebody{\tablebody
                \textbf{\number\n.}& shortText
                \tabularnewline
        }
\repeat

\makeatletter
\let\mcnewpage=\newpage
\newcommand{\TrickSupertabularIntoMulticols}{%
  \renewcommand\newpage{%
    \if@firstcolumn
      \hrule width\linewidth height0pt
      \columnbreak
    \else
      \mcnewpage
    \fi
  }%
}
\makeatother

\subimport{./}{shortcref.tex}

\newcommand{\N}{\mathbb N}
\newcommand{\R}{\mathbb R}

\newcommand{\I}{\set{0, 1}}
\newcommand{\cupdot}{\mathbin{\mathaccent\cdot\cup}} % disjoint union
\newcommand{\set}{\BktC}

\DeclareMathOperator{\lit}{lit}

\newcommand{\where}{:}
\newcommand{\mdoubleplus}{\mathbin{+\mkern-15mu+}}
\newcommand{\const}{\text{const}}

\newcommand{\closesqrt}[1]{\mkern-8mu\mathop{}\sqrt{#1}}

\newcommand{\tupleconcat}{\mdoubleplus}

\newcommand{\AND}{\textsc{And}}
\newcommand{\OR}{\textsc{Or}}

\newcommand{\xor}{\oplus}
\newcommand{\LAND}{\bigwedge}

\newcommand{\aop}{\textsc{And}-\textsc{Or} path}
\newcommand{\AOP}{\textsc{And}-\textsc{Or} Path}
\newcommand{\prgenaopdelayopt}{\textsc{Delay Optimization Problem for Generalized \AOP{}s}}
\newcommand{\prgenaopdepthopt}{\textsc{Depth Optimization Problem for Generalized \AOP{}s}}
\newcommand{\praopdelayopt}{\textsc{Delay Optimization Problem for \AOP{}s}}
\newcommand{\praopdepthopt}{\textsc{Depth Optimization Problem for \AOP{}s}}

\newcommand{\subpath}{sub-path}
\newcommand{\specialsubpath}{special sub-path}

\DeclareMathOperator{\genaopgates}{\Gamma}
\newcommand{\genaopnox}[1]{h(t; \Gamma)_{\widehat{{#1}}}}
\newcommand{\genaopnotx}[1]{h(t; \Gamma)_{\widehat{t_{#1}}}}
\newcommand{\genaopnoti}{h(t; \Gamma)_{\widehat{t_i}}}

\newcommand{\genaopkeepF}[1]{h(t; \Gamma)_{#1}}

\newcommand{\sameins}[1]{S^{#1}}
\newcommand{\diffins}[1]{D^{#1}}

\newcommand{\propgenpart}{segment partition}

\DeclareMathOperator{\PI}{PI}

\DeclareMathOperator{\cktout}{out}
\DeclareMathOperator{\gt}{gt}

\DeclareMathOperator{\depth}{depth}
\DeclareMathOperator{\delay}{delay}
\DeclareMathOperator{\size}{size}
\DeclareMathOperator{\fanout}{fanout}

\newcommand{\totalref}[1]{\cref{#1} (\cpageref{#1})}
\newcount\myloopcounter

\newcommand{\define}[1]{\emph{#1}}

\newcommand{\printktimes}[2]{%
  \myloopcounter0% initialize the loop counter
  \loop\ifnum\myloopcounter < #1 % Test if the loop counter is < #1
  #2%
  \advance\myloopcounter by 1 %
  \repeat % start again
}

\newcommand{\rmDecorate}[2][]{%
 \ifthenelse{\equal{#1}{p}}{%
   \IfDecimal{#2}{\SI[retain-explicit-plus, group-minimum-digits=3]{#2}{\percent}}{#2}%
 }{%
   \IfDecimal{#2}{\num[retain-explicit-plus, group-minimum-digits=3]{#2}}{#2}%
 }
}

\newcommand{\dashrule}[1][black]{%
  \color{#1}\rule[\dimexpr.5ex-.2pt]{4pt}{.4pt}\xleaders\hbox{\rule{4pt}{0pt}\rule[\dimexpr.5ex-.2pt]{4pt}{.4pt}}\hfill\kern0pt%
}

\usepackage{tikz}
\usetikzlibrary{arrows}
\usetikzlibrary{calc}
\usetikzlibrary{shapes}

\usetikzlibrary{positioning}
\usetikzlibrary{shapes.misc}
\usetikzlibrary{fit}
\usetikzlibrary{quotes}
\usetikzlibrary{angles}
\usetikzlibrary{through}
\usetikzlibrary{decorations.pathmorphing}
\usetikzlibrary{shapes.gates.logic.US,shapes.gates.logic.IEC}
\usepackage{pgf}
\usepackage{pgfplots}
\usepgfplotslibrary{colormaps}

\pgfplotsset{compat=1.14}

\colorlet{atcolor}{blue}

\tikzset{and-gate/.style={fill=mid_red,outer sep=0pt, thick, and gate US, draw, rotate=270, text=atcolor}}
\tikzset{or-gate/.style={fill=mid_green,outer sep=0pt, thick, or gate US, draw, rotate=270, text=atcolor}}
\tikzset{nand-gate/.style={fill=mid_red,outer sep=0pt, thick, nand gate US, draw, rotate=270, text=atcolor}}
\tikzset{nor-gate/.style={fill=mid_green,outer sep=0pt, thick, nor gate US, draw, rotate=270, text=atcolor}}
\tikzset{xor-gate/.style={fill=cyan,outer sep=0pt, thick, xor gate US, draw, rotate=270, text=atcolor}}
\tikzset{xnor-gate/.style={fill=cyan,outer sep=0pt, thick, xnor gate US, draw, rotate=270, text=atcolor}}
\tikzset{prefix-gate/.style={outer sep=0pt, circle, scale=2, thick, draw, text=atcolor}}
\tikzset{and3-gate/.style={and-gate, logic gate inputs=nnn}}
\tikzset{and4-gate/.style={and-gate, logic gate inputs=nnnn}}
\tikzset{or3-gate/.style={or-gate, logic gate inputs=nnn}}
\tikzset{or4-gate/.style={or-gate, logic gate inputs=nnnn}}
\tikzset{and5-gate/.style={and-gate, logic gate inputs=nnnnn}}
\tikzset{inv-gate/.style={fill=mid_yellow,outer sep=0pt, thick, not gate US, draw, rotate=270, text=atcolor}}
\tikzset{buf-gate/.style={fill=mid_yellow,outer sep=0pt, thick, buffer gate US, draw, rotate=270, text=atcolor}}
\tikzset{sym-and-gate/.style={fill=mid_yellow,outer sep=0pt, thick, and gate US, draw, rotate=270, text=atcolor}}
\tikzset{sym-or-gate/.style={fill=mid_yellow,outer sep=0pt, thick, or gate US, draw, rotate=270, text=atcolor}}
\tikzset{large-node/.style={scale=1.6}}
\tikzset{uncolored-and-gate/.style={outer sep=0pt, thick, and gate US, draw, rotate=270, text=atcolor}}
\tikzset{uncolored-or-gate/.style={outer sep=0pt, thick, or gate US, draw, rotate=270, text=atcolor}}
\tikzset{concat-and-gate/.style={fill=cyan,outer sep=0pt, thick, and gate US, draw, rotate=270, text=atcolor}}
\tikzset{concat-or-gate/.style={fill=cyan,outer sep=0pt, thick, or gate US, draw, rotate=270, text=atcolor}}
\tikzset{input/.style={scale=1.6}}
\tikzset{smallinput/.style={}}
\tikzset{input-at/.style={atcolor}}
\tikzset{output-at/.style={atcolor}}
\tikzset{output/.style={scale=1.6}}
\tikzset{input/.style={scale=1.6}}
\tikzset{every path/.style={thick, -}}
\tikzset{output-edge/.style={thick, ->}}
\tikzset{marked-edge/.style={very thick, cyan}}
\tikzset{marked-prop/.style={mid_red, draw, thick, scale = 0.9}}
\tikzset{marked-gen/.style={mid_green, draw, thick, scale = 0.9}}

 \definecolor{bn_no_color}{RGB}{0,0,0}
 \definecolor{bn_black}{RGB}{0,0,0}
 \definecolor{bn_cyan}{RGB}{0,255,255}
 \definecolor{bn_magenta}{RGB}{255,0,255}
 \definecolor{bn_yellow}{RGB}{255,255,0}
 \definecolor{bn_blue}{RGB}{0,0,255}
 \definecolor{bn_orange}{RGB}{255,179,0}
 \definecolor{bn_white}{RGB}{255,255,255}
 \definecolor{bn_gray83}{RGB}{212,212,212}
 \definecolor{bn_gray67}{RGB}{171,171,171}
 \definecolor{bn_gray50}{RGB}{128,128,128}
 \definecolor{bn_gray33}{RGB}{86,86,86}
 \definecolor{bn_gray17}{RGB}{43,43,43}
 \definecolor{bn_brown}{RGB}{140,70,20}
 \definecolor{bn_light_cyan}{RGB}{128,255,255}
 \definecolor{bn_light_magenta}{RGB}{255,128,255}
 \definecolor{bn_light_yellow}{RGB}{255,255,128}
 \definecolor{bn_light_blue}{RGB}{128,128,255}
 \definecolor{bn_light_red}{RGB}{255,128,128}
 \definecolor{bn_light_green}{RGB}{128,255,128}
 \definecolor{bn_light_orange}{RGB}{255,210,0}
 \definecolor{bn_light_brown}{RGB}{190,160,138}
 \definecolor{bn_dark_cyan}{RGB}{0,128,128}
 \definecolor{bn_dark_magenta}{RGB}{128,0,128}
 \definecolor{bn_dark_yellow}{RGB}{128,128,0}
 \definecolor{bn_dark_blue}{RGB}{0,0,128}
 \definecolor{bn_dark_red}{RGB}{128,0,0}
 \definecolor{bn_dark_green}{RGB}{0,128,0}
 \definecolor{bn_dark_orange}{RGB}{204,143,0}
 \definecolor{bn_dark_brown}{RGB}{70,35,10}

\begin{document}

\begin{frontmatter}

\journal{Discrete Applied Mathematics}

\title{Constructing Depth-Optimum Circuits for Adders and \AOP{}s}
\author{Ulrich Brenner$^*$, Anna Hermann, Jannik Silvanus}
\address{Research Institute for Discrete Mathematics, University of Bonn \\ Lennéstr. 2, 53113 Bonn, Germany}
\ead{\{brenner,hermann,silvanus\}@dm.uni-bonn.de}
\date{09.12.2020}
\cortext[mycorrespondingauthor]{Corresponding author: Ulrich Brenner}

\begin{abstract}
We examine the fundamental problem of constructing depth-optimum circuits for binary addition.
More precisely, as in literature, we consider the following problem:
Given auxiliary inputs $t_0, \dotsc, t_{m-1}$,
the so-called generate and propagate signals,
construct a depth-optimum circuit over the basis $\set{\AND2, \OR{}2} $
computing all $n$ carry bits of an $n$-bit adder, where $m=2n-1$.
In fact, carry bits are \aop{}s, i.e.,
Boolean functions of the form
${t_0 \lor \bkt{ t_1 \land \bkt{t_2 \lor \bkt{ \dots t_{m-1}} \dots }}}$.
Classical approaches construct so-called prefix circuits
which do not achieve a competitive depth.
For instance, the popular construction by \citet{KS73}
is only a $2$-approximation.
A lower bound on the depth of any prefix circuit
is $1.44 \log_2 m + \const$,
while recent non-prefix circuits
have a depth of $\log_2 m + \log_2 \log_2 m + \const$.
However, it is unknown whether any of these polynomial-time approaches achieves the optimum depth
for all $m \in \N$.

We present a new exponential-time algorithm solving the problem optimally.
The previously best exact algorithm by \citet{Hegerfeld} with a running time of $\mathcal O(2.45^m)$
is viable only for $m\leq29$.
Our algorithm is significantly faster:
We achieve a theoretical running time of $\mathcal O(2.02^m)$ and apply
sophisticated pruning strategies to improve practical running times dramatically.
This allows us to compute optimum circuits for all $m \leq 64$.
Combining these computational results with new theoretical insights, we derive the optimum depths of
$2^k$-bit adder circuits for all $k \leq 13$, previously known only for $k \leq 4$.

In fact, we solve a more general problem,
namely delay optimization of generalized \aop{}s,
which originates from late-stage logic optimization in VLSI design.
Delay is a natural extension of circuit depth to
prescribed input arrival times;
and generalized \aop{}s are a generalization of \aop{}s
where \AND{} and \OR{} do not necessarily alternate.
Our algorithm arises from our new structure theorem which
characterizes delay-optimum generalized \aop{} circuits.

\end{abstract}
           % abstract

\begin{keyword} % max. 6
\MSC[2020]
68R01 \sep % General topics of discrete mathematics in relation to computer science
68W35 \sep % Hardware implementations of nonnumerical algorithms (VLSI algorithms, etc.)
94C99 \sep % None of the above, but in this section (Circuits, networks)
adder circuits \sep And-Or paths \sep depth optimization
\end{keyword}

\end{frontmatter}

\section{Introduction} \label{sec::intro}

In this work, we construct fast circuits for binary addition
and for related Boolean functions, so-called \aop{}s.
An \aop{} is a function of the form
$t_0 \lor \bkt{ t_1 \land \bkt{t_2 \lor \bkt{ \dots t_{m-1}} \dots }}$
for some $m \in \N$;
and a circuit for a Boolean function is a graph-based model for the computation of
the function via elementary building blocks (called gates) on a computer chip.
Here, we use \AND{}2 and \OR{}2 as elementary building blocks,
i.e., logical \AND{} and \OR{} with two inputs each.
Motivated from VLSI design, our objective function is
circuit delay, a generalization of circuit depth to prescribed input arrival times
$a(t_i) \in \N$ for each input $t_i$.
The delay of a circuit is the maximum delay of any input $t_i$,
i.e., $a(t_i)$ plus the maximum length of any directed path
in the circuit starting in $t_i$.
In particular, when $a \equiv 0$,
circuit delay is actually circuit depth,
i.e., the maximum length of any directed path in the circuit.
Given a specific \aop{} with input arrival times,
we want to find a delay-optimum circuit for this Boolean function
using only $\AND2$ and $\OR2$ gates.
Important secondary objective functions include circuit size (i.e., number of gates)
and fanout (i.e., number of successors of a gate).

\aop{}s occur as carry bits in the computation of adder circuits:
Assume we compute the sum
of two $n$-bit binary numbers
$\sum\limits_{i=0}^{n-1}a_i 2^i$ and $\sum\limits_{i=0}^{n-1}b_i 2^i$.
A circuit for this task can be constructed via \define{carry bits}
which are defined recursively by
 $c_0 = 0$
and
$c_{i+1} = g_i \lor \Bkt{p_i \land c_i}$
for $0 \leq i \leq n-1$,
where
$g_i = a_i \land b_i$ and $p_i = a_i \xor b_i $,
see, e.g., \citet{WeinbergerSmith} or \citet{Knowles}.
Using the carry bits, the sum $\sum\limits_{i=0}^{n}s_i 2^i$ can be computed via
$s_i = c_i \xor p_i$
for $i \in \set{0, \dotsc, n-1}$
and
$s_n = c_n$.

\begin{figure}
 \begin{subfigure}{.46\textwidth}
 \vspace{-.cm}
 \centering{
  \adjustbox{max width=.6\textwidth}{
  \centering{\begin{tikzpicture}

\node[and-gate] at (4.6,2) (and1){};
\node[or-gate] at (3.6,1) (or1){};

\node[and-gate] at (2.6,0) (and2){};
\node[or-gate] at (1.6,-1) (or2){};

\node[input] at (6.5, 3.2){$p_0$};
\node[input] at (5.5, 3.2){$g_0$};
\node[input] at (4.5, 3.2){$p_1$};
\node[input] at (3.5, 3.2){$g_1$};
\node[input] at (2.5, 3.2){$p_2$};
\node[input] at (1.5, 3.2){$g_2$};

\node[output] (c_1) at (5.5, -2.5){$c_1$};
\node[output] (c_2) at (3.6, -2.5){$c_2$};
\node[output] (c_3) at (1.6, -2.5){$c_3$};

\draw[thick] (and1.output) -- (or1.input 1);
\draw[thick] (and2.output) -- (or2.input 1);
\draw[thick] (or1.output) -- (and2.input 1);

\draw[thick] (1.5, 3) -- (or2.input 2);
\draw[thick] (2.5, 3) -- (and2.input 2);
\draw[thick] (3.5, 3) -- (or1.input 2);
\draw[thick] (4.5, 3) -- (and1.input 2);
\draw[thick] (5.5, 3) -- (and1.input 1);

\draw[thick, ->] (or2.output) -- (c_3);
\draw[thick, ->] (or1.output) -- (c_2);
\draw[thick, ->] (5.5, 3) -- (c_1);

\end{tikzpicture}}
  }
  }
  \caption{A size-optimum adder circuit.
         All carry bits can be read off
         from a single \aop{}.}
 \label{fig-intro-adders-1}
 \end{subfigure}
 \hfill
 \begin{subfigure}{.46\textwidth}
  \centering{
  \adjustbox{max width=.6\textwidth}{
  \centering{\begin{tikzpicture}

\node[input] (i5) at (1.5, 3.2){$g_2$};
\node[input] (i4) at (2.5, 3.2){$p_2$};
\node[input] (i3) at (3.5, 3.2){$g_1$};
\node[input] (i2) at (4.5, 3.2){$p_1$};
\node[input] (i1) at (5.5, 3.2){$g_0$};
\node[input] (i0) at (6.5, 3.2){$p_0$};

\node[and-gate] at (3,2) (and1){};
\node[and-gate] at (3.8,2) (and2){};
\node[or-gate] at (2.5,1) (or1){};
\node[and-gate] at (4.2,1) (and3){};
\node[or-gate] at (3.5,0) (or2){};

\draw (i3) -- (and1.input 1);
\draw (i4) -- (and1.input 2);
\draw (i2) -- (and2.input 1);
\draw (i4) -- (and2.input 2);
\draw (and1.output) -- (or1.input 1);
\draw (i5) -- (or1.input 2);
\draw (and2.output)   --  (and3.input 2);
\draw (i1) -- (and3.input 1);
\draw (and3.output) -- (or2.input 1);
\draw (or1.output)   -- (or2.input 2);

\node[and-gate] at (6.1,2) (and4){};
\draw (i1) -- (and4.input 1);
\draw (i2) -- (and4.input 2);
\node[or-gate] at (5.1,1) (or3){};
\draw (and4.output) -- (or3.input 1);
\draw (i3) -- (or3.input 2);

\node[output] at ($(or2.output) - (0, 0.7)$) (c_3) {$c_3$};
\node[output] at ($(c_3) + (2, 0)$) (c_1) {$c_1$};
\node[output] at ($(c_1)!0.5!(c_3)$) (c_2) {$c_2$};
\draw[->] (or2.output) -- (c_3);
\draw[->] (i1) -- (c_1);
\draw[->] (or3.output) -- (c_2);

\phantom{\node[or-gate] at (1.6,-1) (or2){};}
\phantom{\node[output] (c_1) at (5.5, -2.5){$c_1$};}
\phantom{\node[output] (c_2) at (3.6, -2.2){$c_2$};}
\phantom{\node[output] (c_3) at (1.6, -2.5){$c_3$};}
\end{tikzpicture}}
  }
  }
  \caption{A depth-optimum adder circuit.
           For each carry bit,
           a depth-optimum \aop{} circuits is constructed separately.}
  \label{fig-intro-adders-2}
 \end{subfigure}
 \caption{Two adder circuits for $3$-bit binary numbers. \AND{} gates are colored red, \OR{} gates green.}
 \label{fig-intro-adders}
\end{figure}
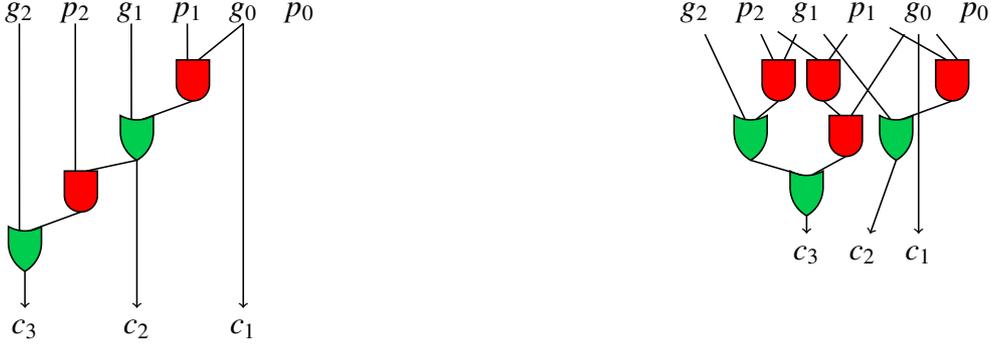

The computation of all $g_i$ and $p_i$ as well as the computation of the sum from the carry bits
only requires a constant depth and a linear number of gates.
Therefore, we call a circuit computing all the \aop{}s
\begin{align}
\begin{split}
 c_{i+1} = g_i \lor \Bkt{p_i \land c_i}
         &= g_i \lor \Bkt{p_i \land \Bkt{g_{i-1} \lor \Bkt{p_{i-1} \land c_{i-1}}}} \\
         &= g_i \lor \Bkt{p_i \land \Bkt{g_{i-1} \lor \Bkt{p_{i-1} \land \Bkt{g_{i-2} \lor \Bkt{p_{i-2} \land \dotsc \Bkt{p_1 \land g_0}}}}}} \label{carry-aop}
\end{split}
\end{align}
for $0 \leq i \leq n-1$ an \define{adder circuit}.
A delay-optimum adder circuit may hence be obtained
by separately computing all carry bits via an \aop{}.
On the other hand, an $n$-bit adder circuit in particular computes the last carry bit $c_n$, so
it contains a circuit for the \aop{} with $2n - 1$ inputs of the form $g_i$ and $p_i$.
Therefore, one can show that, up to a small constant,
the problem of constructing circuits for \aop{}s with minimum depth
can be reduced to the problem of constructing circuits for addition with minimum depth.
In \cref{fig-intro-adders}, we depict two logically equivalent adder circuits
for $n = 3$ bits.

Hence, in the following, we will focus on the construction of fast circuits for \aop{}s.
In fact, we consider generalized \aop{}s, i.e., a generalization of \aop{}s
where \AND{} and \OR{} do not necessarily alternate.
We will see that this more general problem has a rich structure which we will exploit
for our new results.
To the best of our knowledge, we are the first to directly consider this generalized problem.
Delay optimization of generalized \aop{}s can be applied to optimize the delay
of critical combinatorial paths in VLSI design,
but existing approaches (see, e.g., \citet{WerberBonnLogic})
use a simple reduction to \aop{}s which leads to sub-optimal solutions.

% In practice, one would of course regard secondary objective functions like
% circuit size or fanout
% and thus would not construct an adder circuit
% via separate carry bits.

\subsection{Previous Work} \label{sec-prev}

We now review previous results on adder and \aop{} optimization.
Recall from \cref{carry-aop} that an $n$-bit adder
can be obtained from \aop{}s on $1, 3, \dotsc, 2n-1$ inputs,
so the optimum depth of an $n$-bit adder equals the
optimum depth of an $m$-input \aop{} with $m = 2n-1$, i.e., $n = \frac{m+1}{2}$.
Depth bounds for classical adder constructions
are given in terms of $n$ instead of $m$.

Some of the following results only apply to depth optimization,
some also to delay optimization for general arrival times.
For general arrival times,
a lower bound on the delay of a Boolean circuit
on inputs $t_0, \dotsc, t_{m-1}$
is given by $\lceil\log_2 W\rceil$
due to \citet{Golumbic},
where $W:= \sum_{i = 0}^{t-1} 2^{a(t_i)}$.
Note that for $a \equiv 0$, we have $W = m$.
In the subsequent delay bounds, $W$ can also be replaced by $m$
to obtain the corresponding depth bounds.

Depth optimization of adder circuits is a classical and well-studied problem.
Many researchers construct adder circuits
via so-called \define{prefix gates},
e.g.\ % sorted by appearance
\citet{Sklansky60}, % opt. prefix graph, size n log n, fanout n
\citet{KS73}, % opt. prefix graph, size n log n, fanout 2
\citet{LF80}, % tradeoff: prefix depth log n + k, size 2(1 + 2^{-k}n, fanout >= n 2^{-k-1} + 1
or \citet{Royetal2014,Royetal2015}. % depth-opt prefix adder with enumeration for improving size / fanout and thus also practical delay
% \citet{Choi}. % opt prefix adder, but size n^2
% \citet{BrentKung82}, % 2-apx prefix graph, size n, fanout 2
Though, for $n$ bits, these circuits have an optimum depth of $\bktU{\log_2 n}$ in terms
of prefix gates, they have a depth of $2 \bktU{\log_2 n}$ over the basis
$\set{\AND{}2, \OR{}2}$ since each prefix gate has to be realized by
a circuit of depth 2.
% \citet{RWS07}. % prefix adder with delay 2 log W +...
% RWS07: prefix-based aop with delay 1.441 log W +...
\newcommand{\heldspirklfactor}{\mu}
Based on \citet{Werber-Fib-AOPs,RWS07},
\citet{Held-Spirkl-AOPs}
directly optimize the \AND{}-\OR{} delay of their prefix adders
and obtain a delay guarantee
of $\heldspirklfactor \log_2 W + \const$ over $\set{\AND{}2, \OR{}2}$, where $1.44 \leq \heldspirklfactor \leq 1.441$.
However, \citet{Held-Spirkl-AOPs} proved a lower bound of
$\heldspirklfactor \log_2 n - 1$
on the logic gate depth of {\it any} prefix-based adder circuit.
Thus, further progress is only possible
with adders that are not based on prefix gates.

Using non-prefix circuits,
\citet{brent1970addition}, % non-prefix, depth log n + O(sqrt(log n)), size n (single carry bit)
\citet{Khrapchenko-construction}, % non-prefix, depth log n + O(sqrt(log n)), size n
\citet{SpirklMaster}, and
\citet{HSAdders} % non-prefix, log n + O(sqrt(log n)), size n
achieve a guaranteed delay of the form
$\log_2 W + \const \cdot \closesqrt{\log_2 m}$ for \aop{}s on $m$ inputs.
For arbitrary arrival times, the best known delay guarantee of
$\log_2 W + \log_2 \log_2 m + \log_2 \log_2 \log_2 m + 4.3$
was first obtained by \citet{BH-Theory-AOPs};
and for depth optimization,
the best known depth bound of $\log_2 m + \log_2 \log_2 m + 1.58$
is due to \citet{thesis_hermann},
based on \citet{Grinchuk}.

On the other hand, there is a constant $c$ such that
for depth optimization,
a lower bound of $\log_2 m + \log_2 \log_2 m + c$ holds for sufficiently large $m$,
which was shown by \citet{Commentz-Walter-mon}.
\citet{Hitzschke2018} showed that if $c$ is chosen as $c = -5.02$,
then the lower bound holds for $m \geq 2^{2^{18}}$.
\citet{Commentz-Walter-mon} obtains the lower bound via structural
insights on \aop{}s that are the basis for our structure theorem, see \cref{sec-structure}.
%Hitzschke betrachtet n Inputpaare, also gilt m = 2n.
%Fuer n=2^{2^18}, also m=2^{2^18 + 1} ist seine untere
%Schranke 2^18 + 13.99. Es gilt dann
%log m + log log m + c = 2^18 + 1 + 18 + eps + c = 2^18 + 19 + eps + c.
%Dieses eps ist viel kleiner als 0.01, also kann man c =5.02 setzen.

Thus, \citet{Grinchuk} and \citet{thesis_hermann} construct
depth-optimum circuits up to an additive constant,
and the delay-optimization algorithm by \citet{BH-Theory-AOPs}
is, applied for the special case of depth optimization,
optimum up to an additive term of $\mathcal O(\log_2 \log_2 \log_2 m)$.
However, the difference of these upper bounds to the lower bound
is still substantial,
leading to sub-optimal results,
in particular on small instances as they occur in practice in VLSI design.

In order to obtain good empirical results, the algorithm by \citet{BH-practice}
combines the ideas from these theoretical constructions with practical improvements and
generalizations.
Although no better worst-case bound can be shown,
the obtained circuits mostly have better -- often optimal -- delay.
However, there are instances with only $6$ inputs
for which the algorithm does not find an optimum solution,
see Section 6.3 in \citet{thesis_hermann}.
Still, regarding depth optimization,
all instances with known optimum depth
that can be solved by this algorithm
are indeed solved optimally,
and it is open whether this construction is always optimum in the depth case.
There is also a depth-optimization heuristic by \citet{grinchuk2013low}
which might actually be an exact algorithm
(cf.~the depths displayed in the table in Section 5 of \citep{grinchuk2013low}).

There are three previously known exact algorithms for depth optimization of \aop{}s.

Apart from the aforementioned heuristic,
\citet{grinchuk2013low} also provides an exact algorithm for depth optimization
of \aop{}s with a running time of $\Omega(4^m)$.
No explicit empirical results are given, but it is mentioned that the algorithm can only be used for up to 20 or 30 inputs.
The idea of this algorithm is to compute the optimum achievable depth
for all monotone Boolean functions on $m$ inputs in a bottom-up dynamic programming fashion.
Each Boolean function is identified by its truth table, and circuits of larger depth
are obtained by pairwise combinations of existing circuits with \AND{} or \OR{} gates.
Naively, the  dynamic-programming table thus would have $2^{2^m}$ entries.
Grinchuk's main contribution is the observation that a truth table
of size $m$ -- called a \textit{passport} in \citep{grinchuk2013low} --
suffices to identify a monotone \aop{} circuit.
This way, the size of the  dynamic-programming table is reduced to $2^m$,
which implies a running time of $\Omega(4^m)$ to compute all table entries
and hence a depth-optimum \aop{} circuit.

\citet{Hegerfeld} proposes two enumeration algorithms constructing
depth-optimum \aop{} circuits.

In a first algorithm, Hegerfeld constructs all circuits that
are size-optimum among all depth-optimum \aop{} circuits.
This algorithm is based on tree enumeration and is viable for up to $19$ inputs.
The algorithm can also be used to enumerate \aop{} circuits
with non-optimum depth with an increase in running time,
which leads to optimum solutions with respect to delay for certain arrival time profiles.

Hegerfeld's second algorithm is much faster, but is restricted to depth-optimum formula circuits
(i.e., circuits where each gate has fanout~1)
with a certain size guarantee (cf.~\cref{sec::opt-comp}).
It has a provable
running time of $\mathcal O\Bkt{2.45^m}$
and can be applied for up to $29$ inputs.
Hegerfeld does not enumerate formula circuits for \aop{}s directly,
but so-called \define{rectangle-good protocol trees} for \define{Karchmer-Wigderson games}
(see \citet{karchmer1990monotone}) for \aop{}s,
which come from the area of communication complexity.
From these, Hegerfeld derives the optimum formula circuits.
This is the fastest previously known exact algorithm for depth optimization of \aop{}s.

\subsection{Our Contributions}
% generalized

In this work,
we present a new exact algorithm for constructing delay-optimum generalized \aop{} circuits.
In the most prominent special case of depth optimization of \aop{} circuits (and thereby binary adders),
our algorithm is significantly faster than previous approaches, both in theory and in practice.
For the general problem, which occurs in late-stage timing optimization in VLSI design,
we obtain the first known non-trivial exact algorithm.

% structure theorem
We prove a new structure theorem
which characterizes the structure of specific delay-optimum circuits for generalized \aop{}s.
More precisely, we show how optimum solutions for generalized \aop{}s
can be obtained by combining optimum circuits for certain smaller generalized \aop{}s in a recursive fashion,
directly motivating a dynamic programming algorithm.
We stress that an analogous statement does not hold for non-generalized \aop{}s, that is,
generalized \aop{}s also occur as sub-functions of delay-optimum circuits for non-generalized \aop{}s.

% general
In the general case, the running time of our new algorithm is at most $\mathcal O(3^m)$.
For \aop{}s, the bound is improved to $\mathcal O(2.45^m)$.

% depth AOP
For the special case of depth optimization of \aop{}s,
our algorithm has a running time of $\mathcal O(2.02^m)$,
significantly improving upon the previously best running time of $\mathcal O(2.45^m)$
of the formula enumeration algorithm by \citet{Hegerfeld}.
Hegerfeld computes a depth-optimum formula circuit
with a certain size guarantee (cf.~\cref{sec::opt-comp}).
Our algorithm can either compute such a circuit
or an arbitrary delay-optimum circuit,
which can be done much faster.
% runtime comparison
In contrast to Hegerfeld,
in practice, we apply very efficient pruning techniques
that drastically reduce empirical running times.
The largest instance solved by Hegerfeld has $29$ inputs,
while our algorithm with size optimization can solve instances with up to $42$ inputs;
without size optimization even up to $64$ inputs.
Our running times on $26$ inputs are $2.1$ seconds with size optimization
and $0.007$ seconds without size optimization,
while Hegerfeld's running time is $17$ hours.
Our largest running time without size optimization on any of these instances is
roughly $2.7$ hours.

% opt adder depths
From our structure theorem, our computations and
the results computed by the heuristic \aop{} optimization algorithm by \citet{grinchuk2013low},
we deduce the optimum depths of
adder circuits
-- i.e., circuits computing all the carry bits from the $g_i$- and $p_i$-signals
as in \cref{carry-aop} --
over the basis $\set{\AND{}2, \OR{}2}$ on $2^k$ bits, where $k \leq 13$.
To the best of our knowledge, we are the first to obtain such a result.

% paper structure
The rest of the paper is organized as follows:
In \cref{sec-basics}, we formally introduce the problem and basic concepts.
In \cref{sec-structure}, we present and prove our structure theorem.
From this, in \cref{sec-opt-alg}, we derive our exact algorithm,
which is refined for the special case of depth optimization of \aop{}s
in \cref{sec-opt-alg-depth}.
Practical speedups are presented in \cref{sec::opt-impl}.
In \cref{sec::opt-comp}, we show computational results,
i.e., our practical running times and the computed optimum adder depths.

              % 1

\section{Preliminaries} \label{sec-basics}

\subsection{Boolean Functions and Circuits}

Our notation regarding Boolean functions and circuits is based on
\citet{Savage}.
We denote the set of natural numbers including $0$ by $\N$.
For an $n$-tuple $\Bkt{x_0, \dotsc, x_{n-1}}$ and an index $i \in \set{0, \dotsc, n-1}$,
we use the standard notation $\bkt{x_0, \dotsc, \widehat{x_i}, \dotsc, x_{n-1}}$
to denote the $(n-1)$-tuple arising from $x$ by deleting the entry $x_i$.

For us, a \define{Boolean function} is a function
$f \colon \I^m \to \I$ for some $m \in \N$.
We often write $t = \Bkt{t_0, \dotsc, t_{m-1}}$ for
the \define{input variables}, short \define{inputs}, of $f$.
For an input variable $t_i$ of $f$ and a value $\alpha \in \I$,
the \define{restriction} of $f$ to $t_i = \alpha$
is the function $f \mid_{t_i = \alpha} \colon \I^{m-1} \to \I,
(t_0, \dotsc, \widehat{t_i}, \dotsc, t_{m-1}) \mapsto f((t_0, \dotsc, t_{i-1}, \alpha, t_{i+1}, \dotsc, t_{m-1}))$.
An input $t_i$ is \define{essential} for $f$ if
$f\mid_{t_i = 0} \neq f\mid_{t_i = 1}$.

In this work, a \define{circuit} $C$ is a connected acyclic digraph
whose vertices can be partitioned into two sets:
\define{inputs} with no incoming edges that each represent a Boolean variable
or a constant $0$ or $1$, and
\define{gates} with exactly $2$ incoming edges
that each represent an elementary Boolean function among logical \AND{} and \OR{}.
By $\gt(g) \in \set{\AND, \OR} $, we denote the \define{gate type} of a gate $g$,
i.e., its associated Boolean function.
There is a subset of the vertices called \define{outputs}.
All vertices with no outgoing edges are outputs, but there might be others.
If $C$ contains only a single output, we denote it by $\cktout(C)$.
The \define{Boolean function computed at a vertex} of a circuit $C$
can be read off recursively by combining the logical functions
represented by the gates.
If $C$ has a single output which computes a Boolean function $f$,
we also say that $C$ \define{computes} or \define{realizes} $f$
and write $f = f(C)$ for the function computing $C$.

\begin{figure}[!ht]
  \newcommand{\vspacebeforesubcap}{-.4cm}
  \begin{center}
  \begin{subfigure}{0.32\textwidth}
  \begin{center}
  \adjustbox{max width=0.8\textwidth}{
  \centering{\begin{tikzpicture}

\node[or-gate,label={[atcolor]south:2}] at (5,2) (or1){};
\node[and-gate,label={[atcolor]south:3}] at (4,1) (and1){};

\node[or-gate,label={[atcolor]south:4}] at (3,0) (or4){};
\node[and-gate,label={[atcolor]south:5}] at (2,-1) (and2){};

\node[input-at] (ai5) at (1.5, 3.9){$4$};
\node[input-at] (ai4) at (2.5, 3.9){$2$};
\node[input-at] (ai3) at (3.5, 3.9){$1$};
\node[input-at] (ai2) at (4.5, 3.9){$0$};
\node[input-at] (ai1) at (5.5, 3.9){$1$};

\node[input] (i5) at (1.5, 3.2){$t_0$};
\node[input] (i4) at (2.5, 3.2){$t_1$};
\node[input] (i3) at (3.5, 3.2){$t_2$};
\node[input] (i2) at (4.5, 3.2){$t_3$};
\node[input] (i1) at (5.5, 3.2){$t_4$};

\draw (or1.output) -- (and1.input 1);
\draw (i2) -- (or1.input 2);
\draw (i3) -- (and1.input 2);
\draw (i1) -- (or1.input 1);

\draw (or4.output) -- (and2.input 1);

\draw (i5) -- (and2.input 2);

\draw (i4) -- (or4.input 2);
\draw (and1.output) -- (or4.input 1);

\node[output-at] (output) at (2, -2) {};
\draw[output-edge] (and2.output) -- (output) {};

\end{tikzpicture}}
  }
  \end{center}
   \vspace{\vspacebeforesubcap}
  \caption{Circuit $C_1$ with $\depth(C_1) = 4$ and $\delay(C_1) = 5$.}
  \label{ex-aop-delay-1}
 \end{subfigure}
 \hfill
 \begin{subfigure}{0.32\textwidth}
 \begin{center}
   \adjustbox{max width=0.8\textwidth}{
  \centering{\begin{tikzpicture}

\node[input-at] (ai5) at (1.5, 3.9){$4$};
\node[input-at] (ai4) at (2.5, 3.9){$2$};
\node[input-at] (ai3) at (3.5, 3.9){$1$};
\node[input-at] (ai2) at (4.5, 3.9){$0$};
\node[input-at] (ai1) at (5.5, 3.9){$1$};

\node[input] (i5) at (1.5, 3.2){$t_0$};
\node[input] (i4) at (2.5, 3.2){$t_1$};
\node[input] (i3) at (3.5, 3.2){$t_2$};
\node[input] (i2) at (4.5, 3.2){$t_3$};
\node[input] (i1) at (5.5, 3.2){$t_4$};

\node[or-gate,label={[atcolor]south:3}] at (3,2) (or1){};
\draw (i3) -- (or1.input 1);
\draw (i4) -- (or1.input 2);

\node[or-gate,label={[atcolor]south:3}] at (4.2,2) (or2){};
\draw (i2) -- (or2.input 1);
\draw (i4) -- (or2.input 2);

\node[or-gate,label={[atcolor]south:4}] at (4.5,1) (or3){};
\draw (or2.output)   --  (or3.input 2);
\draw (i1) -- (or3.input 1);

\node[and-gate,label={[atcolor]south:5}] at (2.5,1) (and1){};
\draw (or1.output) -- (and1.input 1);
\draw (i5) -- (and1.input 2);

\node[and-gate,label={[atcolor]south:6}] at (3.5,0) (and2){};
\draw (or3.output) -- (and2.input 1);
\draw (and1.output)   -- (and2.input 2);

\node[output-at] (output) at (3.5, -1) {};
\draw[output-edge] (and2.output) -- (output) {};

\phantom{\node[and-gate,label={[atcolor]south:5}] at (2,-1) (and2){};}
\phantom{\node[output-at] (output) at (2, -2) {};}
\phantom{\draw[output-edge] (and2.output) -- (output) {};}

\end{tikzpicture}}
  }
  \end{center}
   \vspace{\vspacebeforesubcap}
   \caption{Circuit $C_2$ with $\depth(C_2) = 3$ and $\delay(C_2) = 6$.}
  \label{ex-aop-delay-2}
  \end{subfigure}
 \hfill
 \begin{subfigure}{0.32\textwidth}
 \begin{center}
   \adjustbox{max width=0.8\textwidth}{
  \centering{\begin{tikzpicture}

\node[input-at] (ai5) at (1.5, 3.9){$4$};
\node[input-at] (ai4) at (2.5, 3.9){$2$};
\node[input-at] (ai3) at (3.5, 3.9){$1$};
\node[input-at] (ai2) at (4.5, 3.9){$0$};
\node[input-at] (ai1) at (5.5, 3.9){$1$};

\node[input] (i5) at (1.5, 3.2){$t_0$};
\node[input] (i4) at (2.5, 3.2){$t_1$};
\node[input] (i3) at (3.5, 3.2){$t_2$};
\node[input] (i2) at (4.5, 3.2){$t_3$};
\node[input] (i1) at (5.5, 3.2){$t_4$};

\node[or-gate,label={[atcolor]south:2}] at (5,2) (or1){};
\draw (i1) -- (or1.input 1);
\draw (i2) -- (or1.input 2);

\node[and-gate,label={[atcolor]south:5}] at (3.5,2) (and2){};
\draw (i3) -- (and2.input 1);
\draw (i5) -- (and2.input 2);

\node[and-gate,label={[atcolor]south:5}] at (2,2) (and3){};
\draw (i4) -- (and3.input 1);
\draw (i5) -- (and3.input 2);

\node[and-gate,label={[atcolor]south:6}] at (4.25,1) (and4){};
\draw (or1.output) -- (and4.input 1);
\draw (and2.output) -- (and4.input 2);

\node[or-gate,label={[atcolor]south:7}] at (3,0) (or5){};
\draw (and4.output) -- (or5.input 1);
\draw (and3.output) -- (or5.input 2);

\draw[output-edge] (or5.output) -- ($(or5.output) - (0, .5)$) {};

\phantom{\node[and-gate,label={[atcolor]south:5}] at (2,-1) (and2){};}
\phantom{\node[output-at] (output) at (2, -2) {};}
\phantom{\draw[output-edge] (and2.output) -- (output) {};}
\end{tikzpicture}}
  }
  \end{center}
   \vspace{\vspacebeforesubcap}
   \caption{Circuit $C_3$ with $\depth(C_2) = 3$ and $\delay(C_2) = 7$.}
  \label{ex-aop-delay-3}
  \end{subfigure}
  \end{center}
  \vspace{-0.5cm}
 \caption{Three circuits with given input arrival times realizing the \aop{} $t_0 \land \Bkt{t_1 \lor \Bkt{t_2 \land \Bkt{t_3 \lor t_4}}}$.}
 \label{ex-aop-delay}
\end{figure}
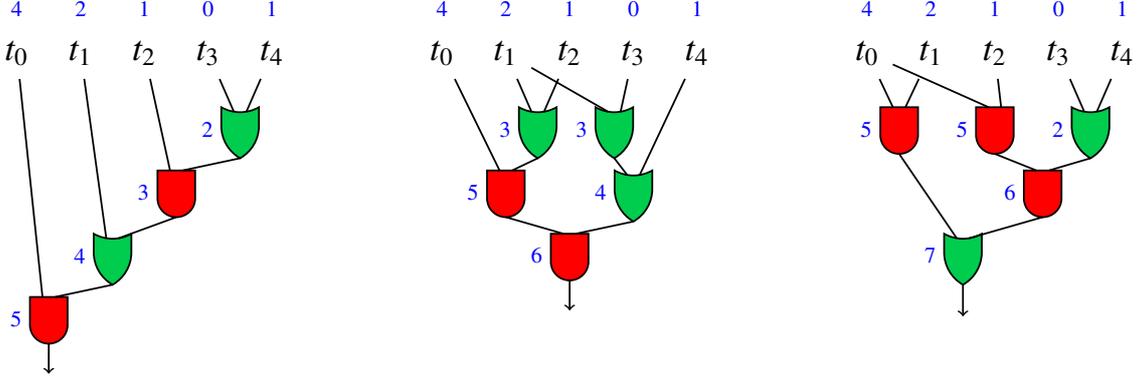

Given a Boolean function $f$, there are numerous circuits realizing $f$.
In order to evaluate the quality of different circuits, we introduce
the following measures:
By $\size(C)$, we denote the \define{size} of a circuit $C$, i.e., its number of gates.
The \define{fanout} of a vertex $v$ of $C$ is defined
by $\fanout(v) := |\delta^+(v)|$.
When each input $t_i$ is associated with an \define{arrival time} $a(t_i) \in \R$
at which the signal $t_i$ is available,
then the \define{delay} of $C$ is
\[\delay(C) := \max_{i \in \set{0, \dotsc, m-1}} \set{a(t_i) + \max_{P} \set{|P|}},\]
where the innermost maximum ranges over all directed paths $P$ in $C$
starting in $t_i$.
Note that when all arrival times are $0$,
the notion of delay coincides with the notion of \define{circuit depth},
i.e.\ the length of a longest directed path in $C$,
and we write $\depth(C)$ instead of $\delay(C)$.
\cref{ex-aop-delay} shows three different circuits computing the same
function with their corresponding depths and delays.
The circuits $C_2$ and $C_3$ are depth-optimum,
while $C_1$ is delay-optimum for the indicated arrival times.
The gates are shown in red ({\sc And} gates) and in green ({\sc Or} gates),
and all edges are directed from top to bottom. The output gate is indicated
by an arrow.

When analyzing the structure of delay-optimum circuits for a certain Boolean function,
prime implicants play an important role.
For a more elaborate introduction, see Section 1.7 of \citet{Crama}.

\begin{definition} \label{def-implicant}
 Let $f \colon \I^m \to \I$ be a Boolean function
 with inputs $t = \Bkt{t_0, \dotsc, t_{m-1}}$.
 A \define{literal} of $f$ is a possibly negated input variable of $f$,
 i.e., $t_i$ or $\overline{t_i}$ for some $i \in \{0, \dotsc, m-1\}$.
 Consider a conjunction
 $\iota\DBkt{t_0, \dotsc, t_{m-1}} = l_{i_1} \land \dotsc \land l_{i_k}$,
 where $l_{i_1}, \dotsc, l_{i_k}$ are literals of $f$.
 We write $\lit(\iota) = \bktCfixed{normal}{l_{i_1}, \dotsc, l_{i_k}}$ for the set of literals of $\iota$.
 We call $\iota$ an \define{implicant} of $f$ if
 for any $\alpha \in \I^m$ with
 $\iota(\alpha ) = 1$,
 we have $f\Bkt{\alpha } = 1$.
 We call $\iota$ a \define{prime implicant} of $f$
 if there is no implicant $\pi$ of $f$ with $\lit(\pi) \subsetneq \lit(\iota)$.
 The set of all prime implicants of $f$ is denoted by $\PI(f)$.
\end{definition}

We need several well-known statements about prime implicants
for which a proof can be found, e.g., in \citet{Crama}.

\begin{qedproposition}[Theorem 1.13 of \citet{Crama}] \label{pi-unique}
 Every Boolean function can be represented by the disjunction of all its prime implicants.
 In particular, a Boolean function $f$ is uniquely determined by $\PI(f)$.
\end{qedproposition}

\begin{qedproposition}[Theorem 1.17 of \citet{Crama}] \label{ess-pi}
 Let $f \colon \I^m \to \I$ be a Boolean function on inputs $t_0, \dotsc, t_{m-1}$.
 Then, $f$ depends essentially on an input $t_i$ if and only if there is a
 prime implicant $\pi$ of $f$
 with $t_i \in \lit(\pi)$
 or $\overline{t_i} \in \lit(\pi)$.
 \end{qedproposition}

Given $\alpha, \beta \in \I^m$, we write $\alpha \leq \beta$
if $\alpha_i \leq \beta_i$ for all $i \in \set{0, \dotsc, m-1}$.
A Boolean function $f \colon \I^m \to \I$ is called \define{monotone}
if for all $\alpha, \beta \in \I^m$ with $\alpha \leq \beta$,
we have $f(\alpha) \leq f(\beta)$.
In this work, we examine a sub-class of the class of monotone functions.

% paragraph break optional here.
For monotone Boolean functions, function decomposition and (prime) implicants behave in a canonical way,
see, e.g., \citet{Commentz-Walter-mon}, proof of Lemma 1.

\begin{lemma}\label{or-implicants}
 Consider Boolean functions $h, f_1, f_2\colon \I^m \to \I$
 on input variables $t_0, \dotsc, t_{m-1}$
 with $h = f_1 \lor f_2$.
 \begin{enumerate}[label=(\alph*)]
 \item Any implicant of $f_1$ or $f_2$ is an implicant of $h$.\label{gen-lb-impl}
 \item If $h, f_1, f_2$ are all monotone,
   then any (prime) implicant of $h$ is a (prime) implicant of $f_1$ or $f_2$. \label{gen-lb-pi}
 \end{enumerate}
 \begin{proof}
  Using the definition of implicants,
  the first statement can be seen easily.

  To show the second statement, assume that $h, f_1, f_2$ are monotone,
  and let $\kappa$ be an implicant of $h$.
  Assume that $\kappa$ is not an implicant of $f_1$ or $f_2$.
  Then, there are $\alpha^{(1)}, \alpha^{(2)} \in \I^m$ with
  \begin{multicols}{2}
  \begin{enumerate}[label=(\roman*)]
  \item $\kappa(\alpha^{(1)}) = \kappa(\alpha^{(2)}) = 1$ and \label{iota true}
  \item $f_1(\alpha^{(1)})     = f_2(\alpha^{(2)})     = 0$. \label{g h false}
  \end{enumerate}
  \end{multicols}
  Define $\alpha \in \I^m$ by $\alpha_i = {\alpha^{(1)}}_i \land {\alpha^{(2)}}_i$ for $i \in \{0, \dotsc, m-1\}$.
  As $\kappa$ is a conjunction of literals, \labelcref{iota true} implies
  $\kappa(\alpha) = 1$.
  Moreover, as $f_1$ and $f_2$ are monotone
  and $\alpha \leq \alpha^{(1)}, \alpha^{(2)}$,
  \labelcref{g h false} implies $f_1(\alpha) = f_2(\alpha) = 0$ and thus
  $h(\alpha) = f_1(\alpha) \lor f_2(\alpha) = 0$,
  which contradicts $\kappa$ being an implicant of $h$.
  Hence, $\kappa$ is an implicant of $f_1$ or $f_2$,
  w.l.o.g.~of~$f_1$.

  Now assume additionally that $\kappa$ is a prime implicant of $h$
  and consider an implicant $\lambda$ of $f_1$ with
  $\lit(\lambda) \subseteq \lit(\kappa)$.
  By the first statement of this \lcnamecref{or-implicants},
  $\lambda$ is an implicant of $h$.
  As $\kappa$ is a prime implicant of $h$, we have $\lambda = \kappa$.
  Thus, $\kappa$ is a prime implicant of $f_1$.
 \end{proof}
\end{lemma}

A \define{formula circuit} is a circuit
where each gate has fanout at most $1$.
It is well known that for any Boolean function~$f$, there
is a formula circuit which has optimum delay
among all circuits for $f$,
see, e.g., \citet{Wegener}, Section 1.4.
Hence, when computing a delay-optimum circuit for $f$,
we may restrict ourselves to formula circuits.

The following lower bound on the delay of general circuits
has first been proven by \citet{Golumbic}.
An alternative proof via Kraft's inequality \citep{Kraft}
is given by \citet{Werber-Fib-AOPs}.

\begin{qedtheorem}[\citet{Golumbic}] \label{thm-general-circuit-lower-bound}
 Consider any Boolean function $f \colon \I^m \to \I$
 on inputs $t_0, \dotsc, t_{m-1}$ with input arrival times $a(t_0), \dotsc, a(t_{m-1}) \in \N$
 that depends essentially on all its inputs.
 Then, any circuit $C$ over $\set{\AND2, \OR2}$ computing $f$ fulfills
 $\delay(C) \geq \bktU{\log_2 W}$, where $W = \sum_{i = 0}^{m-1} 2^{a(t_i)}$.
\end{qedtheorem}

For the special case of depth optimization, \cref{thm-general-circuit-lower-bound} implies
$\depth(C) \geq \bktU{\log_2 m}$
which was first proven by \citet{Winograd}.

For our proofs, we need the concept of reduced circuits.
Consider a circuit $C$ over $\{\AND{}2, \OR{}2\}$
with inputs $t_0, \dotsc, t_{m-1}$ realizing $f$.
The \define{reduced circuit} $C\mid_{t_i = \alpha}$ is a circuit
on inputs $t_0, \dotsc, \widehat{t_i}, \dotsc, t_{m-1}$
realizing the restricted function $f\mid_{t_i = \alpha}$ that arises from replacing $t_i$ by $\alpha$
and then canonically reducing the circuit until it is trivial
or until it does not contain any constants anymore.
More formally, $C\mid_{t_i = \alpha}$ arises from $C$ as follows:
Replace $t_i$ by $\alpha$ and apply the following
to each gate $g$ in topological order:
Assume that there is a predecessor $v \in \delta^-(g)$ which is a constant
(otherwise, do nothing for $g$),
and denote the other predecessor of $g$ by $w$.

\textbf{Case 1:}
Assume that $\gt(g) = \AND{}$ and $v = 0$,
 or $\gt(g) = \OR{}$ and $v = 1$.
 Replace each edge $(g, y) \in \delta^+(g)$ by $(v, y)$.
 If $g$ is an output, then let $v$ be an output.

 \textbf{Case 2:}
Otherwise, replace each edge $(g, y) \in \delta^+(g)$ by $(w, y)$,
 and if $g$ is an output, then let $w$ be an output.

In both cases, remove $g$ from $C$.

At the end, remove all gates from which no output is reachable.

\begin{observation} \label{obs-restr-circuit}
Consider a Boolean function $f \colon \I^m \to \I$
on input variables $t_0, \dotsc, t_{m-1}$ with arrival times
$a(t_0), \dotsc, a(t_{m-1}) \in \N$,
an index $i \in \set{0, \dotsc, m-1}$ and a value $\alpha \in \I$.
Consider a circuit $C$ for $f$ and the reduced circuit $C\mid_{t_i = \alpha}$.
Then, $C\mid_{t_i = \alpha}$ is a circuit for the restricted function $f\mid_{t_i = \alpha}$.
Moreover, we have $\delay(C\mid_{t_i = \alpha}) \leq \delay(C)$ and
$\size(C\mid_{t_i = \alpha}) \leq \size(C)$.
If $\fanout(t_i) > 0$ in $C$,
then we have $\size(C\mid_{t_i = \alpha}) < \size(C)$.
\end{observation}

\subsection{Generalized \AOP{}s}

We now introduce the special class of monotone Boolean functions considered in this work.

\begin{definition} \label{def-gen-aop}
 Let inputs $t = \Bkt{t_0, \dotsc, t_{m-1}}$
 and an $(m-1)$-tuple $\genaopgates = \Bkt{\circ_0, \dotsc, \circ_{m-2}}$
 of gate types $\circ_0, \dotsc, \circ_{m-2} \in \{\AND{}2, \OR{}2\}$
 be given.
 We call a Boolean function of the form
 \begin{align} \label{gen-aop-formula}
  h\Bkt{t; \genaopgates} := t_0 \circ_0 \Bkt{t_1 \circ_1 \Bkt{t_2 \circ_2 \Bkt{ \dotsc \circ_{m-3} \Bkt{t_{m-2} \circ_{m-2} t_{m-1}}}}}
 \end{align}
 a \define{generalized \aop{}}.
 We call the circuit for $h(t; \Gamma)$ arising from \cref{gen-aop-formula}
 the \define{standard circuit}
 for $h(t; \Gamma)$.
\end{definition}

We can now formulate the main problem considered in this work.

\problem{\prgenaopdelayopt}
        {$m \in \N$,
         inputs $t = \Bkt{t_0, \dotsc, t_{m-1}}$,
         gate types $\Gamma = \Bkt{\circ_0, \dotsc, \circ_{m-2}}$,
         arrival times $a(t_0), \dotsc, a(t_{m-1}) \in \N$.}
        {Compute a circuit over $ \set{\AND2, \OR2}$ realizing $h(t; \Gamma)$ with minimum possible delay.}
        {prgenaopdelayopt}

In the special case where all arrival times $a(t_0), \dotsc, a(t_{m-1})$ are identical,
optimizing circuit delay means optimizing circuit depth and we call the
problem \prgenaopdepthopt.

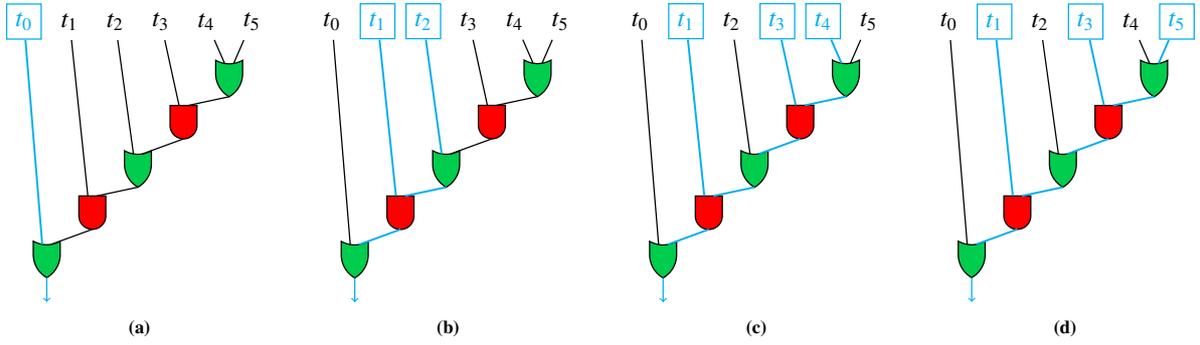
\begin{figure}[ht]
\begin{center}
 \newcommand{\subfigwidth}{0.22\textwidth}
 \begin{subfigure}{\subfigwidth}
  \adjustbox{max width=\textwidth}{
  \centering{\begin{tikzpicture}

\node[input, draw, cyan] (x0) at (-0.5, 3.2){$t_0$};
\node[input] (x1) at (0.5, 3.2){$t_1$};
\node[input] (x2) at (1.5, 3.2){$t_2$};
\node[input] (x3) at (2.5, 3.2){$t_3$};
\node[input] (x4) at (3.5, 3.2){$t_4$};
\node[input] (x5) at (4.5,3.2) {$t_5$};

\node[or-gate] at (4,2) (or1){};
\draw (x5) -- (or1.input 1);
\draw (x4) -- (or1.input 2);

\node[and-gate] at (3,1) (and4){};
\draw (x3) -- (and4.input 2);
\draw (or1.output) -- (and4.input 1);

\node[or-gate] at (2,0) (or2){};
\draw (x2) -- (or2.input 2);
\draw (and4.output) -- (or2.input 1);

\node[and-gate] at (1,-1) (and5){};
\draw (x1) -- (and5.input 2);
\draw (or2.output) -- (and5.input 1);

\node[or-gate] at (0,-2) (or6){};
\draw[marked-edge] (x0) -- (or6.input 2);
\draw (and5.output) -- (or6.input 1);

\draw[->, cyan] (or6.output) -- ($(or6) - (0, 1)$);

\end{tikzpicture}}
  }
  \caption{}
  \label{fig-true-points-1}
 \end{subfigure}
 \quad
 \begin{subfigure}{\subfigwidth}
  \adjustbox{max width=\textwidth}{
  \centering{\begin{tikzpicture}

\node[input] (x0) at (-0.5, 3.2){$t_0$};
\node[input, draw, cyan] (x1) at (0.5, 3.2){$t_1$};
\node[input, draw, cyan] (x2) at (1.5, 3.2){$t_2$};
\node[input] (x3) at (2.5, 3.2){$t_3$};
\node[input] (x4) at (3.5, 3.2){$t_4$};
\node[input] (x5) at (4.5,3.2) {$t_5$};

\node[or-gate] at (4,2) (or1){};
\draw (x5) -- (or1.input 1);
\draw (x4) -- (or1.input 2);

\node[and-gate] at (3,1) (and4){};
\draw (x3) -- (and4.input 2);
\draw (or1.output) -- (and4.input 1);

\node[or-gate] at (2,0) (or2){};
\draw[marked-edge] (x2) -- (or2.input 2);
\draw (and4.output) -- (or2.input 1);

\node[and-gate] at (1,-1) (and5){};
\draw[marked-edge] (x1) -- (and5.input 2);
\draw[marked-edge] (or2.output) -- (and5.input 1);

\node[or-gate] at (0,-2) (or6){};
\draw (x0) -- (or6.input 2);
\draw[marked-edge] (and5.output) -- (or6.input 1);

\draw[->, cyan] (or6.output) -- ($(or6) - (0, 1)$);

\end{tikzpicture}}
  }
  \caption{}
  \label{fig-true-points-2}
 \end{subfigure}
 \quad
 \begin{subfigure}{\subfigwidth}
  \adjustbox{max width=\textwidth}{
  \centering{\begin{tikzpicture}

\node[input] (x0) at (-0.5, 3.2){$t_0$};
\node[input, draw, cyan] (x1) at (0.5, 3.2){$t_1$};
\node[input] (x2) at (1.5, 3.2){$t_2$};
\node[input, draw, cyan] (x3) at (2.5, 3.2){$t_3$};
\node[input, draw, cyan] (x4) at (3.5, 3.2){$t_4$};
\node[input] (x5) at (4.5,3.2) {$t_5$};

\node[or-gate] at (4,2) (or1){};
\draw (x5) -- (or1.input 1);
\draw[marked-edge] (x4) -- (or1.input 2);

\node[and-gate] at (3,1) (and4){};
\draw[marked-edge] (x3) -- (and4.input 2);
\draw[marked-edge] (or1.output) -- (and4.input 1);

\node[or-gate] at (2,0) (or2){};
\draw (x2) -- (or2.input 2);
\draw[marked-edge] (and4.output) -- (or2.input 1);

\node[and-gate] at (1,-1) (and5){};
\draw[marked-edge] (x1) -- (and5.input 2);
\draw[marked-edge] (or2.output) -- (and5.input 1);

\node[or-gate] at (0,-2) (or6){};
\draw (x0) -- (or6.input 2);
\draw[marked-edge] (and5.output) -- (or6.input 1);

\draw[->, cyan] (or6.output) -- ($(or6) - (0, 1)$);

\end{tikzpicture}}
  }
  \caption{}
  \label{fig-true-points-3}
  \end{subfigure}
 \quad
 \begin{subfigure}{\subfigwidth}
  \adjustbox{max width=\textwidth}{
  \centering{\begin{tikzpicture}

\node[input] (x0) at (-0.5, 3.2){$t_0$};
\node[input, draw, cyan] (x1) at (0.5, 3.2){$t_1$};
\node[input] (x2) at (1.5, 3.2){$t_2$};
\node[input, draw, cyan] (x3) at (2.5, 3.2){$t_3$};
\node[input] (x4) at (3.5, 3.2){$t_4$};
\node[input, draw, cyan] (x5) at (4.5,3.2) {$t_5$};

\node[or-gate] at (4,2) (or1){};
\draw[marked-edge] (x5) -- (or1.input 1);
\draw (x4) -- (or1.input 2);

\node[and-gate] at (3,1) (and4){};
\draw[marked-edge] (x3) -- (and4.input 2);
\draw[marked-edge] (or1.output) -- (and4.input 1);

\node[or-gate] at (2,0) (or2){};
\draw (x2) -- (or2.input 2);
\draw[marked-edge] (and4.output) -- (or2.input 1);

\node[and-gate] at (1,-1) (and5){};
\draw[marked-edge] (x1) -- (and5.input 2);
\draw[marked-edge] (or2.output) -- (and5.input 1);

\node[or-gate] at (0,-2) (or6){};
\draw (x0) -- (or6.input 2);
\draw[marked-edge] (and5.output) -- (or6.input 1);

\draw[->, cyan] (or6.output) -- ($(or6) - (0, 1)$);

\end{tikzpicture}}
  }
  \caption{}
  \label{fig-true-points-4}
  \end{subfigure}
 \caption{All prime implicants of the \aop{} $t_0 \lor \Bkt{t_1 \land \Bkt{t_2 \lor \Bkt{t_3 \land \Bkt{t_4 \lor t_5}}}}$.
                   \crefrange{fig-true-points-1}{fig-true-points-4}
                   illustrate one prime implicant each.
                   The corresponding inputs are boxed.}
 \label{fig-true-points}
 \end{center}
\end{figure}

Due to the close connection to adder circuits (see \cref{sec::intro}),
our main interest lies in the optimization of \define{\aop s},
i.e., generalized \aop{}s where the gate types alternate between \AND{} and \OR{}.
Here, we call the respective optimization problems
\praopdelayopt{} and \praopdepthopt.
\totalref{ex-aop-delay} shows two logically equivalent circuits
for the \aop{}
$t_0 \land \bkt{ t_1 \lor \bkt{t_2 \land \bkt{ t_3 \lor t_4}}}$
on $5$ inputs:
the standard circuit and a circuit with a better depth.
However, for the indicated blue input arrival times,
the standard circuit has a better delay than the other circuit.
Standard circuits for other generalized \aop{}s are shown, e.g.,
in \totalref{fig-true-points-1} and \totalref{gen-aop-lb-aop-1}.

\begin{figure}[!ht]
  \begin{subfigure}{0.48\textwidth}
  \adjustbox{max width=\textwidth}{
  \centering{\begin{tikzpicture}

% modify the command below to remove ids from plots
\newcommand{\showid}[2]{#1}
\newcommand{\showwiredelay}[1]{#1}
\newcommand{\showindex}[1]{#1}
% end of GateTikzPainter preamble

 \node[input, scale=2] (i0) at (-0.0000, 0.0000)   {$t_{13}$};
 \node[input, scale=2] (i2) at (-2.0000, 0.0000)   {$t_{12}$};
 \node[input, scale=2] (i4) at (-4.0000, 0.0000)   {$t_{11}$};
 \node[input, scale=2] (i6) at (-6.0000, 0.0000)   {$t_{10}$};
 \node[input, scale=2] (i8) at (-8.0000, 0.0000)   {$t_9$};
 \node[input, scale=2] (i10) at (-10.0000, 0.0000) {$t_8$};
 \node[input, scale=2] (i12) at (-12.0000, 0.0000) {$t_7$};
 \node[input, scale=2] (i14) at (-14.0000, 0.0000) {$t_6$};
 \node[input, scale=2] (i16) at (-16.0000, 0.0000) {$t_5$};
 \node[input, scale=2] (i18) at (-18.0000, 0.0000) {$t_4$};
 \node[input, scale=2] (i20) at (-20.0000, 0.0000) {$t_3$};
 \node[input, scale=2] (i22) at (-22.0000, 0.0000) {$t_2$};
 \node[input, scale=2] (i24) at (-24.0000, 0.0000) {$t_1$};
 \node[input, scale=2] (i26) at (-26.0000, 0.0000) {$t_0$};
 \node[large-node, and-gate] at (-1.0000, -2.0000) (AND28){\rotatebox{90}{\showid{}{1}}};
 \draw (i0) -- (AND28.input 1) ;
 \draw (i2) -- (AND28.input 2) ;
 \node[large-node, or-gate] at (-6.0, -2.0) (OR30){\rotatebox{90}{\showid{}{1}}};
 \draw (i4) -- (OR30.input 1) ;
 \draw (i8) -- (OR30.input 2) ;
 \node[large-node, or-gate] at (-8.0, -2.0) (OR32){\rotatebox{90}{\showid{}{1}}};
 \draw (i6) -- (OR32.input 1) ;
 \draw (i8) -- (OR32.input 2) ;
 \node[large-node, and-gate] at (-12.0, -2.0) (AND34){\rotatebox{90}{\showid{}{1}}};
 \draw (i10) -- (AND34.input 1) ;
 \draw (i14) -- (AND34.input 2) ;
 \node[large-node, and-gate] at (-14.0, -2.0) (AND36){\rotatebox{90}{\showid{}{1}}};
 \draw (i12) -- (AND36.input 1) ;
 \draw (i14) -- (AND36.input 2) ;
 \node[large-node, or-gate] at (-19.0, -2.0) (OR38){\rotatebox{90}{\showid{}{1}}};
 \draw (i18) -- (OR38.input 1) ;
 \draw (i20) -- (OR38.input 2) ;
 \node[large-node, or-gate] at (-22.4, -2.0) (OR40){\rotatebox{90}{\showid{}{1}}};
 \draw (i20) -- (OR40.input 1) ;
 \draw (i24) -- (OR40.input 2) ;
 \node[large-node, or-gate] at (-3.5, -4.0) (OR42){\rotatebox{90}{\showid{}{2}}};
 \draw (AND28.output) -- (OR42.input 1) ;
 \draw (OR30.output) -- (OR42.input 2) ;
 \node[large-node, and-gate] at (-9.5, -4.0) (AND44){\rotatebox{90}{\showid{}{2}}};
 \draw (OR32.output) -- (AND44.input 1) ;
 \draw (AND34.output) -- (AND44.input 2) ;
 \node[large-node, or-gate] at (-19.0, -4.0) (OR46){\rotatebox{90}{\showid{}{2}}};
 \draw (i16) -- (OR46.input 1) ;
 \draw (OR40.output) -- (OR46.input 2) ;
 \node[large-node, and-gate] at (-21.0, -4.0) (AND48){\rotatebox{90}{\showid{}{2}}};
 \draw (OR38.output) -- (AND48.input 1) ;
 \draw (i22) -- (AND48.input 2) ;
 \node[large-node, and-gate] at (-6.5, -6.0) (AND50){\rotatebox{90}{\showid{}{3}}};
 \draw (OR42.output) -- (AND50.input 1) ;
 \draw (AND44.output) -- (AND50.input 2) ;
 \node[large-node, or-gate] at (-16.0, -6.0) (OR52){\rotatebox{90}{\showid{}{3}}};
 \draw (AND36.output) -- (OR52.input 1) ;
 \draw (OR46.output) -- (OR52.input 2) ;
 \node[large-node, or-gate] at (-22.2, -6.0) (OR54){\rotatebox{90}{\showid{}{3}}};
 \draw (AND48.output) -- (OR54.input 1) ;
 \draw (i24) -- (OR54.input 2) ;
 \node[large-node, or-gate] at (-11.2, -8.0) (OR56){\rotatebox{90}{\showid{}{4}}};
 \draw (AND50.output) -- (OR56.input 1) ;
 \draw (OR52.output) -- (OR56.input 2) ;
 \node[large-node, and-gate] at (-24.1, -8.0) (AND58){\rotatebox{90}{\showid{}{4}}};
 \draw (OR54.output) -- (AND58.input 1) ;
 \draw (i26) -- (AND58.input 2) ;
 \node[large-node, and-gate] at (-17.7, -10.6) (AND60){\rotatebox{90}{\showid{}{5}}};
 \draw (OR56.output) -- (AND60.input 1) ;
 \draw (AND58.output) -- (AND60.input 2) ;

 \draw[->] (AND60.output) -- ($(AND60) - (0, 1.3)$);
 \phantom{\node[large-node, and-gate] at (-17.7, -10.6) (AND60){\rotatebox{90}{\showid{}{5}}};}
 \end{tikzpicture}}
  }
  \caption{A depth-optimum formula circuit with optimum size $17$ which is not
  strongly depth-optimum.}
  \label{fig-non-size-opt-1}
 \end{subfigure}
 \hfill
 \begin{subfigure}{0.48\textwidth}
   \adjustbox{max width=\textwidth}{
  \centering{\begin{tikzpicture}

% modify the command below to remove ids from plots
\newcommand{\showid}[2]{#1}
\newcommand{\showwiredelay}[1]{#1}
\newcommand{\showindex}[1]{#1}
% end of GateTikzPainter preamble

 \node[input, scale=2] (i0) at (-0.0000, 0.0000)   {$t_{13}$};
 \node[input, scale=2] (i2) at (-2.0000, 0.0000)   {$t_{12}$};
 \node[input, scale=2] (i4) at (-4.0000, 0.0000)   {$t_{11}$};
 \node[input, scale=2] (i6) at (-6.0000, 0.0000)   {$t_{10}$};
 \node[input, scale=2] (i8) at (-8.0000, 0.0000)   {$t_9$};
 \node[input, scale=2] (i10) at (-10.0000, 0.0000) {$t_8$};
 \node[input, scale=2] (i12) at (-12.0000, 0.0000) {$t_7$};
 \node[input, scale=2] (i14) at (-14.0000, 0.0000) {$t_6$};
 \node[input, scale=2] (i16) at (-16.0000, 0.0000) {$t_5$};
 \node[input, scale=2] (i18) at (-18.0000, 0.0000) {$t_4$};
 \node[input, scale=2] (i20) at (-20.0000, 0.0000) {$t_3$};
 \node[input, scale=2] (i22) at (-22.0000, 0.0000) {$t_2$};
 \node[input, scale=2] (i24) at (-24.0000, 0.0000) {$t_1$};
 \node[input, scale=2] (i26) at (-26.0000, 0.0000) {$t_0$};

\node[or-gate, large-node] at (-23.5,-2) (or1){};
\draw (i22) -- (or1.input 1);
\draw (i24) -- (or1.input 2);

\node[or-gate, large-node] at (-22,-2) (or2){};
\draw (i20) -- (or2.input 1);
\draw (i24) -- (or2.input 2);

\node[or-gate, large-node] at (-20,-4) (or3){};
\draw (or2.output)   --  (or3.input 2);
\draw (i18) -- (or3.input 1);

\node[and-gate, large-node] at (-24.5,-4) (and1){};
\draw (or1.output) -- (and1.input 1);
\draw (i26) -- (and1.input 2);

\node[and-gate, large-node] at (-22,-6) (and2){};
\draw (or3.output) -- (and2.input 1);
\draw (and1.output)   -- (and2.input 2);

 \node[large-node, and-gate] at (-1.0000, -2.0000) (AND28){\rotatebox{90}{\showid{}{1}}};
 \draw (i0) -- (AND28.input 1) ;
 \draw (i2) -- (AND28.input 2) ;
 \node[large-node, or-gate] at (-6.0, -2.0) (OR30){\rotatebox{90}{\showid{}{1}}};
 \draw (i4) -- (OR30.input 1) ;
 \draw (i8) -- (OR30.input 2) ;
 \node[large-node, or-gate] at (-8.0, -2.0) (OR32){\rotatebox{90}{\showid{}{1}}};
 \draw (i6) -- (OR32.input 1) ;
 \draw (i8) -- (OR32.input 2) ;
 \node[large-node, and-gate] at (-12.0, -2.0) (AND34){\rotatebox{90}{\showid{}{1}}};
 \draw (i10) -- (AND34.input 1) ;
 \draw (i14) -- (AND34.input 2) ;
 \node[large-node, and-gate] at (-14.0, -2.0) (AND36){\rotatebox{90}{\showid{}{1}}};
 \draw (i12) -- (AND36.input 1) ;
 \draw (i14) -- (AND36.input 2) ;
 \node[large-node, or-gate] at (-20.5, -2.0) (OR40){\rotatebox{90}{\showid{}{1}}};
 \draw (i20) -- (OR40.input 1) ;
 \draw (i24) -- (OR40.input 2) ;
 \node[large-node, or-gate] at (-3.5, -4.0) (OR42){\rotatebox{90}{\showid{}{2}}};
 \draw (AND28.output) -- (OR42.input 1) ;
 \draw (OR30.output) -- (OR42.input 2) ;
 \node[large-node, and-gate] at (-9.5, -4.0) (AND44){\rotatebox{90}{\showid{}{2}}};
 \draw (OR32.output) -- (AND44.input 1) ;
 \draw (AND34.output) -- (AND44.input 2) ;
 \node[large-node, and-gate] at (-6.5, -6.0) (AND50){\rotatebox{90}{\showid{}{3}}};
 \draw (OR42.output) -- (AND50.input 1) ;
 \draw (AND44.output) -- (AND50.input 2) ;
 \node[large-node, or-gate] at (-18.0, -4.0) (OR46){\rotatebox{90}{\showid{}{2}}};
 \draw (i16) -- (OR46.input 1) ;
 \draw (OR40.output) -- (OR46.input 2);
 \node[large-node, or-gate] at (-16.0, -6.0) (OR52){\rotatebox{90}{\showid{}{3}}};
 \draw (AND36.output) -- (OR52.input 1) ;
 \draw (OR46.output) -- (OR52.input 2) ;
 \node[large-node, or-gate] at (-11.2, -8.0) (OR56){\rotatebox{90}{\showid{}{4}}};
 \draw (AND50.output) -- (OR56.input 1) ;
 \draw (OR52.output) -- (OR56.input 2) ;
 \node[large-node, and-gate] at (-17.7, -10.6) (AND60){\rotatebox{90}{\showid{}{5}}};
 \draw (OR56.output) -- (AND60.input 1) ;
 \draw (and2.output) -- (AND60.input 2) ;

 \draw[->] (AND60.output) -- ($(AND60) - (0, 1.3)$);

 \phantom{\node[large-node, and-gate] at (-17.7, -10.6) (AND60){\rotatebox{90}{\showid{}{5}}};}
\end{tikzpicture}}
  }
   \caption{A size-optimum circuit among all strongly depth-optimum formula circuits for the given \aop{} with size $18$.}
  \label{fig-non-size-opt-2}
  \end{subfigure}
 \caption{Two formula circuits for the \aop{} $t_0 \land \Bkt{t_1 \lor \Bkt{ \dotsc t_{13}}}$
         with optimum depth $5$.
         They only differ in the left sub-circuit of the final output.}
 \label{fig-non-size-opt}
\end{figure}
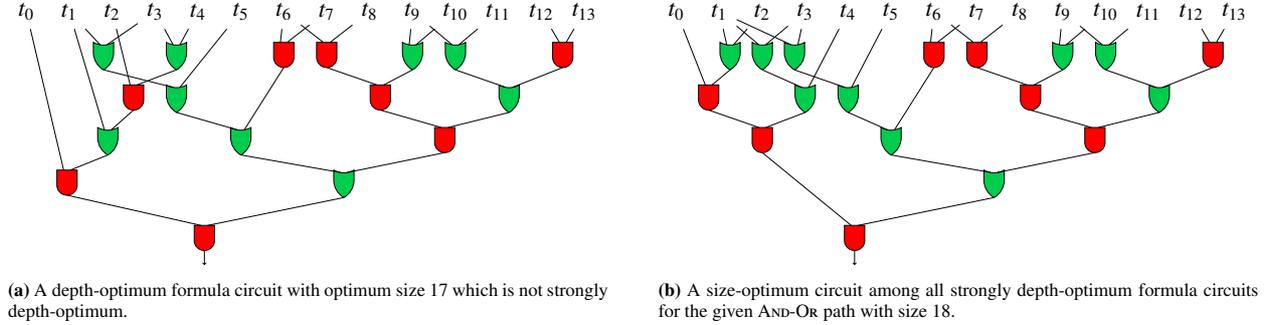

Given a Boolean function $f$ and a circuit $C$ for $f$
with prescribed input arrival times,
we call $C$ \define{strongly delay-optimum}
(or, in case of $a \equiv 0$, \define{strongly depth-optimum})
if for each vertex $v$,
the Boolean function computed at $v$ is
realized by a delay-optimum circuit in $C$.
In \cref{fig-non-size-opt}, we show two depth-optimum formula circuits
for the \aop{} $t_0 \land \Bkt{t_1 \lor \Bkt{ \dotsc t_{13}}}$.
The circuit in \cref{fig-non-size-opt-1} is a circuit with optimum depth
and, among all depth-optimum circuits, optimum size,
while the circuit in \cref{fig-non-size-opt-2}
is at least size-optimum among all strongly depth-optimum formula circuits.
Note that in \cref{fig-non-size-opt-1},
the left predecessor of the output gate computes an \aop{} on $5$
inputs with a non-optimum depth of $4$ and a size of $5$.
In \cref{fig-non-size-opt-2},
we instead use a depth-optimum circuit with depth $3$ and size $5$.

Note that we assume the input arrival times to be natural numbers.
As shown in \citet{thesis_hermann}, this is not a restriction:
When the arrival times are arbitrary fractional numbers,
we can still solve the \prgenaopdelayopt{} optimally
using a certain type of binary search
on specific instances with all arrival times being natural numbers
(see Theorem 5.1.5 of \citet{thesis_hermann}).
This increases the running time by at most a factor of $\mathcal O(\log_2 m)$.
Hence, during this work, we may assume all arrival times to be natural numbers.

In order to understand (generalized) \aop{}s more thoroughly,
it is helpful to divide the inputs into groups.
Here, we use the notation $x \tupleconcat y$ for the concatenation
of two tuples $x$ and $y$.

\begin{definition} \label{gen-aop-input-groups}
Let $h(t; \Gamma)$ be a generalized \aop{}
with $m \geq 1$ inputs.
For $i \in \set{0, \dotsc, m-2}$,
we call $\circ_i = \gt(t_i)$ the \define{gate type} of $t_i$,
and we call $t_i$ a \define{$\circ_i$-signal}.
Given a gate type $\circ \in \set{\AND, \OR}$,
we denote the set of all $\circ$-signals plus $t_{m-1}$
by $\sameins{\circ}$.
We call $\sameins{\circ}$ the \define{same-gate input set} of the
generalized \aop{} $h(t; \Gamma)$ and the gate type $\circ$.
By $D^{\circ} := \set{t_0, \dotsc, t_{m-1}} \backslash S^{\circ}$,
we denote the \define{diff-gate input set} of $h(t; \Gamma)$ and $\circ$.
The \define{\propgenpart{}} of $h\Bkt{t; \Gamma}$
is the unique partition
$\Bkt{t_0, \dotsc, t_{m-1}} = P_0 \tupleconcat \dotsc \tupleconcat P_c$
of the inputs into maximal consecutive sub-tuples $P_0, \dotsc, P_c$
called \define{input segments} such that for each $b \in \set{0, \dotsc, c}$,
we have $P_b \subseteq \sameins{\AND}$ or $P_b \subseteq \sameins{\OR}$.
\end{definition}

\totalref{gen-aop-lb-aop-1} visualizes a generalized \aop{} and its \propgenpart{}.
Note that the last input $t_{m-1}$ does not have a gate type
and, for $m \geq 2$, always is in the same input segment as $t_{m-2}$.
In this example, we have
$\sameins{\AND} = \set{t_0, t_5, t_6, t_8, t_9, t_{10}, t_{11}}$,
$\diffins{\AND} = \set{t_1, t_2, t_3, t_4, t_7}$,
$\sameins{\OR} = \set{t_1, t_2, t_3, t_4, t_7, t_{11}}$,
$\diffins{\OR} = \set{t_0, t_5, t_6, t_8, t_9, t_{10}}$.

It is an easy exercise to characterize the prime implicants of generalized \aop{}s
as in the following proposition.
\cref{fig-true-points} shows all prime implicants for an \aop{} on $6$ inputs.

\begin{qedproposition}\label{aop-prime-implicants}
 The set of prime implicants of $h(t; \Gamma)$ is given by
 \[\left\{ t_i \land \LAND_{j < i, \gt(t_j) = \AND{}} t_j
    \where{} t_i \in \sameins{\OR} \right\}\,. \tag*{\qedhere}\]
\end{qedproposition}

Together with \cref{ess-pi}, this \lcnamecref{aop-prime-implicants}
implies the following important statement.

\begin{qedcorollary}\label{prop-aop-all-essential}
Any generalized \aop{} depends essentially on all of its inputs.
\end{qedcorollary}

We can now give another basic lower bound on the delay
of any circuit over $\set{\AND{}2, \OR{}2}$ realizing a given generalized \aop{}.

\begin{proposition} \label{min-1-2-lb}
 Let $m \in \N$ with $m \geq 2$.
  Let inputs $t = \Bkt{t_0, \dotsc, t_{m-1}}$
 with arrival times $a(t_0), \dotsc, a(t_{m-1}) \in \N$
 and gate types $\Gamma = \Bkt{\circ_0, \dotsc, \circ_{m-2}}$ be given.
 Let $\Bkt{t_0, \dotsc, t_{m-1}} = P_0 \tupleconcat \dotsc \tupleconcat P_c$
 be the \propgenpart{} of $h(t; \Gamma)$.
 Consider a circuit $C$ over $\set{\AND2, \OR2}$ realizing $h(t; \Gamma)$ .
 Then, we have
\[\delay(C) \geq \max \set{\max_{t_i \in P_0} a(t_i) + 1,
                           \max_{t_i \in P_b : b > 0} a(t_i) + 2}\,. \qedhere\]
\begin{proof}
Using \cref{prop-aop-all-essential}
and $m \geq 2$,
we immediately see that each input has depth at least $1$ in $C$.
Thus, it suffices
to prove that each input $t_i$ contained in $P_b$
for some $b > 0$ has depth at least $2$ in $C$.
For such an input, one can show easily that
for any $\alpha \in \set{0, 1}$, the function $f(C\mid_{t_i = \alpha})$
depends essentially on $t_0$ (note that $i \neq 0$).
This is not the case if some directed path from $t_i$ to $\cktout(C)$
contains only gates of the same type, so any path from $t_i$ to $\cktout(C)$
contains at least one \AND{} gate and one \OR{} gate.
Hence, $t_i$ has depth at least $2$ in $C$.
\end{proof}
\end{proposition}

Note that for fixed $m \in \N$, there are exactly $2$ \aop{}s on $m$ inputs.
They can be turned into each other by exchanging all \AND{} and \OR{} operations.
Even more, any circuit for the first \aop{} can be turned into a circuit for the second
\aop{} with the same delay by exchanging all \AND{} and \OR{} gates.
This process is called \define{dualization} and the statement is proven in a more general setting,
for instance, in \citet{Crama}, Theorem 1.3.
             % 2

\section{Structure Theorem} \label{sec-structure}

Our structure theorem and our algorithm presented in \cref{sec-opt-alg}
both reduce the problem of optimizing a
given generalized \aop{} to smaller instances of a specific form.

\begin{definition} \label{def-subpath}
Consider a generalized \aop{} $h(t; \Gamma)$
with $m \geq 1$ inputs.
Given indices $0 \leq i_0 < \dotsc < i_{k-1} \leq m - 1$,
the generalized \aop{}
$
  t_{i_0} \circ_{i_0} \Bkt{t_{i_1}\circ_{i_1} \Bkt{
   \dotsc \circ_{i_{k-3}} \Bkt{t_{i_{k-2}} \circ_{i_{k-2}} t_{i_{k-1}}}}}
$
is called a \define{\subpath{}} of $h(t; \Gamma)$.
Now, let a gate type $\circ \in \set{\AND, \OR}$ and a set
$\sameins{\circ}_1$ with
$\emptyset \neq \sameins{\circ}_1 \subseteq \sameins{\circ}$ be given,
and let $i$ be maximum with $t_i \in \sameins{\circ}_1$.
Then, the sub-path
of $h(t; \Gamma)$ containing all
signals from $\sameins{\circ}_1$ and all signals $t_j \in \diffins{\circ}$ with
$j < i$ is denoted by $\genaopkeepF{\sameins{\circ}_1}$
and called a \define{\specialsubpath{}} of $h(t; \Gamma)$.
\end{definition}

\cref{gen-aop-lb-aop} shows a generalized \aop{} and gives several examples
for \specialsubpath{}s.

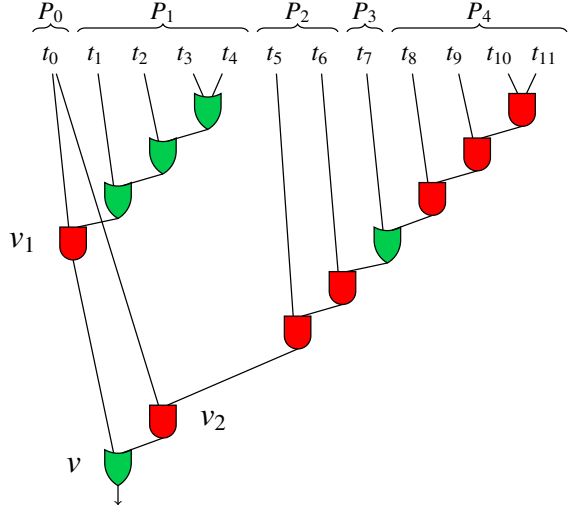
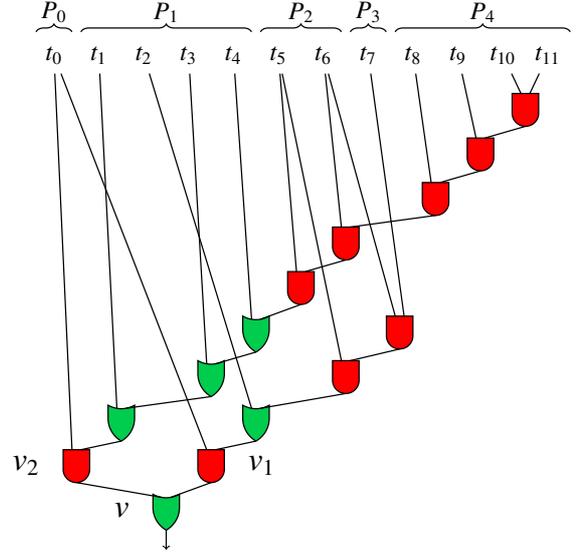
\begin{figure}
 \begin{subfigure}[t]{.46\textwidth}
  \adjustbox{max width=\textwidth}{
  \centering{\begin{tikzpicture}

\node[input] (x0)  at (-1.5, 6.2){$t_0$};
\node[input] (x1)  at (-0.5, 6.2){$t_1$};
\node[input] (x2)  at (0.5,  6.2){$t_2$};
\node[input] (x3)  at (1.5,  6.2){$t_3$};
\node[input] (x4)  at (2.5,  6.2){$t_4$};
\node[input] (x5)  at (3.5,  6.2){$t_5$};
\node[input] (x6)  at (4.5,  6.2){$t_6$};
\node[input] (x7)  at (5.5,  6.2){$t_7$};
\node[input] (x8)  at (6.5,  6.2){$t_8$};
\node[input] (x9)  at (7.5,  6.2){$t_9$};
\node[input] (x10) at (8.5,  6.2){$t_{10}$};
\node[input] (x11) at (9.5,  6.2){$t_{11}$};

\draw[thick, decorate,decoration={brace, amplitude=5pt, raise=15pt}] (x0.west) -- (x0.east)  node [midway, above, sloped, yshift=15pt, scale=1.6] {$P_0$};
\draw[thick, decorate,decoration={brace, amplitude=5pt, raise=15pt}] (x1.west) -- (x4.east)  node [midway, above, sloped, yshift=15pt, scale=1.6] {$P_1$};
\draw[thick, decorate,decoration={brace, amplitude=5pt, raise=15pt}] (x5.west) -- (x6.east)  node [midway, above, sloped, yshift=15pt, scale=1.6] {$P_2$};
\draw[thick, decorate,decoration={brace, amplitude=5pt, raise=15pt}] (x7.west) -- (x7.east)  node [midway, above, sloped, yshift=15pt, scale=1.6] {$P_3$};
\draw[thick, decorate,decoration={brace, amplitude=5pt, raise=15pt}] (x8.west) -- (x11.east) node [midway, above, sloped, yshift=15pt, scale=1.6] {$P_4$};

\node[and-gate] at (9,5) (or8){};
\draw (x10) -- (or8.input 2);
\draw (x11) -- (or8.input 1);

\node[and-gate] at (8,4) (and7){};
\draw (or8.output) -- (and7.input 1);
\draw (x9) -- (and7.input 2);

\node[and-gate] at (7,3) (or1){};
\draw (and7.output) -- (or1.input 1);
\draw (x8) -- (or1.input 2);

\node[or-gate] at (6,2) (and1){};
\draw (or1.output) -- (and1.input 1);
\draw (x7) -- (and1.input 2);

\node[and-gate] at (5, 1) (and20){};
\draw (x6) -- (and20.input 2);
\draw (and1.output) -- (and20.input 1);

\node[and-gate] at (4,0) (and2){};
\draw (and20.output) -- (and2.input 1);
\draw (x5) -- (and2.input 2);

\node[and-gate, label={[label distance=.4cm]north:{\huge $v_2$}}] at (1,-2) (or11){};
\draw (and2.output) -- (or11.input 1);
\draw (x0) -- (or11.input 2);

% prefix gen aop
\node[or-gate] at (2,5) (and6){};
\draw (x4) -- (and6.input 1);
\draw (x3) -- (and6.input 2);

\node[or-gate] at (1,4) (or9){};
\draw (x2) -- (or9.input 2);
\draw (and6.output) -- (or9.input 1);

\node[or-gate] at (0,3) (and10){};
\draw (or9.output) -- (and10.input 1);
\draw (x1) -- (and10.input 2);

\node[and-gate, label={[label distance=.4cm]south:{\huge $v_1$}}] at (-1,2) (and11){};
\draw (and10.output) -- (and11.input 1);
\draw (x0) -- (and11.input 2);

% final gates

\node[or-gate, label={[label distance=.4cm]south:{\huge $v$}}] at (0, -3) (or12){};
\draw (or11.output) -- (or12.input 1);
\draw (and11.output) -- (or12.input 2);

\draw[->] (or12.output) -- ($(or12) - (0, .9)$);

\phantom{\node[or-gate, label={[label distance=.4cm]south:{\huge $v$}}] at (1, -4) (or13){};}
\phantom{\draw (or11.output) -- (or13.input 1);}
\phantom{\draw (and11.output) -- (or13.input 2);}
\phantom{\draw[->] (or13.output) -- ($(or13) - (0, .9)$);}

\phantom{\node[and-gate, label={[label distance=.4cm]south:{\huge $v_2$}}] at (-1,-3) (and11){};}
\end{tikzpicture}}
  }
  \caption{In this case, we have $f(C_{v_1}) = \genaopkeepF{\sameins{\circ}_1}$
        and $f(C_{v_2}) = \genaopkeepF{\sameins{\circ}_2}$
        with $\sameins{\OR{}}_1 = \set{t_1, \dotsc, t_4}$
        and $\sameins{\OR{}}_2 = \set{t_7, t_{11}}$.}
 \end{subfigure}
 \hfill
 \begin{subfigure}[t]{.46\textwidth}
  \adjustbox{max width=\textwidth}{
  \centering{\begin{tikzpicture}

\node[input] (x0)  at (-1.5, 6.2){$t_0$};
\node[input] (x1)  at (-0.5, 6.2){$t_1$};
\node[input] (x2)  at (0.5,  6.2){$t_2$};
\node[input] (x3)  at (1.5,  6.2){$t_3$};
\node[input] (x4)  at (2.5,  6.2){$t_4$};
\node[input] (x5)  at (3.5,  6.2){$t_5$};
\node[input] (x6)  at (4.5,  6.2){$t_6$};
\node[input] (x7)  at (5.5,  6.2){$t_7$};
\node[input] (x8)  at (6.5,  6.2){$t_8$};
\node[input] (x9)  at (7.5,  6.2){$t_9$};
\node[input] (x10) at (8.5,  6.2){$t_{10}$};
\node[input] (x11) at (9.5,  6.2){$t_{11}$};

\draw[thick, decorate,decoration={brace, amplitude=5pt, raise=15pt}] (x0.west) -- (x0.east)  node [midway, above, sloped, yshift=15pt, scale=1.6] {$P_0$};
\draw[thick, decorate,decoration={brace, amplitude=5pt, raise=15pt}] (x1.west) -- (x4.east)  node [midway, above, sloped, yshift=15pt, scale=1.6] {$P_1$};
\draw[thick, decorate,decoration={brace, amplitude=5pt, raise=15pt}] (x5.west) -- (x6.east)  node [midway, above, sloped, yshift=15pt, scale=1.6] {$P_2$};
\draw[thick, decorate,decoration={brace, amplitude=5pt, raise=15pt}] (x7.west) -- (x7.east)  node [midway, above, sloped, yshift=15pt, scale=1.6] {$P_3$};
\draw[thick, decorate,decoration={brace, amplitude=5pt, raise=15pt}] (x8.west) -- (x11.east) node [midway, above, sloped, yshift=15pt, scale=1.6] {$P_4$};

\node[and-gate] at (9,5) (or8){};
\draw (x10) -- (or8.input 2);
\draw (x11) -- (or8.input 1);

\node[and-gate] at (8,4) (and7){};
\draw (or8.output) -- (and7.input 1);
\draw (x9) -- (and7.input 2);

\node[and-gate] at (7,3) (or1){};
\draw (and7.output) -- (or1.input 1);
\draw (x8) -- (or1.input 2);

\node[and-gate] at (5,2) (and21){};
\draw (or1.output) -- (and21.input 1);
\draw (x6) -- (and21.input 2);

\node[and-gate] at (4,1) (and22){};
\draw (and21.output) -- (and22.input 1);
\draw (x5) -- (and22.input 2);

\node[and-gate] at (6.2, 0) (and20){};
\draw (x6) -- (and20.input 2);
\draw (x7) -- (and20.input 1);

\node[and-gate] at (5,-1) (and2){};
\draw (and20.output) -- (and2.input 1);
\draw (x5) -- (and2.input 2);

\node[or-gate] at (3,-2) (or20){};
\draw (and2.output) -- (or20.input 1);
\draw (x2) -- (or20.input 2);

\node[and-gate, label={[label distance=.4cm]north:{\huge $v_1$}}] at (2,-3) (or11){};
\draw (or20.output) -- (or11.input 1);
\draw (x0) -- (or11.input 2);

% prefix gen aop
\node[or-gate] at (3,0) (or21){};
\draw (and22.output) -- (or21.input 1);
\draw (x4) -- (or21.input 2);

\node[or-gate] at (2,-1) (and6){};
\draw (or21.output) -- (and6.input 1);
\draw (x3) -- (and6.input 2);

\node[or-gate] at (0,-2) (and10){};
\draw (and6.output) -- (and10.input 1);
\draw (x1) -- (and10.input 2);

\node[and-gate, label={[label distance=.4cm]south:{\huge $v_2$}}] at (-1,-3) (and11){};
\draw (and10.output) -- (and11.input 1);
\draw (x0) -- (and11.input 2);

% final gates

\node[or-gate, label={[label distance=.4cm]south:{\huge $v$}}] at (1, -4) (or12){};
\draw (or11.output) -- (or12.input 1);
\draw (and11.output) -- (or12.input 2);

\draw[->] (or12.output) -- ($(or12) - (0, .9)$);

\end{tikzpicture}}
  }
  \caption{In this case, we have
        $f(C_{v_1}) = \genaopkeepF{\sameins{\circ}_1}$
        and $f(C_{v_2}) = \genaopkeepF{\sameins{\circ}_2}$
        with $\sameins{\OR{}}_1 = \set{t_2, t_7}$
        and $\sameins{\OR{}}_2 = \set{t_1, t_3, t_4, t_{11}}$.}
 \end{subfigure}
 \caption{Two possible circuits arising from \cref{structure-thm}
        applied to the generalized \aop{} from \cref{gen-aop-lb-aop-1}.}
 \label{fig-gen-aop-partitions}
\end{figure}

In \cref{structure-real-thm}, we will see that
for any generalized \aop{} $h(t; \Gamma)$ with input arrival times,
there is always a delay-optimum formula circuit $C$
which arises from combining two circuits for \specialsubpath{}s
$\genaopkeepF{\sameins{\circ}_1}$ and $\genaopkeepF{\sameins{\circ}_2}$
with a $\circ$-gate,
where $\circ \in \set{\AND2, \OR2}$
and $\sameins{\circ}= \sameins{\circ}_1 \cupdot \sameins{\circ}_2$.
Our proof idea is based on Lemma 1 from \citet{Commentz-Walter-mon}.
However, only \aop{}s are considered there, and not generalized \aop{}s,
and only a partial description of the structure of \aop{} circuits is given,
not a complete characterization.
The main objects considered in both Commentz-Walter's and our proof are prime implicants.
An easy consequence of \cref{aop-prime-implicants} is the following \lcnamecref{pi-disjoint}.

\begin{observation} \label{pi-disjoint}
 Let a generalized \aop{} $h(t; \Gamma)$ with $t = \Bkt{t_0, \dotsc, t_{m-1}}$ be given.
 Consider any two prime implicants $\pi \neq \rho \in \PI(h(t; \Gamma))$.
 Choose $i \in \set{0, \dotsc, m-1}$ maximum with $t_i \in \lit(\pi)$.
 Then, we have $t_i \notin \lit(\rho)$.
\end{observation}

The following \lcnamecref{structure-thm}
is the main ingredient of our structure theorem, \cref{structure-real-thm}.
We consider a formula circuit~$C$ implementing a generalized \aop{} $h(t; \Gamma)$ with
$\cktout(C) = \OR{}$.
If $C$ is delay-optimum for the given arrival times and,
among all delay-optimum circuits, size-optimum,
then we will show that the two sub-circuits of $\cktout(C)$
compute special sub-paths of $h(t; \Gamma)$,
where each input of $\sameins{\OR}$ is contained in exactly one of
the two sub-circuits.
\cref{fig-gen-aop-partitions} shows two examples for such circuits for the
generalized \aop{} from \cref{gen-aop-lb-aop-1}.
\cref{fig-str-thm} illustrates the proof of the \lcnamecref{structure-thm}.

\begin{figure}[p]
 \begin{subfigure}{.46\textwidth}
 \vspace{-.1cm}
  \adjustbox{max width=\textwidth}{
  \centering{\begin{tikzpicture}

\node[input] (x0)  at (-1.5, 6.2){$t_0$};
\node[input] (x1)  at (-0.5, 6.2){$t_1$};
\node[input] (x2)  at (0.5,  6.2){$t_2$};
\node[input] (x3)  at (1.5,  6.2){$t_3$};
\node[input] (x4)  at (2.5,  6.2){$t_4$};
\node[input] (x5)  at (3.5,  6.2){$t_5$};
\node[input] (x6)  at (4.5,  6.2){$t_6$};
\node[input] (x7)  at (5.5,  6.2){$t_7$};
\node[input] (x8)  at (6.5,  6.2){$t_8$};
\node[input] (x9)  at (7.5,  6.2){$t_9$};
\node[input] (x10) at (8.6,  6.2){$t_{10}$};
\node[input] (x11) at (9.8,  6.2){$t_{11}$};

\draw[thick, decorate,decoration={brace, amplitude=5pt, raise=15pt}] (x0.west) -- (x0.east)  node [midway, above, sloped, yshift=15pt, scale=1.6] {$P_0$};
\draw[thick, decorate,decoration={brace, amplitude=5pt, raise=15pt}] (x1.west) -- (x4.east)  node [midway, above, sloped, yshift=15pt, scale=1.6] {$P_1$};
\draw[thick, decorate,decoration={brace, amplitude=5pt, raise=15pt}] (x5.west) -- (x6.east)  node [midway, above, sloped, yshift=15pt, scale=1.6] {$P_2$};
\draw[thick, decorate,decoration={brace, amplitude=5pt, raise=15pt}] (x7.west) -- (x7.east)  node [midway, above, sloped, yshift=15pt, scale=1.6] {$P_3$};
\draw[thick, decorate,decoration={brace, amplitude=5pt, raise=15pt}] (x8.west) -- (x11.east) node [midway, above, sloped, yshift=15pt, scale=1.6] {$P_4$};

\node[and-gate] at (9,5) (or8){};
\node[and-gate] at (8,4) (and7){};
\node[and-gate] at (7,3) (or1){};

\node[or-gate] at (6,2) (and1){};

\node[and-gate] at (5,1) (or4){};
\node[and-gate] at (4,0) (and2){};

\node[or-gate] at (3,-1) (or5){};
\node[or-gate] at (2,-2) (and6){};
\node[or-gate] at (1,-3) (or9){};
\node[or-gate] at (0,-4) (and10){};

\node[and-gate] at (-1,-5) (or11){};

\draw (or1.output) -- (and1.input 1);
\draw (x8) -- (or1.input 2);
\draw (x7) -- (and1.input 2);
\draw (x9) -- (and7.input 2);

\draw (or4.output) -- (and2.input 1);
\draw (or8.output) -- (and7.input 1);

\draw (x5) -- (and2.input 2);
\draw (x10) -- (or8.input 2);
\draw (x11) -- (or8.input 1);

\draw (x6) -- (or4.input 2);
\draw (and1.output) -- (or4.input 1);

\draw (and2.output) -- (or5.input 1);
\draw (or5.output) -- (and6.input 1);

\draw (x4) -- (or5.input 2);
\draw (x3) -- (and6.input 2);

\draw (x2) -- (or9.input 2);
\draw (and6.output) -- (or9.input 1);
\draw (and7.output) -- (or1.input 1);

\draw (or9.output) -- (and10.input 1);
\draw (x1) -- (and10.input 2);
\draw (and10.output) -- (or11.input 1);
\draw (x0) -- (or11.input 2);

\draw[->] (or11.output) -- ($(or11) - (0, .9)$);

\end{tikzpicture}}
  }
  \caption{Standard circuit for $h(t; \Gamma)$
  with $\sameins{\AND} = \set{t_0, t_5, t_6, t_8, t_9, t_{10}, t_{11}}$.}
  \label{gen-aop-lb-aop-1}
 \end{subfigure}
 \hfill
 \begin{subfigure}{.46\textwidth}
  \adjustbox{max width=\textwidth}{
  \centering{\begin{tikzpicture}

\node[input, blue, draw] (x0)  at (-1.5, 6.2){$t_0$};
\node[input] (x1)  at (-0.5, 6.2){$t_1$};
\node[input] (x2)  at (0.5,  6.2){$t_2$};
\node[input] (x3)  at (1.5,  6.2){$t_3$};
\node[input] (x4)  at (2.5,  6.2){$t_4$};
\node[input, blue, draw] (x5)  at (3.5,  6.2){$t_5$};
\node[input] (x6)  at (4.5,  6.2){$t_6$};
\node[input] (x7)  at (5.5,  6.2){$t_7$};
\node[input, blue, draw] (x8)  at (6.5,  6.2){$t_8$};
\node[input, blue, draw] (x9)  at (7.5,  6.2){$t_9$};
\node[input, blue, draw] (x10) at (8.6,  6.2){$t_{10}$};
\node[input, blue, draw] (x11) at (9.8,  6.2){$t_{11}$};

\draw[thick, decorate,decoration={brace, amplitude=5pt, raise=15pt}] (x0.west) -- (x0.east)  node [midway, above, sloped, yshift=15pt, scale=1.6] {$P_0$};
\draw[thick, decorate,decoration={brace, amplitude=5pt, raise=15pt}] (x1.west) -- (x4.east)  node [midway, above, sloped, yshift=15pt, scale=1.6] {$P_1$};
\draw[thick, decorate,decoration={brace, amplitude=5pt, raise=15pt}] (x5.west) -- (x5.east)  node [midway, above, sloped, yshift=15pt, scale=1.6] {$P_2$};
\draw[thick, decorate,decoration={brace, amplitude=5pt, raise=15pt}] (x7.west) -- (x7.east)  node [midway, above, sloped, yshift=15pt, scale=1.6] {$P_3$};
\draw[thick, decorate,decoration={brace, amplitude=5pt, raise=15pt}] (x8.west) -- (x11.east) node [midway, above, sloped, yshift=15pt, scale=1.6] {$P_4$};

\node[and-gate] at (9,5) (or8){};
\node[and-gate] at (8,4) (and7){};
\node[and-gate] at (7,3) (or1){};

\node[or-gate] at (6,2) (and1){};

\node[and-gate] at (4,0) (and2){};

\node[or-gate] at (3,-1) (or5){};
\node[or-gate] at (2,-2) (and6){};
\node[or-gate] at (1,-3) (or9){};
\node[or-gate] at (0,-4) (and10){};

\node[and-gate] at (-1,-5) (or11){};

\draw (or1.output) -- (and1.input 1);
\draw (x8) -- (or1.input 2);
\draw (x7) -- (and1.input 2);
\draw (x9) -- (and7.input 2);

\draw (and1.output) -- (and2.input 1);
\draw (or8.output) -- (and7.input 1);

\draw (x5) -- (and2.input 2);
\draw (x10) -- (or8.input 2);
\draw (x11) -- (or8.input 1);

\draw (and2.output) -- (or5.input 1);
\draw (or5.output) -- (and6.input 1);

\draw (x4) -- (or5.input 2);
\draw (x3) -- (and6.input 2);

\draw (x2) -- (or9.input 2);
\draw (and6.output) -- (or9.input 1);
\draw (and7.output) -- (or1.input 1);

\draw (or9.output) -- (and10.input 1);
\draw (x1) -- (and10.input 2);
\draw (and10.output) -- (or11.input 1);
\draw (x0) -- (or11.input 2);

\draw[->] (or11.output) -- ($(or11) - (0, .9)$);

\end{tikzpicture}}
  }
  \caption{Standard circuit for $\genaopkeepF{\sameins{\AND{}}_1}$
  with $\sameins{\AND}_1 = \set{t_0, t_5, t_8, t_9, t_{10}, t_{11}}$.}
  \label{gen-aop-lb-aop-2}
 \end{subfigure}
 \begin{subfigure}{.46\textwidth}
 \vspace{.3cm}
  \adjustbox{max width=\textwidth}{
  \centering{\begin{tikzpicture}

\node[input] (x0)  at (-1.5, 6.2){$t_0$};
\node[input, blue, draw] (x1)  at (-0.5, 6.2){$t_1$};
\node[input, blue, draw] (x2)  at (0.5,  6.2){$t_2$};
\node[input, blue, draw] (x3)  at (1.5,  6.2){$t_3$};
\node[input, blue, draw] (x4)  at (2.5,  6.2){$t_4$};
\node[input] (x5)  at (3.5,  6.2){$t_5$};
\node[input] (x6)  at (4.5,  6.2){$t_6$};
\node[input] (x7)  at (5.5,  6.2){$t_7$};
\node[input] (x8)  at (6.5,  6.2){$t_8$};
\node[input] (x9)  at (7.5,  6.2){$t_9$};
\node[input] (x10) at (8.6,  6.2){$t_{10}$};
\node[input] (x11) at (9.8,  6.2){$t_{11}$};

\draw[thick, decorate,decoration={brace, amplitude=5pt, raise=15pt}] (x0.west) -- (x0.east)  node [midway, above, sloped, yshift=15pt, scale=1.6] {$P_0$};
\draw[thick, decorate,decoration={brace, amplitude=5pt, raise=15pt}] (x1.west) -- (x4.east)  node [midway, above, sloped, yshift=15pt, scale=1.6] {$P_1$};

\node[or-gate] at (2,5) (and6){};
\node[or-gate] at (1,4) (or9){};
\node[or-gate] at (0,3) (and10){};

\node[and-gate] at (-1,2) (or11){};

\draw (x4) -- (and6.input 1);

\draw (x3) -- (and6.input 2);

\draw (x2) -- (or9.input 2);
\draw (and6.output) -- (or9.input 1);

\draw (or9.output) -- (and10.input 1);
\draw (x1) -- (and10.input 2);
\draw (and10.output) -- (or11.input 1);
\draw (x0) -- (or11.input 2);

\draw[->] (or11.output) -- ($(or11) - (0, .9)$);

\phantom{\node[and-gate] at (-1,-5) (or11){};}

\end{tikzpicture}}
  }
  \vspace{-.1cm}
  \caption{Standard circuit for $\genaopkeepF{\sameins{\OR{}}_1}$
  with $\sameins{\OR}_1 = \set{t_1, t_2, t_3, t_4}$.}
  \label{gen-aop-lb-aop-3}
  \end{subfigure}
 \hfill
 \begin{subfigure}{.46\textwidth}
 \vspace{.3cm}
  \adjustbox{max width=\textwidth}{
  \centering{\begin{tikzpicture}

\node[input] (x0)  at (-1.5, 6.2){$t_0$};
\node[input] (x1)  at (-0.5, 6.2){$t_1$};
\node[input, blue, draw] (x2)  at (0.5,  6.2){$t_2$};
\node[input] (x3)  at (1.5,  6.2){$t_3$};
\node[input] (x4)  at (2.5,  6.2){$t_4$};
\node[input] (x5)  at (3.5,  6.2){$t_5$};
\node[input] (x6)  at (4.5,  6.2){$t_6$};
\node[input] (x7)  at (5.5,  6.2){$t_7$};
\node[input] (x8)  at (6.5,  6.2){$t_8$};
\node[input] (x9)  at (7.5,  6.2){$t_9$};
\node[input] (x10) at (8.6,  6.2){$t_{10}$};
\node[input, blue, draw] (x11) at (9.8,  6.2){$t_{11}$};

\draw[thick, decorate,decoration={brace, amplitude=5pt, raise=15pt}] (x0.west) -- (x0.east)  node [midway, above, sloped, yshift=15pt, scale=1.6] {$P_0$};
\draw[thick, decorate,decoration={brace, amplitude=5pt, raise=15pt}] (x2.west) -- (x2.east)  node [midway, above, sloped, yshift=15pt, scale=1.6] {$P_1$};
\draw[thick, decorate,decoration={brace, amplitude=5pt, raise=15pt}] (x5.west) -- (x11.east) node [midway, above, sloped, yshift=15pt, scale=1.6] {$P_2$};

\node[and-gate] at (9,5) (or8){};
\node[and-gate] at (8,4) (and7){};
\node[and-gate] at (7,3) (or1){};

\node[and-gate] at (5,1) (and10){};

\node[and-gate] at (4,0) (and2){};

\node[and-gate] at (-1,-5) (or11){};

\draw (or1.output) -- (and10.input 1);
\draw (x8) -- (or1.input 2);
\draw (x9) -- (and7.input 2);

\draw (x6) -- (and10.input 2);
\draw (and10.output) -- (and2.input 1);

\draw (or8.output) -- (and7.input 1);

\draw (x5) -- (and2.input 2);
\draw (x10) -- (or8.input 2);
\draw (x11) -- (or8.input 1);

\draw (and7.output) -- (or1.input 1);
\node[or-gate] at (1,-3) (or10){};
\draw (and2.output) -- (or10.input 1);
\draw (x2) -- (or10.input 2);
\draw (or10.output) -- (or11.input 1);
\draw (x0) -- (or11.input 2);

\draw[->] (or11.output) -- ($(or11) - (0, .9)$);

\end{tikzpicture}}
  }
  \caption{Standard circuit for $\genaopkeepF{\sameins{\OR{}}_1}$
  with $\sameins{\OR}_1 = \set{t_2, t_{11}}$.}
  \label{gen-aop-lb-aop-4}
  \end{subfigure}
 \caption{A generalized \aop{} $h(t; \Gamma)$ and three \specialsubpath{}s
 as in \cref{def-subpath}.
 We also show the respective \propgenpart{}s,
 and the respective input set $\sameins{\circ}_1$ is marked blue.}
 \label{gen-aop-lb-aop}
\end{figure}

\begin{figure}
 \begin{subfigure}[t]{.28\textwidth}
 \begin{center}
  \adjustbox{max width=0.8\textwidth}{
  \centering{\begin{tikzpicture}

\node[and-gate] at (5,2) (and1){};
\node[or-gate] at (4,1) (or1){};

\node[and-gate] at (3,0) (and4){};
\node[or-gate] at (2,-1) (or2){};
\node[input] at ($(or2) - (0,1.5)$) (or2text){$h$};

\node[input-at] (ai5) at (1.5, 3.9){$3$};
\node[input-at] (ai4) at (2.5, 3.9){$2$};
\node[input-at] (ai3) at (3.5, 3.9){$1$};
\node[input-at] (ai2) at (4.5, 3.9){$1$};
\node[input-at] (ai1) at (5.5, 3.9){$1$};

\node[input] (i5) at (1.5, 3.2){$t_0$};
\node[input] (i4) at (2.5, 3.2){$t_1$};
\node[input] (i3) at (3.5, 3.2){$t_2$};
\node[input] (i2) at (4.5, 3.2){$t_3$};
\node[input] (i1) at (5.5, 3.2){$t_4$};

\draw (and1.output) -- (or1.input 1);
\draw (i2) -- (and1.input 2);
\draw (i3) -- (or1.input 2);
\draw (i1) -- (and1.input 1);

\draw (and4.output) -- (or2.input 1);

\draw (i5) -- (or2.input 2);

\draw (i4) -- (and4.input 2);
\draw (or1.output) -- (and4.input 1);

\node[output-at] (output) at (2, -2) {};
\draw[output-edge] (or2.output) -- (output) {};

\phantom{\node[or-gate,label={[black]south:$t_0$}] at (1.5,0) (or7){};}

\end{tikzpicture}}
  }
  \end{center}
  \vspace{-.02cm}
  \caption{We have $\PI(h) = \set{t_0, t_1 \land t_2, t_1 \land t_3 \land t_4}$.}
  \label{fig-str-thm-1}
 \end{subfigure}
 \hfill
 \begin{subfigure}[t]{.28\textwidth}
  \begin{center}
  \adjustbox{max width=0.8\textwidth}{
  \centering{\begin{tikzpicture}

% right path
\node[and-gate] at (5.5,2) (and1){};
\node[or-gate] at (4.5,1) (or1){};

\node[and-gate] at (3.5,0) (and4){};
\node[input] at ($(and4) - (0,1)$) (and4text){$f_2$};
\node[or-gate] at (2.5,-1) (or2){};
\node[input] at ($(or2) - (0,1.5)$) (or2text){$h$};

% left path
\node[and-gate] at (3.5,2) (and5){};
\node[and-gate] at (2.5,1) (and6){};

\node[or-gate] at (1.5,0) (or7){};
\node[input] at ($(or7) - (0,1)$) (or7text){$f_1$};

\node[input-at] (ai5) at (1.5, 3.9){$3$};
\node[input-at] (ai4) at (2.5, 3.9){$2$};
\node[input-at] (ai3) at (3.5, 3.9){$1$};
\node[input-at] (ai2) at (4.5, 3.9){$1$};
\node[input-at] (ai1) at (5.5, 3.9){$1$};

\node[input] (i5) at (1.5, 3.2){$t_0$};
\node[input, cyan, draw] (i4) at (2.5, 3.2){$t_1$};
\node[input, cyan, draw] (i3) at (3.5, 3.2){$t_2$};
\node[input, cyan, draw] (i2) at (4.5, 3.2){$t_3$};
\node[input] (i1) at (5.5, 3.2){$t_4$};

\draw (and1.output) -- (or1.input 1);
\draw (i2) -- (and1.input 2);
\draw (i3) -- (or1.input 2);
\draw (i1) -- (and1.input 1);

\draw (i4) -- (and4.input 2);
\draw (or1.output) -- (and4.input 1);

\draw (i2) -- (and5.input 1);
\draw (i3) -- (and5.input 2);

\draw(and5.output) -- (and6.input 1);
\draw(i4) -- (and6.input 2);

\draw(and6.output) -- (or7.input 1);
\draw(i5) -- (or7.input 2);

\draw (and4.output) -- (or2.input 1);
\draw (or7.output) -- (or2.input 2);

\node[output-at] (output) at (2.5, -2) {};
\draw[output-edge] (or2.output) -- (output) {};

\phantom{\node[or-gate,label={[black]south:$t_0$}] at (1.5,0) (or7){};}

\end{tikzpicture}}
  }
  \end{center}
  \caption{We have $\PI(f_1) = \set{t_0, t_1 \land t_2 \land t_3}$ and
           $\PI(f_2) = \set{t_1 \land t_2, t_1 \land t_3 \land t_4}$.}
 \label{fig-str-thm-2}
 \end{subfigure}
 \hfill
 \begin{subfigure}[t]{.28\textwidth}
  \begin{center}
\adjustbox{max width=0.8\textwidth}{
  \centering{\begin{tikzpicture}

% right path
\node[and-gate] at (5.5,2) (and1){};
\node[or-gate] at (4.5,1) (or1){};

\node[and-gate] at (3.4,0) (and4){};
\node[input] at ($(and4) - (0,1)$) (and4text){$f_2$};
\node[or-gate] at (2.5,-1) (or2){};
\node[input] at ($(or2) - (0,1.5)$) (or2text){$h$};

% left path
\node[and-gate,label={[black]south:0}] at (3.5,2) (and5){};
\node[and-gate,label={[black]south:0}] at (2.5,1) (and6){};

\node[or-gate,label={[black]south:$t_0$}] at (1.5,0) (or7){};
\node[input] at ($(or7) - (0,1)$) (or7text){$f_1$};

\node[input-at] (ai5) at (1.5, 3.9){$3$};
\node[input-at] (ai4) at (2.5, 3.9){$2$};
\node[input-at] (ai3) at (3.5, 3.9){$1$};
\node[input-at] (ai2) at (4.5, 3.9){$1$};
\node[input-at] (ai1) at (5.5, 3.9){$1$};

\node[input] (i5) at (1.5, 3.2){$t_0$};
\node[input, cyan, draw] (i4) at (2.5, 3.2){$t_1$};
\node[input, cyan, draw] (i3) at (3.5, 3.2){$t_2$};
\node[input, cyan, draw] (i2) at (4.5, 3.2){$t_3$};
\node[input] (i1) at (5.5, 3.2){$t_4$};

\draw (and1.output) -- (or1.input 1);
\draw (i2) -- (and1.input 2);
\draw (i3) -- (or1.input 2);
\draw (i1) -- (and1.input 1);

\draw (i4) -- (and4.input 2);
\draw (or1.output) -- (and4.input 1);

\draw (i2) -- (and5.input 1);
\draw (i3) -- (and5.input 2);

\draw(and5.output) -- (and6.input 1);
\draw(i4) -- (and6.input 2);

\draw(and6.output) -- (or7.input 1);
\draw(i5) -- (or7.input 2);

\draw (and4.output) -- (or2.input 1);
\draw (or7.output) -- (or2.input 2);

\node[output-at] (output) at (2.5, -2) {};
\draw[output-edge] (or2.output) -- (output) {};

\end{tikzpicture}}
  }
  \end{center}
  \caption{Propagating $0$ into $t_2$ for $C_1$ in \cref{fig-str-thm-2}, we obtain this circuit,
  for which the reduced circuit is again the circuit from \cref{fig-str-thm-1}.}
  \label{fig-str-thm-3}
  \end{subfigure}
 \caption{Example illustration for the proof of \cref{structure-thm}.
          \cref{fig-str-thm-2} depicts a circuit $C$ for the \aop{} in \cref{fig-str-thm-1};
          the prime implicant $\rho = t_1 \land t_2 \land t_3$ of $f_1$ is highlighted.
          It contains the prime implicant $\pi = t_1 \land t_2$ of $h$,
          for which the highest index is $i = 2$.
          In \cref{fig-str-thm-3}, we show an intermediate step to reduce the circuit
          w.r.t. $t_2 = 0$.
          The reduced circuit is again the circuit in \cref{fig-str-thm-1}.}
 \label{fig-str-thm}
\end{figure}

\begin{lemma} \label{structure-thm}
 Let $m \in \N_{\geq 2}$, inputs $t = \Bkt{t_0, \dotsc, t_{m-1}}$
 with arrival times $a(t_0), \dotsc, a(t_{m-1}) \in \N$
 and gate types $\Gamma = \Bkt{\circ_0, \dotsc, \circ_{m-2}}$ be given.
 Consider a delay-optimum formula circuit $C$ for $h(t; \Gamma)$ with minimum number of gates.
 Assume that $\gt(\cktout(C)) = \OR{}$.
 Denote the predecessors of $v := \cktout(C)$ by $v_1$ and $v_2$.
 Write $h := h(t; \Gamma)$, and $f_1 := f(C_{v_1})$, and $f_2 := f(C_{v_2})$.
 Then, the following statements are fulfilled:
 \begin{enumerate}
  \item We have $\PI(h) = \PI(f_1) \cupdot \PI(f_2)$.
  \item There exists a partition $\sameins{\OR} = \sameins{\OR}_1 \cupdot \sameins{\OR}_2$
   with $\sameins{\OR}_1, \sameins{\OR}_2 \neq \emptyset$
   such that for each $k \in \set{1, 2}$,
   the function $f_k$ depends es\-sen\-tial\-ly
   on all inputs of $\sameins{\OR}_k$ and on no input of $\sameins{\OR}_{3-k}$.
  \item Let $k\in \{1, 2\}$.
   Consider the special sub-path
   $h_k := \genaopkeepF{\sameins{\OR{}}_k}$ of $h(t; \Gamma)$.
   Then, we have $f_k = h_k$.
   \end{enumerate}
 \begin{proof}
  We apply \cref{or-implicants} to the monotone functions $h, f_1, f_2$.
  \cref{gen-lb-pi} implies $\PI(h) \subseteq \PI(f_1) \cup \PI(f_2)$.
  In order to show the first statement,
  it remains to prove that $\PI(f_k) \subseteq \PI(h)$ for each $k \in \set{1, 2}$
  and that $\PI(f_1) \cap \PI(f_2) = \emptyset$.

  By \cref{gen-lb-impl}, any prime implicant $\rho$ of $f_1$
  is an implicant of $h$
  and must hence contain a prime implicant $\pi$ of $h$.
  By \cref{gen-lb-pi} and the definition of prime implicants,
  we have $\rho = \pi$,
  or $\pi$ is a prime implicant of $f_2$.
  Note that $\rho = \pi$ would imply $\rho \in \PI(h)$).
  Hence, to prove the first statement, it suffices to show
  the following claim.

  \begin{claim_no_num}
   If there are $\rho \in \PI(f_1)$ and $\pi \in \PI(h)$
   with $\lit(\pi) \subseteq \lit(\rho)$ and $\pi \in \PI(f_2)$,
   then $C$ is not a size-minimum delay-optimum circuit for $h$.
   \begin{proof_of_claim}
  Choose $i \in \set{0, \dotsc, m-1}$ maximum such that $t_i$ is contained in $\pi$.
  As $C$ is a formula circuit, $C_1$ and $C_2$ do not share any gates.
  Consider the circuit $B$ arising from $C$ by
  replacing $C_1$ with the reduced circuit $C_1 \mid_{t_i = 0}$.
  Note that $B$ is again a formula circuit.
  Write $g := f(B)$, and for $k \in \set{1, 2}$, write
  $B_k := B_{v_k}$ and $g_k := f(B_k)$.
  As $\rho$ contains $t_i$, by \cref{ess-pi}, $f_1$
  depends essentially on $t_i$.
  Hence, by \cref{obs-restr-circuit},
  we have $\delay(B) \leq \delay(C)$ and $\size(B) < \size(C)$.
  It remains to show that $B$ and $C$ are logically equivalent.

  Let $\alpha \in \I^m$.
  As $B$ is monotone and arises from $C$ by fixing an input to $0$,
  we have $g(\alpha) = 0$ whenever $h(\alpha) = 0$.
  Thus, assume that $h(\alpha) = 1$.
  Then, there is $\psi \in \PI(h)$ with $\psi(\alpha) = 1$.

  \textbf{Case 1:} We have $\psi \in \PI(f_2)$.

  Here, as $B_2 = C_2$, we have $g_2(\alpha) = f_2(\alpha) = 1$
  and thus $g(\alpha) = 1$.

  \textbf{Case 2:} We have $\psi \notin \PI(f_2)$.

  By \cref{gen-lb-pi}, we have $\psi \in \PI(f_1)$.
  As $\pi \in \PI(f_2)$, we must have $\psi \neq \pi$.
  By the choice of $t_i$,
  \cref{pi-disjoint} implies that $t_i \notin \lit(\psi)$.
  As any implicant $\iota$ of $f_1$ with $t_i \notin \lit(\iota)$
  is an implicant of $g_1$, we have $\psi \in \PI(g_1)$.
  This implies $g(\alpha) = 1$.

  Thus, $B$ is a delay-optimum formula circuit for $h$
  with better size than $C$.
  \end{proof_of_claim}
  \end{claim_no_num}

  Now, we show the second statement.
  For each $k \in \set{1, 2}$,
  let $\sameins{\OR{}}_k$ consist of the inputs among $\sameins{\OR{}}$
  that $f_k$ depends on essentially.
  By \cref{ess-pi}, a Boolean function depends essentially on an input $t_i$
  if and only if $t_i$ is contained in any of its prime implicants.
  By \cref{pi-disjoint}, for each input $t_i \in \sameins{\OR}$,
  there is exactly one prime implicant of $h$ containing $t_i$.
  Thus, the first statement implies $\sameins{\OR} = \sameins{\OR}_1 \cupdot \sameins{\OR}_2$.

  Now, assume the conditions of the third statement.
  By \cref{aop-prime-implicants} and the first two statements, the prime implicants of $f_k$ are
  $\set{t_i \land \LAND_{j < i, \gt(t_j) = \AND{}} t_j \where
   t_i \in \sameins{\OR}_k}\,$;
  and by \cref{aop-prime-implicants} and the definition of $h_k$,
  these are precisely the prime implicants of $h_k$.
  By \cref{pi-unique},
  we deduce $h_k = f_k$,
  hence the third statement.
\end{proof}
\end{lemma}

\begin{theorem}[Structure theorem] \label{structure-real-thm}
 Let $m \in \N_{\geq 2}$,
 inputs $t = \Bkt{t_0, \dotsc, t_{m-1}}$
 with arrival times $a(t_0), \dotsc, a(t_{m-1}) \in \N$
 and gate types $\Gamma = \Bkt{\circ_0, \dotsc, \circ_{m-2}}$ be given.
 Consider a delay-optimum formula circuit $C$ for $h(t; \Gamma)$ with minimum number of gates.
 Let $\circ := \gt(\cktout(C))$.
 Denote the predecessors of $v := \cktout(C)$ by $v_1$ and $v_2$.
 Write $f_1 := f(C_{v_1})$, and $f_2 := f(C_{v_2})$.
 Then,
 there is a partition $\sameins{\circ} = \sameins{\circ}_1 \cupdot \sameins{\circ}_2$
 into non-empty subsets such that
 for each $k \in \{1, 2\}$,
 the function $f_k$ depends essentially on the inputs of $\sameins{\circ}_k$,
 but not on those of $\sameins{\circ}_{3-k}$, and we have
 \[f_k = \genaopkeepF{\sameins{\circ}_k}\,.\]
 \begin{proof}
  By duality, it suffices to consider the case $\gt(\cktout(C)) = \OR$.
  In this case, the statements hold by \cref{structure-thm}.
 \end{proof}
\end{theorem}

As a consequence of this \lcnamecref{structure-real-thm},
we can derive an upper bound on the maximum number of inputs an \aop{}
may have such that an \aop{} circuit with depth $d$ exists.
For this, we need the following \lcnamecref{def-gen-aop-details}.

\begin{notation} \label{def-gen-aop-details}
 Let $h(t; \Gamma)$ with $t = \Bkt{t_0, \dotsc, t_{m-1}}$ and
 $\Gamma = \Bkt{\circ_0, \dotsc, \circ_{m-2}}$
 be a generalized \aop{}.
 Given $i \in \set{0, \dotsc, m-1}$,
 we define $\genaopnoti{}$ as the generalized \aop{} arising from $h(t; \Gamma)$
 by removing $t_i$, i.e.,
 \[ \genaopnoti{} := \begin{cases}
h\big((t_0, \dotsc, \widehat{t_i}, \dotsc, t_{m-1});
                       \Bkt{\circ_0, \dotsc, \widehat{\circ_i}, \dotsc, \circ_{m-2}}\big)
  & \text{ if } i \leq m - 2\,, \\ \vphantom{2^{2^{2^{2^2}}}}
h\big((t_0, \dotsc, t_{m-2}));
                       \Bkt{\circ_0, \dotsc, \circ_{m-3}}\big)
  & \text{ if } i = m - 1\,.
                     \end{cases}
 \]
We extend this notation to removal of a subset $F = \set{t_{i_0}, \dotsc, t_{i_{f-1}}}$ of inputs by
$\genaopnox{F} := \bktRfixed{Big}{\bktRfixed{big}{\genaopnotx{i_0}}_{\widehat{t_{i_1}}}}_{\dotsc}$.
\end{notation}

The upper bound presented in the following \lcnamecref{max-m-for-d}
can be seen easily;
we presume that much stronger bounds
can be derived from \cref{structure-real-thm}.

\begin{corollary} \label{max-m-for-d}
  If an \aop{} $h(t)$ on $m \geq 3$ inputs can be realized by a circuit $C$ with depth $d + 1$,
  then an \aop{} on $\tilde m := \BktU{\frac{m}{2}}$ inputs can by realized by a circuit with depth $d$.
 \begin{proof}
  W.l.o.g., we may assume that $C$
  is a depth-optimum formula circuit with depth $d+1$
  with minimum number of gates for $h(t)$.
  Dualization allows us to assume that $\cktout(C) = \OR{}$.
  By \cref{structure-real-thm},
  there are circuits $C_1$ and $C_2$
  with depth at most $d$ each
  that realize generalized \aop{}s $f_1$ and $f_2$, respectively, such that
  $C = C_1 \lor C_2$.
  Consider the corresponding partition $S^{\OR} = S^{\OR}_1 \cupdot S^{\OR}_2$
  of the same-gate signals of $h(t)$ as in \cref{structure-real-thm}.

  As $h(t)$ is an \aop{} and $m \geq 3$, we have $D^{\OR} \neq \emptyset$.
  As $h(t)$ is an \aop{} and $t_{m-1} \in \sameins{\OR}$,
  for every $t_i \in D^{\OR}$, we have $t_{i+1} \in \sameins{\OR}$.
  Hence, the function
  \[\vartheta \colon D^{\OR} \to S^{\OR}, \quad
  t_i \mapsto t_{i+1}\]
  is well-defined.
  For $k \in \set{1, 2}$, let $D^{\OR}_k := \vartheta^{-1}\Bkt{S^{\OR}_k}$.
  Note that $D^{\OR} = D^{\OR}_1 \cupdot D^{\OR}_2$.

  Now, for each $k \in \set{1, 2}$,
  let $B_k$ denote the reduced circuit arising from $C_k$ by
  fixing all inputs $t_i \in D_{3-k}$ to $\alpha := 1$,
  and let $g_k := f(B_k)$.
  Then, as all inputs in $D_{3-k}$ are \AND{}-signals,
  by considering the standard circuit for $h(t)$,
  we observe that $g_k = \Bkt{f_k}_{\widehat{D^{\OR}_{3-k}}}$.
  By construction, the essential variables of $g_k$
  are the variables of $S^{\OR}_k$ and $D^{\OR}_k$.
  Let $t_{j_k}$ be the essential variable of $g_k$
  with $j_k$ maximum.

  Consider $k \in \set{1, 2}$.
  We show that $g_k$ is an \aop{}:
  First note that by \cref{def-subpath} and the choice of $\alpha$,
  every input of $g_k$ except for $t_{j_k}$
  is an \AND{}-signal (\OR{}-signal) of $g_k$
  if and only if it is an \AND{}-signal (\OR{}-signal of $h(t)$.
  By definition of $\vartheta$,
  for any two \OR{}-signals $t_i, t_j$ of $g_k$
  with $i < j < j_k$,
  the \AND{}-signal $t_{j-1} = \vartheta^{-1}(t_j)$ of $h(t)$
  is an input of $g_k$.
  Furthermore, for any two \AND{}-signals $t_i \neq t_j$
  of $g_k$ with $i < j < j_k$,
  the \OR{}-signal $t_{i+1} = \vartheta(t_i)$ of $h(t)$
  is an input of $g_k$.
  Hence, the inputs of $g_k$
  (except for $t_{j_k})$
  are alternatingly \AND{}- and \OR{}-signals
  and $g_k$ is an \aop{}.

  Let $m_1$ and $m_2$ be the numbers of inputs of $B_1$ and $B_2$, respectively.
  As
  \[\set{t_0, \dotsc, t_{\widetilde m-1}}
    = \sameins{\OR} \cupdot D^{\OR}
    = \sameins{\OR}_1 \cupdot \sameins{\OR}_2 \cupdot D^{\OR}_1 \cupdot D^{\OR}_2\,,\]
  we have $m_1 + m_2 = m$.
  Choose $i \in \set{1, 2}$ such that $m_i$ is maximum.
  Then, we have $m_i \geq \tilde m$.
  As $B_i$ is an \aop{} circuit on at least $\tilde m$ inputs with depth at most $d$,
  the \lcnamecref{max-m-for-d} is proven.
 \end{proof}
\end{corollary}

Note that this \lcnamecref{max-m-for-d} is much weaker than our structure theorem
(\cref{structure-real-thm}):
We only use that the delay of the two sub-circuits of $C$ is at least
the delay of two specific \aop{}s,
not the concrete structure of $C$.
Hence, instead of deriving \cref{max-m-for-d} from our structure theorem,
we could also have generalized Lemma 1 from \citet{Commentz-Walter-mon}
-- which is proven only for \aop{}s with an even number of inputs --
to \aop{}s with an arbitrary number of inputs.
From this generalized result, the \lcnamecref{max-m-for-d} also follows.

For the special case when all input arrival times are equal,
we conjecture that partitions of the same-gate inputs
into two ``non-overlapping`` sets are always best for the delay.

\begin{conjecture} \label{structure-conj}
 Consider \cref{structure-real-thm} for the case of depth optimization
 and let $\sameins{\circ} = \sameins{\circ}_1 \cupdot \sameins{\circ}_2$
 be a partition as in the theorem. Then, there is a $k \in \{1,2\}$ such that for all inputs
 $t_i \in \sameins{\circ}_k$ and $t_j \in \sameins{\circ}_{3-k}$,
 we have $i < j$.
\end{conjecture}

Note that in \totalref{fig-non-size-opt-2}, \cref{structure-conj}
is fulfilled.
For instance, for the outermost partition, we have
$\sameins{\AND} = \set{t_0, t_2, t_4} \cupdot \set{t_6, t_8, t_{10}, t_{12}, t_{13}}$.
As mentioned in \cref{sec-prev},
we conjecture that the polynomial-time delay-optimization algorithm for \aop{}s by \citet{BH-practice}
is actually an exact algorithm for the special case of depth optimization,
and in this algorithm, precisely those partitions described in \cref{structure-conj} are considered.
  % 3

\section{General Algorithm} \label{sec-opt-alg}

The structure theorem from the previous \lcnamecref{sec-structure}
motivates an exact algorithm for
the \prgenaopdelayopt{}:
Consider a generalized \aop{} $h(t; \Gamma)$ with prescribed input arrival times.
Assume that we know a delay-optimum formula circuit
for all strict sub-paths of $h(t; \Gamma)$.
Then, by \cref{structure-real-thm},
there are $\circ \in \set{\AND{}, \OR{}}$ and a partition
$\sameins{\circ} = \sameins{\circ}_1 \cupdot \sameins{\circ}_2$
such that a delay-optimum circuit $C$ for $h(t; \Gamma)$
can be obtained from
delay-optimum circuits $C_1$ for $\genaopkeepF{\sameins{\circ}_1}$ and
$C_2$ for $\genaopkeepF{\sameins{\circ}_2}$ via
$C = C_1 \circ C_2$.
\cref{alg::opt_dp} describes the arising recursive algorithm
for computing the optimum delay; by backtracking,
an optimum formula circuit can be computed.
The sub-paths of $h(t; \Gamma)$ arising during the algorithm
are identified with non-empty subsets $I$ of $\set{t_0, \dotsc, t_{m-1}}$
via an injective map $\kappa$
which maps each sub-path to its essential inputs.
It is not hard to see that $\kappa$ is actually a bijection.
\cref{alg::opt_dp}
recursively applies \cref{structure-real-thm}
and stores the computed delays $d(I)$
for non-empty subsets $I \subseteq \set{t_0, \dotsc, t_{m-1}}$
in a  dynamic-programming table of size at most $2^{m} - 1$.

\begin{figure}[ht]
 \begin{algorithm}[H]
   \DontPrintSemicolon
    \SetKwFunction{funcopt}{compute\_opt}
    \SetKwProg{myproc}{procedure}{}{}
   \KwIn{Inputs $t = \Bkt{t_0, \dotsc, t_{m-1}}$
    with arrival times $a(t_0), \dotsc, a(t_{m-1}) \in \N$,
    and gate types $\Gamma = \Bkt{\circ_0, \dotsc, \circ_{m-2}}$.}
    \KwOut{Optimum delay of any circuit over $\set{\AND{}2, \OR{}2}$ computing $h(t; \Gamma)$.}
    \BlankLine
    \ForEach{$\emptyset \neq I \subseteq \set{t_0, \dotsc, t_{m-1}}$}
    {
     Set $d(I) := \infty$.\;
     }
     \KwRet{\funcopt{$\set{t_0, \dotsc, t_{m-1}}$}}\;
     \myproc{\funcopt{$I$}}
    {
    Assume that $\emptyset \neq I = \set{t_{i_0}, \dotsc, t_{i_{r-1}}}$ with
    $0 \leq i_0 < \dotsc < i_{r-1} \leq m-1$
    and let $\Gamma' := \Bkt{\circ_{i_0}, \dotsc, \circ_{i_{r-2}}}$.\;
    \If{$d(I) < \infty$}
    {
     \KwRet{$d(I)$}\;
    }
    \If{$r = 1$}
    {
       Set $d(I) =a(t_{i_{r-1}})$.\;
       \KwRet{$d(I)$}\;
    }
    \ForEach{$\circ \in \set{\AND{}, \OR{}}$}
    {
      Let $\sameins{\circ} \subseteq I$ consist of all signals $t_{i_j}$
      with $\circ_{i_j} = \circ$ and the signal $t_{i_{r-1}}$.\;
      \ForEach{partition $\sameins{\circ} = \sameins{\circ}_1 \cupdot \sameins{\circ}_2$
       with $\sameins{\circ}_1, \sameins{\circ}_2 \neq \emptyset$ \label{alg-enum-part}}
      {\label{alg-partition}
         \ForEach{$k \in \set{1, 2}$}
         {\label{opt-alg-delay-comp}
            Let $I_{k}$ denote the input set of $h\Bkt{\Bkt{t_{i_0}, \dotsc, t_{i_{r-1}}}; \Gamma'}_{\sameins{\circ}_k}$.\;\label{alg-line-IK}
            Let $d_k :=$ \funcopt{$I_{k}$}.\;
         }
         Set $d(I) = \min \set{d(I), \max\set{d_1, d_2} + 1}$\,. \label{opt-alg-set-d}
      }
    }
    \KwRet{$d(I)$}\;
    }
    \caption{Exact algorithm for delay optimization of generalized \aop{}s}
   \label{alg::opt_dp}
 \end{algorithm}
\end{figure}

In the following \lcnamecref{thm-opt-alg},
we estimate the running time of \cref{alg::opt_dp}.

\begin{theorem} \label{thm-opt-alg}
 Let inputs $t = \Bkt{t_0, \dotsc, t_{m-1}}$
 with arrival times ${a(t_0), \dotsc, a(t_{m-1}) \in \N}$
 and gate types $\Gamma = \Bkt{\circ_0, \dotsc, \circ_{m-2}}$ be given.
 Then, \cref{alg::opt_dp} computes the optimum delay
 of any circuit realizing the generalized \aop{} $h(t; \Gamma)$.
 The  dynamic-programming table needed to store the delay of all
 sub-paths considered during the computation has exactly $2^m - 1$ entries.
 Denoting by $a$ and $o$ the number of \AND{}-signals and \OR{}-signals
 among $t_0, \dotsc, t_{m-2}$,
 the algorithm can be implemented to run in time $\mathcal O(3^a 2^o + 2^a 3^o)$.
 In particular, if $h(t; \Gamma)$ is an \aop{},
 then the running time is $\mathcal O\Bkt{\Bkt{\closesqrt{6}}^m}$.
 By backtracking, we can obtain a delay-optimum formula circuit for $h(t; \Gamma)$.
\begin{proof}
We have already argued that a sub-path of $h(t; \Gamma)$ arising from recursive
application of \cref{structure-real-thm} can be identified with
the set of its essential inputs via a bijection $\kappa$.
Hence, by induction on $m$ and \cref{structure-real-thm},
it is easy to see that \cref{alg::opt_dp} computes the optimum delay of any formula
circuit -- and thus of any circuit -- for $h(t; \Gamma)$.

We assume a random-access machine model with unit costs
that allows us to perform basic operations on integers of arbitrary size in constant time.
This way, we can represent a set of inputs $I$ by an integer
and access its table entry $d(I)$ in constant time.

First, we show that the algorithm can be implemented such that each execution of \cref{alg-line-IK} takes constant time:
Before starting to enumerate partitions in \cref{alg-enum-part}, for each input $t_i$,
we precompute the set $\diffins{\circ}_i \subseteq \diffins{\circ}$
of diff-gate inputs $t_j$
of $h\bkt{\bkt{t_{i_0}, \dotsc, t_{i_{r-1}}}; \Gamma'}$ with $j < i$.
Then, in \cref{alg-line-IK}, the set $I_k$ can be computed as
$I_k = \sameins{\circ}_{k} \cup \diffins{\circ}_{i_q}$,
where $q := \max\BktC{i \mid t_i \in \sameins{\circ}_{k}}$.
Here, we assume that $i_q$ is known by keeping track
of the largest index contained in each occurring same-gate set while enumerating partitions.
The precomputation only takes polynomial time and is dominated by the following
exponential-time partition enumeration.

Now, let $T := \set{t_0, \dotsc, t_{m-1}}$.
By the above, the running time of \cref{alg::opt_dp}
is dominated by enumerating all partitions of the respective set $\sameins{\circ}$
in \cref{alg-partition} for each $\circ \in \set{\AND{}, \OR{}}$
and for all subsets $\emptyset \neq I \subseteq T$.
A partition of $\sameins{\OR{}}$ into $2$ non-empty subsets corresponds to choosing a subset
$\sameins{\OR{}}_1 \subseteq \sameins{\OR{}} \backslash \set{t_{i_{r-1}}}$
and setting $\sameins{\OR{}}_2 := \sameins{\OR{}} \backslash \sameins{\OR{}}_1$.
By \cref{def-subpath}, the \specialsubpath{}s $\genaopkeepF{\sameins{\OR}_1}$
and $\genaopkeepF{\sameins{\OR}_2}$
are uniquely determined by $I$, $\sameins{\OR{}}$ and $\sameins{\OR{}}_1$.

Hence, it remains to bound the number of sets
$\sameins{\OR}_1 \subsetneq \sameins{\OR} \subsetneq I$ considered during the algorithm.
For fixed $I$, $\sameins{\OR{}}$ and $\sameins{\OR{}}_1$,
the following holds:
An \AND{}-signal of $h(t; \Gamma)$ may by in $I$ or in $T \backslash I$.
Each \OR{}-signal of $h(t; \Gamma)$ has three options:
it is contained in $\sameins{\OR{}}_1$, in $\sameins{\OR{}} \backslash \sameins{\OR{}}_1$
or in $\set{t_0, \dotsc, t_{m-1}} \backslash \sameins{\OR{}}$.
By convention, $t_{m-1}$ is either contained in $\sameins{\OR{}} \backslash \sameins{\OR{}}_1$
or in $\set{t_0, \dotsc, t_{m-1}} \backslash \sameins{\OR{}}$.
Hence, there are at most $2 \cdot 3^o 2^a$ partitions for the case that the split gate is an \OR{}.

Similarly, when $\circ = \AND{}$, we have $2 \cdot 3^a 2^o$ partitions.
Summing up yields the running time bound.

If $h(t; \Gamma)$ is an \aop{},
we have $a, o \in \set{\BktD{\frac{m-1}{2}}, \BktU{\frac{m-1}{2}}}$,
so the running time follows directly.
\end{proof}
\end{theorem}

Note that the formula circuit constructed by our algorithm is strongly delay-optimum.
In our implementation of the algorithm,
as an option, we can compute a size-optimum circuit among
all strongly delay-optimum formula circuits by storing both delay and size
for each sub-path in \cref{opt-alg-set-d} and updating them accordingly.
The algorithm could also be adapted to compute a
size-optimum circuit among all delay-optimum formula circuits
by storing a set of non-dominated candidate circuits
regarding delay and size for each sub-path.
As we see in \totalref{depth-opt-runtimes},
already our restricted size optimization increases the running time a lot,
so we did not implement this extension.
   % 4

\section{Improved Algorithm for Depth Optimization of \AOP{}s} \label{sec-opt-alg-depth}

In this \lcnamecref{sec-opt-alg-depth},
we speed up \cref{alg::opt_dp} for the special case
of depth optimization of \aop{}s.
For this, we partition all sub-paths considered during the algorithm
into so-called \define{$sp$-equivalence classes}, where
two sub-paths with \propgenpart{}s
$P_0 \tupleconcat \dotsc \tupleconcat P_c$
and $P'_0 \tupleconcat \dotsc \tupleconcat P'_{c'}$
are considered as \define{$sp$-equivalent} if and only if
$c = c'$ and $|P_b| = |P'_b|$ for all $b \in \set{0, \dotsc, c}$.
Then, up to renaming of the input variables,
any two $sp$-equivalent \aop{}s are either logically equivalent
or dual to each other,
i.e., the delays of optimum circuits for them coincide.
Thus, for each $sp$-equivalence class,
it suffices to compute
the optimum delay only for one sub-path.
Recall that we identify each sub-path with its essential inputs.
Whenever we compute an optimum circuit for
inputs $t'$ during the algorithm,
we instead compute an optimum solution for its \define{$sp$-representative} $\tilde t$.
We define $\tilde t$ by mapping
each input of $t'$ to a certain input in $\set{t_0, \dotsc, t_{m-1}}$.

Assume that $t' = \bkt{t_{i_0}, \dotsc, t_{i_r}}$ with
$0 \leq i_0 < \dotsc < i_r \leq m-1$.
We always map $t_{i_0}$ to $t_0$.
For $0 \leq j < r - 1$, assuming that input $t_{i_j}$ is mapped to $t_k$,
the input $t_{i_{j+1}}$ is mapped to $t_{k+1}$ if the gate types of
$t_{i_j}$ and $t_{i_{j+1}}$ are different, and to $t_{k+2}$ otherwise.
As always, the last input $t_{i_r}$ has no specified gate type
and is mapped to the next free input.
For instance, given an \aop{} on inputs $t = \Bkt{t_0, \dotsc, t_{10}}$ with gate types
$\Gamma = \set{\land, \lor, \dotsc, \land}$,
the sub-path on inputs $t' := \Bkt{t_2, t_5, t_6, t_8, t_9, t_{10}}$
is mapped to its $sp$-representative $\tilde t := \Bkt{t_0, t_1, t_2, t_4, t_5, t_6}$.

Furthermore, during partitioning, we avoid generating redundant partitions:
For instance, consider again the $sp$-representative $\tilde t$ from above,
and let $P_0 \tupleconcat \dotsc \tupleconcat P_c$ be its \propgenpart{}.
Note that $c = 3$ with
$P_0 = (t_0)$, $P_1 = (t_1)$, $P_2 = (t_2, t_4)$, $P_3 = (t_5, t_6)$.
We have $\sameins{\AND{}} = \set{t_0, t_2, t_4, t_6}$.
Here, the partitions $\sameins{\AND{}} = \set{t_0, t_2} \cupdot \set{t_4, t_6}$
and $\sameins{\AND{}} = \set{t_0, t_4} \cupdot \set{t_2, t_6}$
lead to the sub-paths $f_1$ on $(t_0, t_1, t_2)$ and $f_2$ on $(t_1, t_4, t_5, t_6)$,
or $g_1$ on $(t_0, t_1, t_4)$ and $g_2$ on $(t_1, t_2, t_5, t_6)$, respectively.
For both $f_1$ and $g_1$, the $sp$-representative is the sub-path on $(t_0, t_1, t_2)$,
and for both $f_2$ and $g_2$, it is the sub-path on $(t_0, t_1, t_2, t_3)$.
More generally, two partitions
$\sameins{\circ{}} = \sameins{\circ{}}_1 \cupdot \sameins{\circ{}}_2$
and
$\sameins{\circ{}} = T^{\circ{}}_1 \cupdot T^{\circ{}}_2$
lead to the same $sp$-representatives if
$|\sameins{\circ{}}_1 \cap P_b| = |T^{\circ{}}_1 \cap P_b|$
for every $b \in \set{0, \dotsc, c}$.
Hence, it suffices to consider those partitions
$\sameins{\circ{}} = \sameins{\circ{}}_1 \cupdot \sameins{\circ{}}_2$
for which
$\sameins{\circ{}}_1 \cap P_b$ is a prefix of $P_b$ for all $b \in \set{0, \dotsc, c}$
(in the example, this is the partition $\sameins{\AND{}} = \set{t_0, t_2} \cupdot \set{t_4, t_6}$).
We call such partitions \define{$sp$-conform}.

The procedure to restrict all computations
to $sp$-representatives and $sp$-conform partitions
is called \define{depth normalization}.
We will see that running \cref{alg::opt_dp} with depth normalization leads to
much better theoretical and practical running times.
As a first step, we now estimate the number of $sp$-representatives.

\begin{theorem} \label{opt-aop-depth-runtime}
Let $m \geq 1$ and an \aop{} $h(t)$ on inputs $t = \Bkt{t_0, \dotsc, t_{m-1}}$
with arrival times $a \equiv 0$ be given.
Then, the number of $sp$-representatives is
exactly $F_{m+1} \in \mathcal O(\varphi^m)$,
where $F_{m+1}$ is the $(m+1)$-th Fibonacci number and
$\varphi := \frac{1 + \sqrt{5}}{2} \approx 1.618$ is the golden ratio.
\begin{proof}
We show that the indicator vectors $(x_0, \dotsc, x_{m-1})$ of $sp$-representatives
are exactly the $0$-$1$ strings of length $m$ with the following properties:
\begin{enumerate}
 \item We have $x_0 = 1$.
 \item Whenever for some $i \in \set{2, \dotsc, m-1}$, we have
 $x_i = 1$, then $x_{i-1} \neq 0$ or $x_{i-2} \neq 0$.
 \item Choose $i \in \set{0, \dotsc, m-1}$ maximum with $x_i = 1$.
 Then, we have $i = 0$ or $x_{i - 1} = 1$.
\end{enumerate}

For any $sp$-representative, all conditions are fulfilled:
The first property is valid as all $sp$-representatives have at least $1$ input
and the first input is always mapped to $t_0$.
The second property holds as otherwise, $t_{i-2}$
could have been chosen instead of $t_{i}$ during normalization.
The last property holds as
each $sp$-representative has at least one entry and as
the last input is always mapped to the next free position,
ignoring gate types.

For the other direction, it is easy to see that for any
input vector $t'$ whose indicator vector satisfies the conditions above,
the $sp$-representative is again $t'$.

Denote the set of $0$-$1$ strings of length $m$
with the properties above by $G_m$.
By induction, we will show that $|G_m| = F_{m+1}$.

We have $|G_1| = |\set{(1)}| = 1 = F_2$
and $|G_2| = |\set{(1, 0), (1, 1)}| = 2 = F_3$.

Now, consider $m \geq 3$.
We construct $G_m$ by adding prefixes to elements of $G_{m-1}$ and $G_{m-2}$:
By prepending $(1)$ to any element of $G_{m-1}$,
we obtain all elements of $G_m$ that start with $(1, 1)$.
By prepending $(1, 0)$ to elements of $G_{m-2}$,
we obtain all elements of $G_m$ starting with $(1, 0, 1)$
and, additionally, the invalid element $(1, 0, 1, 0, 0, \ldots, 0)$
which violates the third rule.
Moreover, so far, we miss all elements of $G_m$ starting with $(1, 0, 0)$.
But by the second condition, the only such valid element is $(1, 0, 0, \dotsc, 0)$.
Together with the induction hypothesis, we obtain
\[|G_m| = |G_{m-1}| + \Bkt{|G_{m-2}| - 1} + 1 = F_{m} + F_{m-1} = F_{m+1}\,. \tag*{\qedhere}\]
 \end{proof}
\end{theorem}

Hence, using depth normalization, we reduce the number
of sub-paths for which an optimum circuit
is computed during \cref{alg::opt_dp}
from $2^m - 1$ to $F_{m+1} \in \mathcal O(1.619^m)$.
Using similar, but more involved techniques,
we will estimate the running time of \cref{alg::opt_dp}
with depth normalization in \cref{thm-depth-table-padovan}.
For this, we will prove
that the $sp$-conform partitions partitions considered in the algorithm
essentially correspond to elements of the set $Q_n$
defined as follows (with $n \approx \frac{m}{2}$).

\begin{table}
 \centering
 \begin{tabularx}{\textwidth}{llLLLL}
 \toprule
  $(x_{2i}, x_{2i+1})$ & $(0, 1)$ & $(1, 0)$ & $(1, 1)$ & $(2, 0)$ & $(2, 1)$ \\
  Allowed values for $(x_{2i-2}, x_{2i-1})$
 & all but $(1, 0)$, $(2, 0)$
 & all
 & all
 & all but $(1, 0)$
 & all but $(1, 0)$     \\
 \bottomrule
 \end{tabularx}
 \caption{Extension rules for $Q_n$.}
 \label{table-cond-Q}
\end{table}

\begin{definition}
 Let $Q_1 := \set{(0, 1), (1, 0), (1, 1), (2, 0), (2, 1)}$,
 and for $n \in \N_{\geq 2}$, let
 $Q_n$ be the set of $0$-$1$-$2$ strings
 $(x_0, \dotsc, x_{2n-1})$ of length $2n$
 such that for all $i \in \set{0, \dotsc, n-1}$, the following
 conditions are fulfilled:
 \begin{enumerate}
  \item We have $(x_{2i}, x_{2i+1}) \in Q_1$.
  \item If $i > 0$,
  the entries $(x_{2i-2}, x_{2i-1}, x_{2i}, x_{2i+1})$
  satisfy the extension rules from \cref{table-cond-Q}.
 \end{enumerate}
\end{definition}

For instance, we have $(2, 1, 1, 0) \in Q_2$ and $(1, 0, 2, 0) \notin Q_{2}$.
Note that the rules imply that there are no consecutive zeroes;
and that for $x = \Bkt{x_0, \dotsc, x_{2n-1}} \in Q_{n-1}$ and $y = \Bkt{y_0, y_1} \in Q_1$,
we have $x \tupleconcat y \in Q_n$ if and only if
$\Bkt{x_{2n-2}, x_{2n-1}, y_0, y_1}$ fulfill the extension rules.

\begin{lemma} \label{lemma-Q-recursion}
 Let $n \in \N_{\geq 1}$.
 Then, we have $|Q_n| \in \mathcal O(\beta_1^n)$,
 where
\[\beta_1 =
\frac{1}{3} \Bkt{5 + \sqrt[3]{\frac{1}{2} \Bkt{97 - 3 \sqrt{69}}} + \sqrt[3]{\frac{1}{2}\Bkt{97 + 3 \sqrt{69}}}}
< 4.08
\]
is the unique real root of the polynomial $\pi(x) := x^3 - 5 x^2 + 4x - 1$.
\begin{proof}
 We will show that for $n \geq 4$, we have
 \begin{equation} \label{eq-q-recursion}
  |Q_n| = 5 |Q_{n-1}| - 4 |Q_{n-2}| + |Q_{n-3}|\,.
 \end{equation}
 From this, the statement can be deduced as follows:
 The characteristic polynomial $\pi(x)$ of $|Q_n|$ has three distinct roots
 $\beta_1, \beta_2, \beta_3$
 with
   $\beta_1 \approx 4.08$,
   $\beta_2 \approx 0.46 - 0.18 i$,
   $\beta_3 \approx 0.46 + 0.18 i$.
 Using a well-known statement about linear recurrence relations
 (see, e.g., Theorem 6.7.8 in \citet{conradie2015logic}),
 it follows that there are coefficients
 $\lambda_1, \lambda_2, \lambda_3 \in \mathbb{C}$
 such that for all $n \in \N_{\geq 1}$, we have
 $|Q_n| = \lambda_1 \beta_1^n + \lambda_2 \beta_2^n + \lambda_3 \beta_3^n$.
 Since $|\beta_2| = |\beta_3| < \beta_1$,
 we obtain $|Q_n| \in \mathcal O(\beta_1^n)$.

 Now, we show \cref{eq-q-recursion}.
 The idea is to construct $Q_{n}$ from $Q_{n-1}$, $Q_{n-2}$, and $Q_{n-3}$
 by adding suffixes.

 Let $F := \set{(1, 0, 0, 1), (2, 0, 0, 1), (1, 0, 2, 0), (1, 0, 2, 1)}$.
 The extension rules imply
 $Q_n = \Bkt{Q_{n-1} \times Q_1} \backslash \Bkt{Q_{n-2} \times F}$.
 Apart from elements of $Q_{n-1} \times Q_1$,
 the set $Q_{n-2} \times F$ also contains the set $Q_{n-3} \times \set{(1, 0, 2, 0, 0, 1)}$,
 which is not contained in $Q_{n-1} \times Q_1$ as the extension rules
 for $Q_{n-1}$ do not allow $(1, 0, 2, 0)$.
 Since the extension rules allow any element before $(1, 0)$,
 we obtain
 \[|Q_n| = \left|Q_{n-1} \times Q_1\right|
   - \Bkt{\left|Q_{n-2} \times F\right| - \left|Q_{n-3} \times \set{(1, 0, 2, 0, 0, 1)}\right|}\,.\]
 This implies \cref{eq-q-recursion} and thus the \lcnamecref{lemma-Q-recursion}.
 \end{proof}
\end{lemma}

In order to use the sets $Q_n$
for estimating our running time,
we need the following intermediate construction.

\begin{definition} \label{def-R}
For $n \in \N_{\geq 1}$,
let $R_n := \bigcup_{i = 1}^{n} \{a \tupleconcat 0^{(2(n-i))} \where a \in Q_i\}$
be the set of strings of length $2n$ arising from an element of any $Q_i$
with $i \in \set{1, \dotsc, n}$
by appending $2(n-i)$ zeroes.
\end{definition}

\begin{observation} \label{obs-Rn}
Note that by \cref{def-R} and \cref{lemma-Q-recursion}, we have
\[|R_n| = \sum_{i = 1}^n |Q_i| \in \mathcal O\Bkt{\sum_{i = 1}^n \beta_1^i}
= \mathcal O\Bkt{\beta_1^n}\,.\]
\end{observation}

\begin{theorem} \label{thm-depth-table-padovan}
Let an \aop{} $h(t)$ on $m \geq 1$ inputs $t = \Bkt{t_0, \dotsc, t_{m-1}}$
with arrival times $a \equiv 0$ be given.
Assume that \totalref{alg::opt_dp} with depth normalization
is applied to compute a depth-optimum circuit for $h(t)$.
Then, the resulting running time
is at most $\mathcal O\Bkt{\alpha^m}$,
where
\[\alpha := \sqrt{\beta_1} =
\sqrt{\frac{1}{3} \Bkt{5 + \sqrt[3]{\frac{1}{2} \Bkt{97 - 3 \sqrt{69}}} + \sqrt[3]{\frac{1}{2}\Bkt{97 + 3 \sqrt{69}}}}}
\leq 2.02
\]
and $\beta_1$ is defined as in \cref{lemma-Q-recursion}.
\begin{proof}
We again use a random-access machine model with unit costs.

We assume that for each sub-path,
the $sp$-representative can be determined in constant time
using a precomputed look-up table.
The table can be computed in time $\mathcal O(m \cdot 2^m)$,
which is dominated by the claimed running time.
Thus, as in the proof of \cref{thm-opt-alg},
it suffices to bound the number
of $sp$-conform partitions of $\sameins{\circ}$ considered by the algorithm for $sp$-representatives.
As in \cref{opt-aop-depth-runtime}, we encode this situation in strings
with certain properties and then estimate the number of these strings.

Consider an $sp$-representative $h(t'; \Gamma')$ with
input set $I = \{t_{i_0}, \dotsc, t_{i_{r-1}}\}$
such that $0 \leq i_0 < \dotsc < i_{r-1} \leq m-1$.
Let $I = P_0 \tupleconcat \dotsc \tupleconcat P_c$ be the \propgenpart{} of $h(t'; \Gamma')$,
and consider an $sp$-conform partition $\sameins{\circ} = \sameins{\circ}_1 \cupdot \sameins{\circ}_2$
of its same-gate inputs $\sameins{\circ} \subseteq I$.
By definition of $sp$-conformity,
$\sameins{\circ{}}_1 \cap P_b$ is a prefix of $P_b$ for all $b \in \set{0, \dotsc, c}$.
Let $x$ denote the $0$-$1$-$2$ string arising from $t$
by mapping each input that is not contained in $I$ to $0$,
each input $t_i \in \sameins{\circ}_1$ to $2$
and each input $t_i \in \sameins{\circ}_2$ to $1$,
and each other input of $I$ to~$1$.
Note that this is the same proof idea as in \cref{opt-aop-depth-runtime},
where there were $3$ possible states for the \OR{}-signals
and $2$ possible states for the \AND{}-signals.
We define
\[x' := \begin{cases}
                     x \tupleconcat (0) & \text{ if $t_0 \in \sameins{\circ}$,} \\
                     (0) \tupleconcat x & \text{ otherwise,}
                 \end{cases}
\quad \text{and} \quad x'' := \begin{cases}
                     x' & \text{ if $m$ odd,} \\
                     x' \tupleconcat (0) & \text{ otherwise.}
                 \end{cases}
\]
Now, $x''$ has $2n$ entries, where $n := \BktU{\frac{m+1}{2}}$.
We will show that $x'' \in R_n$.
From this, the result follows:
The mapping $(I, \sameins{\circ}, \sameins{\circ}_1, \sameins{\circ}_2)
\mapsto x''$ is clearly injective.
Hence, $|R_n|$ is an upper bound on the number of partitions considered.
We obtain a total running time of
\[\mathcal O(|R_n|) \overset{\shortcref{obs-Rn}}{\subseteq} \mathcal O(\beta_1^n) = \mathcal O(\beta_1^{m/2})
= \mathcal O(\alpha^m)\,.\]
Thus, it suffices to prove the following claim.
\begin{claim_no_num}
 We have $x'' \in R_n$.
\begin{proof_of_claim}
All elements of $\sameins{\circ} \backslash \{t_{i_{r-1}}\}$
correspond to even entries of $x''$.
As $x$ arises from an $sp$-representative,
$x$ does not contain two consecutive zeroes
except for trailing zeroes.
As $t_0 \in I$ by normalization and hence $x_0 \neq 0$,
the same holds for $x'$ and $x''$.

Now, it remains to show that for any $i \in \set{1, \dotsc, n-1}$
with $x_{2i} \neq 0$ or $x_{2i+1} \neq 0$,
the extension rules are fulfilled for
$(x_{2i - 2}, x_{2i-1}, x_{2i}, x_{2i+1})$.
The case that $(x_{2i - 2}, x_{2i-1}, x_{2i}, x_{2i+1})
= (x_{2i - 2}, 0, 0, 1)$ with $x_{2i - 2} \neq 0$
is already excluded as there must not be consecutive zeroes before a $1$.
The case $(x_{2i - 2}, x_{2i-1}, x_{2i}, x_{2i+1}) = (1, 0, 2, x_{2i+1})$
with $x_{2i+1}$ arbitrary
cannot occur as here, the entries in $t'$ corresponding to $x_{2i - 2}$
and $x_{2i}$ are in the same input segment $P_b'$ for some $b \in \set{0, \dotsc, c}$,
hence, by normalization, we have $x_{2i-2} \geq x_{2i}$.
All other configurations are permitted.

Hence, we have $x'' \in R_n$.
\end{proof_of_claim}
\end{claim_no_num}
This proves the \lcnamecref{thm-depth-table-padovan}.
\end{proof}
\end{theorem}

Apparently, the sequence $(|Q_n|)_{n \in \N}$
is given by sequence A012814 in the OEIS \citep{oeis},
which consists of every $5$th entry of the \define{Padovan sequence},
see sequence A000931 in the OEIS.
The growth rate of the Padovan sequence is given by
$\rho := \sqrt[5]{\beta_1}$
which is also known as the \define{plastic number}.
Hence, the running time of our algorithm with depth normalization
can also be expressed as
$\mathcal O\Bkt{\Bkt{\rho^{5/2}}^m}$.
 % 5

\section{Practical Implementation} \label{sec::opt-impl}

We implemented \totalref{alg::opt_dp} in a C++ program,
using $64$-bit bit sets to encode the sub-paths via the bijection $\kappa$
to subsets of $\set{t_0, \dotsc, t_{m-1}}$.
In order to obtain good practical running times,
we implemented several speedup techniques.
On most instances, these in particular
imply that we compute the delay for only a fraction of the sub-paths
from our  dynamic-programming table, see also \totalref{table-runtimes-pruning}.
Hence, we store the table
in a hash set, which violates the worst-case running time guarantee of \cref{alg::opt_dp},
but is much faster in practice and, more important, much less memory-consuming.

For describing our speedup techniques,
assume that we apply \cref{alg::opt_dp} to a generalized
\aop{} $h(t; \Gamma)$ with $m$ inputs and arrival times
$a(t_0), \dotsc, a(t_{m-1})$.
Moreover, when the procedure \funcopt is applied to a subset
$I \subseteq \set{t_0, \dotsc, t_{m-1}}$
with $I = \bktC{t_{i_0}, \dotsc, t_{i_{r-1}}}$ for
$0 \leq i_0 < \dotsc < i_{r-1} \leq m-1$,
we denote the corresponding sub-path by $h(t'; \Gamma')$
and its \propgenpart{} by
$P_0 \tupleconcat \dotsc \tupleconcat P_c$.

When we apply our algorithm for depth optimization,
we use the depth normalization as in \cref{opt-aop-depth-runtime}.
Most of the other speedups techniques are based on lower bounds and upper bounds
on the delay.
For a sub-path $h(t'; \Gamma')$,
we maintain not only the best delay of a circuit for $h(t'; \Gamma')$ computed so far
(and, in size-optimization mode, the best possible size for the best possible delay),
but also a lower bound on its delay.
Furthermore, when calling the procedure \funcopt, we assume that we are
given an additional parameter $D$ and are supposed to find a circuit with best delay
among all solutions with delay at most $D$.
When applying \funcopt recursively to the sub-functions considered in partitioning,
we hence may use an upper bound of $D-1$.

Now, it is possible that we do not find a solution when applying the procedure \funcopt.
In this case, we may update the lower bound to $D + 1$.
On the other hand, if during partitioning, we find a solution with delay $d \leq D$,
in case of the non-size-optimization mode, we are only interested
in another solution if it has delay strictly smaller than $d$,
and in case of the size-optimization mode, if it has delay at most $d$.
Hence, we may update the upper bound $D$ to $d - 1$ or $d$ in the respective mode
for the remaining partitions to be considered.
Note that if we allowed fractional arrival times,
we would not be able to subtract $1$ here.

We shall see later how we set $D$ for the outermost call of the algorithm
(cf.~\textit{delay probing}).
We call the mechanism that handles upper bounds during the algorithm
\define{upper bound propagation}.
Having very good lower and upper bounds has a high impact on the running time,
so we carefully use any information available to update our bounds.

Assume now that we apply \funcopt to compute a table entry, i.e.,
to find an optimum circuit for the sub-path $h(t'; \Gamma')$
with input set $I$
with delay at most $D$.
Before starting our partitioning process (see \cref{sec-opt-part}),
we compute several lower bounds as in \cref{sec-opt-lb}.
If any of these is larger than $D$, then we know that there is no
circuit with delay at most $D$ for $h(t'; \Gamma')$
and need not start the partitioning process.

\subsection{Lower Bounds} \label{sec-opt-lb}

A \define{basic lower bound} that can be computed quickly
for any generalized \aop{} $h(t'; \Gamma')$ arises from
the lower bounds in \cref{thm-general-circuit-lower-bound,min-1-2-lb}, i.e.,
\[ \max\set{\bktUfixed{Bigg}{\log_2{\sum_{j = i_0}^{i_{r-1}-1} 2^{a(t_{i_j})}}},
                     \max \bktCfixed{bigg}{\max_{t_{i_j} \in P_0} a(t_{i_j}) + 1,
                           \max_{t_{i_j} \in P_b : b > 0} a(t_{i_j}) + 2}}\,.
\]
Note that the first lower bound requires the arrival times to be natural numbers.

We use two other \define{strong lower bounds} that
each consider a specific restricted sub-path
$h(t''; \Gamma'')$ of $h(t'; \Gamma')$
with similar structural complexity.
For $h(t''; \Gamma'')$, we recursively apply the algorithm with depth bound $D$.
Either there is no solution, in which case $D+1$ is a lower bound
on the optimum delay for $h(t''; \Gamma'')$, thus also for $h(t'; \Gamma')$;
otherwise, we know the optimum delay for $h(t''; \Gamma'')$,
which is a lower bound for $h(t'; \Gamma')$.
This usually yields a strong lower bound,
but is very time-consuming.

First, only in the special case of depth optimization, we
consider the sub-path $h(t''; \Gamma'')$ arising from $h(t'; \Gamma')$ by keeping
only the largest input segment in the \propgenpart{} completely
and condensing each other input segment to a single input
(except for the last segment, which keeps $2$ inputs).
In the case of depth optimization, only the input-segment sizes matter,
so there are only $\mathcal O(m^3)$ of these sub-paths,
and it is not harmful to solve them optimally.

Secondly, also in the case of delay optimization,
we consider a restricted sub-path $h(t''; \Gamma'')$
that arises from removing a single input of $h(t'; \Gamma')$ in a way that
hopefully the optimum delay of any circuit for $h(t''; \Gamma'')$
is the same as for $h(t'; \Gamma')$.
Hence, among all inputs with the minimum arrival time,
we remove an input of the largest input segment.
Empirically, we see that in the case of depth optimization,
this lower bound is tight in $97 \%$ of its applications.
This matches the observation that if
we iteratively apply this lower bound $m$ times,
starting with a generalized \aop{} with optimum depth $d$,
the optimum depth changes only $d$ times,
where $d \ll m$.

\subsection{Partitioning the Same-Gate Inputs} \label{sec-opt-part}

For determining a solution with delay $D$ for a sub-path $h(t'; \Gamma')$
-- if it exists --,
we enumerate partitions
$\sameins{\circ} = \sameins{\circ}_1 \cupdot \sameins{\circ}_2$
of its same-gate input set $\sameins{\circ}$ for all $\circ \in \set{\AND, \OR}$
in \cref{alg-partition} of \cref{alg::opt_dp}.
In our implementation, we first choose $\circ := \circ_0$
because empirically, this more often yields a good circuit,
and afterwards the other gate type.
For both, we enumerate partitions of $\sameins{\circ}$ and recursively try
to find a solution with delay at {most $D$}.

We avoid generating too many partitions of a set $\sameins{\circ}$
by enumerating partitions in a specific order
and skipping certain partitions that provably do not lead to a better solution.
In a recursive approach, one by one, we assign the inputs
to $\sameins{\circ}_1$ or to $\sameins{\circ}_2$.
Here, just as in standard branch-and-bound algorithms,
we follow the idea to make the most important decisions first.
Recall from the proof of \cref{thm-opt-alg} that by convention,
the last input $t_{i_{r-1}}$ is always contained in $\sameins{\circ}_2$.

Now, we first enumerate the highest input index $i_l$
for which input $t_{i_l}$ is assigned to the other part, $\sameins{\circ}_1$.
Once $t_{i_l}$ is fixed, we have completely determined
which of the diff-gate inputs are contained in
in both $h(t'; \Gamma')_{\sameins{\circ}_1}$ and $h(t'; \Gamma')_{\sameins{\circ}_2}$,
or only in $h(t'; \Gamma')_{\sameins{\circ}_2}$.
Based on this, we compute another lower bound,
the \define{cross-partition lower bound},
by applying \cref{thm-general-circuit-lower-bound} to
all inputs of $h(t';\Gamma')$, where those inputs that are contained in both sub-functions
are counted twice,
and may stop when this lower bound exceeds $D$.

As $t_{i_l}$ is the input with the highest index in $\sameins{\circ}_1$,
we already know that all inputs $t_i \in \sameins{\circ}$
with $i > i_l$ must be in $\sameins{\circ}_2$.
It remains to enumerate those $t_i \in \sameins{\circ}$ with $i < i_l$.
They are assigned to the sets $\sameins{\circ}_1$ and $\sameins{\circ}_2$
recursively, in the order of decreasing arrival time,
and in case of ties, inputs with larger indices are considered first.
For each input, we first assign it to $\sameins{\circ}_2$ and recursively
continue with the other inputs;
and then assign it to $\sameins{\circ}_1$ and go into recursion.
This way, we in particular prioritize the construction of consecutive
sets $\sameins{\circ}_1$ and $\sameins{\circ}_2$, which often allows
finding an optimum solution quickly (cf.~\totalref{table-runtimes-pruning}).

Now, assume that we try to compute a circuit for $h(t'; \Gamma')$
with delay at most $D$
via a fixed partition
$\sameins{\circ} = \sameins{\circ}_1 \cupdot \sameins{\circ}_2$.
Before computing a solution, we evaluate all lower bounds available for the
two sub-instances, and stop if any of the lower bounds exceeds $D - 1$.
Otherwise, we recursively compute the table entries
of $h(t'; \Gamma')_{\sameins{\circ}_1}$ and $h(t'; \Gamma')_{\sameins{\circ}_2}$
with delay bound $D - 1$.
As already mentioned, based on whether we did find a solution or not,
we may update the lower bound for $h(t'; \Gamma')$.

Note that the lower bound $L$ on the best delay
achievable for $h(t'; \Gamma')$ is also a lower bound for all
sub-paths on a superset of the inputs $I$ of $h(t'; \Gamma')$.
Hence, if we have updated $L$ for $h(t'; \Gamma')$,
in \define{lower bound propagation},
we also update the lower bound for certain sub-paths whose inputs
are a superset of $I$.
Doing this for all supersets would be too costly;
so we only update lower bounds of supersets
which are already contained in our  dynamic-programming table
and arise from adding a single input.
For those sets whose lower bounds are improved, we recursively repeat this procedure.

If we did not find a solution with delay at most $D$ for the current partition,
we might discard a part of our enumeration tree
in \define{subset enumeration pruning}:
Consider the inputs of $\sameins{\circ}$ in the order $t_{j_0}, \dotsc, t_{j_p}$
in which we enumerate whether to assign them to $\sameins{\circ}_1$ or $\sameins{\circ}_2$;
i.e., when considering input $t_{j_k}$,
we have already assigned the inputs $t_{j_0}, \dotsc, t_{j_{k-1}}$ to one of the two
subsets.
If we add $t_{j_k}$ to $\sameins{\circ}_2$, the set $\sameins{\circ}_2$
is minimal among all sets that will arise from enumerating assignments for
the elements $t_{j_{k+1}}, \dotsc, t_{j_p}$.
The first assignment that will be tried for $t_{j_{k+1}}, \dotsc, t_{j_p}$
is to put them all into $\sameins{\circ}_1$.
Hence, when the computation of a solution
for this sub-path with delay at most $D$
was not successful because the \aop{} $h(t'; \Gamma')_{\sameins{\circ}_2}$
had too large delay, we already know that all other partitions with
$t_{j_0}, \dotsc, t_{j_k}$ unchanged will also not lead to delay at most $D$.
Hence, we can skip this part of our enumeration tree.
The same holds when adding $t_{j_k}$ to $\sameins{\circ}_1$.

Finally, we note that the running time for the computation
of a table entry highly depends on $D$.
Hence, when computing a table entry with a lower bound of $L$,
in \define{delay probing},
we in fact loop over all possible delays ${d \in \set{L, \dotsc, D}}$
with increasing $d$
and try to find a solution with delay $d$.
The first value $d$ for which a solution is found
is then the optimum delay of any circuit for $h(t'; \Gamma')$.
            % 6 (Practical Implementation)
\section{Computational Results} \label{sec::opt-comp}
In \cref{sec-del-opt-runtime},
we analyze results for delay optimization of \aop{}s and generalized \aop{}s.
Then, in \cref{sec-opt-runtime}, we consider the \praopdepthopt{}.
In particular, here we analyze all speedup techniques in detail, including their individual impact on the
empirical running time.
% In \cref{sec-del-opt-runtime,sec-opt-runtime},
% we examine the empirical running time of our algorithm,
% in particular the dramatic impact of our speedup techniques.
These speedups allow us to solve all instances of the  \praopdepthopt{} with up to $64$ inputs.
For this problem, we also compare our running times with those of
the previously best algorithm by \citet{Hegerfeld}
which only allows to solve instances with up to $29$ inputs.
In \cref{sec-opt-adder-depths},
% from our computational results, \cref{max-m-for-d}
% and the results by the heuristic from \citet{grinchuk2013low},
we derive the optimum depths of $n$-bit adder circuits
for $n$ that are a power of two
for up to $n = 8192$ bits.

All our tests ran on a machine with
two Intel(R) Xeon(R) CPU E5-2687W v3 processors,
using a single thread.

\subsection{Delay Optimization of \AOP{}s and Generalized \AOP{}s} \label{sec-del-opt-runtime}

% \begin{sidewaystable}[p]
\begin{table}[t!]

\centering
\begin{tabular}{ rrrrr}
\toprule
          & \multicolumn{2}{c}{\aop{}s [s]} & \multicolumn{2}{c}{Generalized \aop{}s [s]} \\
\cmidrule(lr){2-3} \cmidrule(lr){4-5}
\# inputs & With size opt. & No size opt. & With size opt. & No size opt.\\
\midrule
10 & 0.001    & 0.000  & 0.001    & 0.000   \\
20 & 0.674    & 0.002  & 0.701    & 0.001   \\
30 & 1628.027 & 0.023  & 1922.000 & 0.011   \\
40 &          & 12.944 &          & 0.336   \\
\bottomrule
\end{tabular}
\caption[Running times of the exact algorithm for delay optimization of \aop{}s.]{
Average running times for \cref{alg::opt_dp}
with speedups from \cref{sec::opt-impl}.
For each number of inputs,
we tested $100$ instances with randomly chosen integral arrival times and, in case of generalized \aop{}s, random gate types.}
\label{table-runtimes-ats-opt}
% \end{sidewaystable}

\end{table}

% First, we consider the case of delay optimization of \aop{}s and generalized \aop{}s.
In \cref{table-runtimes-ats-opt},
we state the average running times of our algorithm
on the following testbed:
For each number $m \in \set{10, 20, 30, 40}$ of inputs,
we created $100$ \aop{} instances with random integral arrival times
uniformly distributed among $\set{0, \dotsc, m-1}$
and $100$ generalized \aop{} instances where, additionally,
the gate types are chosen uniformly among \AND{}2 and \OR{}2.
We show the respective running times of our algorithm
both for the computation
of the optimum delay (``No size opt.''),
and of the optimum delay and optimum size of a strongly delay-optimum formula circuit
(``With size opt.'').
For lines where no running time is shown,
the memory limit of $300$ GB was attained on at least one instance.

We see that running times are much higher
when size optimization is enabled,
that is, our speedup techniques are particularly effective when size optimization is not required.
For $m=20$ inputs, average running times to solve \aop{} instances are $0.674$ seconds with size optimization
and only $0.002$ seconds without size optimization.
Without any speedups, our algorithm takes $13.4$ seconds on such instances -- cf.\ scenario 1 in \totalref{table-runtimes-pruning} --,
demonstrating the dramatic impact of our speedup techniques.
Not surprisingly, the speedup is even more substantial for larger $m$.

Finally, we note that the effectiveness of our pruning strategies varies drastically
with the arrival time profile,
thus also our running times.
For instance, for the \aop{} runs with size optimization,
the running times on instances with $30$ inputs vary from
$0.2$ seconds up to $5.4$ hours.
By examining instances with high running times,
we could most likely further improve our speedup techniques.

\subsection{Depth Optimization of \AOP{}s} \label{sec-opt-runtime}
Now, we consider the \praopdepthopt{}.
Note that up to duality, for this, there is exactly one instance for
a fixed number of inputs.
In \cref{depth-opt-runtimes}, we give a comparison of our algorithm,
i.e., \cref{alg::opt_dp} with depth normalization and
speedups from \cref{sec::opt-impl},
with the formula enumeration algorithm by \citet{Hegerfeld}.

Hegerfeld's algorithm finds size-optimum formula circuits
among all strongly depth-optimum formula circuits.
Note that Hegerfeld erroneously states that the algorithm
computes a size-optimum formula circuit
among all depth-optimum formula circuits.
For instance, for the \aop{} on $14$ inputs,
Hegerfeld reports a size of $18$ (see \totalref{depth-opt-runtimes}),
but in \totalref{fig-non-size-opt-1}, we see a depth-optimum
formula circuit with size~$17$.
This circuit is not strongly delay-optimum.

\begin{landscape}
\begin{table}
\begin{centering}
\begin{multicols}{2}
\newcommand{\mydashedline}{\multicolumn{6}{@{}c@{}}{\makebox[0.9\columnwidth]{\dashrule}}}
\TrickSupertabularIntoMulticols
  \tablefirsthead{%
  m & d & s & \citep{Hegerfeld} [s] & \multicolumn{2}{c}{\cref{alg::opt_dp} [s]} \\
  \cmidrule(lr){4-4}\cmidrule(lr){5-6}
            &       &      &    With size opt.          & With size opt.  & No size opt. \\
  \midrule
  }
  \tablehead{%
  m & d & s & \citep{Hegerfeld} [s] & \multicolumn{2}{c}{\cref{alg::opt_dp} [s]} \\
  \cmidrule(lr){4-4}\cmidrule(lr){5-6}
            &       &      &    With size opt.          & With size opt.  & No size opt. \\
  \midrule
  }
\begin{supertabular}{rrrrrr}
%   1  & 0   & 0    & 0                  &  0.000                   & 0.000\\
%   \mydashedline \\
%   2  & 1   & 1    & 0                  &  0.000                   & 0.000       \\
%   \mydashedline \\
%   3  & 2   & 2    & 0                  &  0.000                   & 0.000       \\
%   \mydashedline \\
%   4  & 3   & 3    & 0                  &  0.000                   & 0.000       \\
  5  & 3   & 5    & 0                  &  0.000                   & 0.000       \\
  6  & 3   & 6    & 0                  &  0.000                   & 0.000       \\
  \mydashedline \\
  7  & 4   & 7    & 0                  &  0.000                   & 0.000       \\
  8  & 4   & 9    & 0                  &  0.000                   & 0.000       \\
  9  & 4   & 10   & 0                  &  0.000                   & 0.000       \\
  10 & 4   & 13   & 0                  &  0.000                   & 0.000       \\
  \mydashedline \\
  11 & 5   & 13   & 0                  & 0.001                & 0.000       \\
  12 & 5   & 14   & 0                  & 0.002                & 0.000       \\
  13 & 5   & 16   & 0                  & 0.004                & 0.000       \\
  14 & 5   & 18   & 0                  & 0.005                & 0.000       \\
  15 & 5   & 20   & 1                  & 0.007                & 0.000       \\
  16 & 5   & 21   & 2                  & 0.008                & 0.000       \\
  17 & 5   & 24   & 4                  & 0.008                & 0.000    \\
  18 & 5   & 25   & 11                 & 0.009                & 0.000  \\
  19 & 5   & 29   & 27                 & 0.015                & 0.002   \\
  \mydashedline \\
  20 & 6   & 27   & 71                 & 0.234                & 0.005   \\
  21 & 6   & 28   & 180                & 0.358                & 0.007   \\
  22 & 6   & 31   & 463                & 0.588                & 0.008   \\
  23 & 6   & 32   & 1035               & 0.923                & 0.008   \\
  24 & 6   & 35   & 2893               & 1.259                & 0.007   \\
  25 & 6   & 36   & 7214               & 1.631                & 0.007   \\
  26 & 6   & 38   & 22661              & 2.097                & 0.007   \\
  27 & 6   & 40   & 60598              & 2.401                & 0.007   \\
  28 & 6   & 42   & $\leq$ 480960      & 2.680                & 0.007   \\
  29 & 6   & 44   & $\leq$ 2775000     & 2.763                & 0.007   \\
  30 & 6   & 47   &                    & 2.927                & 0.008   \\
  31 & 6   & 49   &                    & 2.991                & 0.008   \\
  32 & 6   & 53   &                    & 3.068                & 0.009   \\
  33 & 6   & 57   &                    & 3.159                & 0.010   \\
  \mydashedline \\
  \phantom{1231243} \\
  34 & 7   & 51   &                    &   1822\,\;~~~~       & 0.300  \\
  35 & 7   & 53   &                    &   2921\,\;~~~~       & 0.861  \\
  36 & 7   & 55   &                    &   5145\,\;~~~~       & 0.978  \\
  37 & 7   & 57   &                    &   8064\,\;~~~~       & 0.958  \\
  38 & 7   & 59   &                    &  13949\,\;~~~~       & 0.961   \\
  39 & 7   & 61   &                    &  19539\,\;~~~~       & 0.957    \\
  40 & 7   & 63   &                    &  33778\,\;~~~~       & 0.974    \\
  41 & 7   & 65   &                    &  53287\,\;~~~~       & 0.954    \\
  42 & 7   & 67   &                    &  87514\,\;~~~~       & 0.945      \\
  43 & 7   & 70   &                    & 143409\,\;~~~~       & 0.939     \\
  44 & 7   & $\leq$ 73  &              &                       & 0.945   \\
  45 & 7   & $\leq$ 76  &              &                       & 0.958    \\
  46 & 7   & $\leq$ 77  &              &                       & 0.941     \\
  47 & 7   & $\leq$ 83  &              &                       & 1.285     \\
  48 & 7   & $\leq$ 84  &              &                       & 1.406   \\
  49 & 7   & $\leq$ 84  &              &                       & 1.399   \\
  50 & 7   & $\leq$ 85  &              &                       & 1.404   \\
  51 & 7   & $\leq$ 89  &              &                       & 1.410   \\
  52 & 7   & $\leq$ 90  &              &                       & 1.405  \\
  53 & 7   & $\leq$ 93  &              &                       & 1.407    \\
  54 & 7   & $\leq$ 94  &              &                       & 1.409    \\
  55 & 7   & $\leq$ 98  &              &                       & 1.415      \\
  56 & 7   & $\leq$ 99  &              &                       & 1.410     \\
  57 & 7   & $\leq$ 104  &             &                      & 1.406     \\
  58 & 7   & $\leq$ 105  &             &                      & 1.395    \\
  59 & 7   & $\leq$ 109  &             &                      & 1.413     \\
  60 & 7   & $\leq$ 110  &             &                      & 1.425  \\
  \mydashedline \\
  61 & 8   & $\leq$ 111  &             &                      & 4574\,\;~~~~~ \\
  62 & 8   & $\leq$ 113  &             &                      & 8468\,\;~~~~~ \\
  63 & 8   & $\leq$ 114  &             &                      & 9729\,\;~~~~~ \\
  64 & 8   & $\leq$ 117  &             &                      & 9037\,\;~~~~~ \\
\end{supertabular}
\end{multicols}
\vspace{-.8cm}
\caption{Single-threaded running times of our algorithm
(with and without size optimization)
and Hegerfeld's formula enumeration algorithm \citep{Hegerfeld}.
Each line also shows the computed optimum depth $d$ of
any \aop{} circuit on $m$ inputs
and the optimum size $s$ of any strongly-depth optimum formula circuit.
Hegerfeld's running times are taken from \citep{Hegerfeld};
for $28$ and $29$ inputs, Hegerfeld ran his algorithm in parallel
and displayed wall time multiplied by number of threads.
Dashed lines separate instances with different~$d$.}
\label{depth-opt-runtimes}
\end{centering}
\end{table}
\end{landscape}

\makeatletter
{
\newcommand{\mydashedline}{\multicolumn{16}{@{}c@{}}{\makebox[0.93\linewidth]{\dashrule}}}
\begin{sidewaystable}[p!]
\small

\centering
\begin{tabular}{rrrrrrrrrrrrrrrr}
~\\
~\\
\toprule
& \multicolumn{3}{c}{Scenario 1} & \multicolumn{3}{c}{Scenario 2} & \multicolumn{3}{c}{Scenario 3} & \multicolumn{3}{c}{Scenario 4} & \multicolumn{3}{c}{Scenario 5}  \\
\cmidrule(lr){2-4}\cmidrule(lr){5-7}\cmidrule(lr){8-10}\cmidrule(lr){11-13}\cmidrule(lr){14-16}
$m$ & ~ $\log_2 E$ & $\log_2 P$ & $T$ [s] & ~ $\log_2 E$ & $\log_2 P$ & $T$ [s] & ~ $\log_2 E$ & $\log_2 P$ & $T$ [s] & ~ $\log_2 E$ & $\log_2 P$ & $T$ [s] & ~ $\log_2 E$ & $\log_2 P$ & $T$ [s]\\
\midrule
20     & 20     & 27     & 13.4      & 13     & 22     & 0.8        &   10     & 18     & 0.1       & 10     & 15     & 0.0    &  8     & 13      & 0.0   \\
21     & 21     & 29     & 35.4      & 14     & 23     & 1.6        &   10     & 19     & 0.1       & 10     & 15     & 0.0    &  9     & 13      & 0.0   \\
22     & 22     & 30     & 94.6      & 15     & 24     & 3.5        &   11     & 19     & 0.1       & 10     & 16     & 0.0    &  9     & 14      & 0.0   \\
23     & 23     & 31     & 237.9     & 16     & 25     & 6.9        &   11     & 20     & 0.2       & 11     & 16     & 0.0    &  9     & 14      & 0.0   \\
24     & 24     & 32     & 630.6     & 16     & 26     & 15.4       &   11     & 20     & 0.2       & 11     & 16     & 0.0    &  9     & 14      & 0.0   \\
25     & 25     & 34     & 1540.4    & 17     & 27     & 32.1       &   11     & 20     & 0.2       & 11     & 16     & 0.0    &  9     & 14      & 0.0   \\
26     & 26     & 35     & 4055.7    & 18     & 28     & 69.7       &   11     & 20     & 0.2       & 11     & 16     & 0.0    &  9     & 14      & 0.0   \\
27     & 27     & 36     & 10034.2   & 18     & 29     & 142.4      &   11     & 21     & 0.3       & 11     & 17     & 0.0    &  9     & 14      & 0.0   \\
28     & 28     & 38     & 25055.1   & 19     & 30     & 315.9      &   11     & 21     & 0.4       & 11     & 17     & 0.0    &  9     & 14      & 0.0   \\
29     &        &        &           & 20     & 31     & 642.0      &   12     & 22     & 0.8       & 11     & 17     & 0.0    &  9     & 14      & 0.0   \\
30     &        &        &           & 20     & 32     & 1406.2     &   12     & 23     & 1.4       & 11     & 17     & 0.0    &  9     & 14      & 0.0   \\
31     &        &        &           & 21     & 33     & 2939.7     &   12     & 24     & 2.6       & 11     & 17     & 0.0    &  9     & 14      & 0.0   \\
32     &        &        &           & 22     & 34     & 6445.6     &   12     & 25     & 6.8       & 11     & 17     & 0.0    &  9     & 14      & 0.0   \\
33     &        &        &           & 22     & 35     & 13062.4    &   12     & 26     & 14.3      & 11     & 17     & 0.0    &  9     & 14      & 0.0   \\
\mydashedline \\
34     &        &        &           &        &        &            &   17     & 31     & 623.0     & 16     & 26     & 14.1   &  11     & 19     & 0.3   \\
35     &        &        &           &        &        &            &   18     & 33     & 1666.7    & 17     & 27     & 34.2   &  12     & 21     & 0.9   \\
36     &        &        &           &        &        &            &   19     & 33     & 3066.7    & 18     & 27     & 45.4   &  12     & 21     & 1.0   \\
37     &        &        &           &        &        &            &   19     & 34     & 6013.4    & 18     & 28     & 60.9   &  12     & 21     & 1.0   \\
38     &        &        &           &        &        &            &   20     & 35     & 9211.6    & 19     & 28     & 75.8   &  12     & 21     & 1.0   \\
39     &        &        &           &        &        &            &   20     & 36     & 15861.5   & 19     & 28     & 92.2   &  12     & 21     & 1.0   \\
40     &        &        &           &        &        &            &   21     & 36     & 22140.3   & 19     & 29     & 109.8  &  12     & 21     & 1.0   \\
41     &        &        &           &        &        &            &          &        &           & 20     & 29     & 135.7  &  12     & 21     & 1.0   \\
42     &        &        &           &        &        &            &          &        &           & 20     & 29     & 145.1  &  12     & 21     & 0.9   \\
\vphantom{$\Bkt{2^{2^{2^2}}}$} $\dotsb$ & \multicolumn{9}{c}{} & & $\dotsb$ &  & & $\dotsb$ \\
59     &        &        &           &        &        &            &          &        &           & 21     & 29     & 224.7  &  12     & 22     & 1.4   \\
60     &        &        &           &        &        &            &          &        &           & 21     & 29     & 226.6  &  12     & 22     & 1.4   \\
\mydashedline \\
61     &        &        &           &        &        &            &          &        &           &        &        &        &  18     & 32     & 4574.9   \\
62     &        &        &           &        &        &            &          &        &           &        &        &        &  18     & 33     & 8468.1   \\
63     &        &        &           &        &        &            &          &        &           &        &        &        &  18     & 33     & 9729.8   \\
64     &        &        &           &        &        &            &          &        &           &        &        &        &  18     & 33     & 9037.3   \\
\bottomrule
\end{tabular}
\caption[Running times of exact algorithm for different speedup scenarios.]{
 Comparison of speedup scenarios for computing depth-optimum
 \aop{} circuits with $m = 20, \dotsc, 64$ inputs.
 The number of table entries computed is denoted by $E$,
 the number of partitions computed by $P$,
 and the running time in seconds by $T$.
 Dashed lines separate ranges of instances with the same optimum depth.}
\label{table-runtimes-pruning}
\end{sidewaystable}
}

\FloatBarrier

For our algorithm, \cref{depth-opt-runtimes} again shows running times
for the computation of the optimum depth and optimum size of a strongly depth-optimum formula circuit
(``With size opt.''),
and for the computation of the optimum depth only (``No size opt.'').
Hegerfeld's running times are taken from \citep{Hegerfeld}.
On any instance solved both by Hegerfeld's algorithm and our algorithm,
the computed optimum depths coincide;
and using our size-optimization, we verified that Hegerfeld
computes the optimum size of any strongly depth-optimum circuit on each instance.

For up to $14$ inputs, Hegerfeld's algorithm runs less than a second,
and the largest solved instance has $29$ inputs.
Our algorithm with size optimization
solves instances with up to
$33$ inputs within $3.2$ seconds; the largest instance we
can solve has $43$ inputs.
When we disregard circuit size and construct any depth-optimum circuit,
we can solve any instance with up to $64$ inputs within $3$ hours;
and any instance with up to $60$ inputs even in up to $1.5$ seconds.
Note that the running time increases drastically with increasing depth.
As our implementation uses $64$-bit bit sets to
encode
the sub-paths, we currently
cannot consider instances with more than $64$ inputs.
By adjusting the bit sets used,
this technicality can be overcome.
However, we do not expect to solve an instance with $110$ inputs,
where the next change in depth is likely,
see \cref{table-optimum-m}.

In order to examine the impact of our speedup techniques,
we define $5$ scenarios:
in scenario $1$, we run the basic algorithm without any enhancements;
in scenario $5$, we enable all speedup techniques from \cref{sec::opt-impl}.
The intermediate scenarios all add a selection of speedups to the previous scenario:
% \begin{multicols}{2}
\begin{itemize}
 \item Scenario $1$: No speedups.
 \item Scenario $2$: Add depth normalization.
 \item Scenario $3$: Add upper bound propagation, basic lower bound.
 \item Scenario $4$: Add cross-partition lower bound, subset enumeration pruning.
 \item Scenario $5$: Add strong lower bounds, lower bound propagation, delay probing.
\end{itemize}
% \end{multicols}

For each scenario, we ran the algorithm on all depth optimization instances
with at least $20$ inputs --
we only state results for an instance-scenario pair
if the running time is at most $8$ hours.
For each run, we store the number $E$ of
table entries for which
the partitioning process has been started
and the number $P$ of partitions considered.
In \cref{table-runtimes-pruning},
we show the logarithms of these numbers, rounded to the nearest integer,
and the running times.

In general, for fixed $m$,
the number of entries and partitions and the running time
reduces significantly with increasing scenario number.
From scenario $3$ on, we can solve the instance with $34$ inputs
within the running time limit of $8$ hours,
which is the first instance with an optimum depth of $7$.
Only when using all pruning techniques in scenario $5$,
we can solve the instance with $61$ inputs.
In particular, note that in contrast to scenarios $1$ - $4$,
in scenario~$5$, the running time does not necessarily increase
with increasing $m$.
In a range of inputs where the optimum depth does not increase
(e.g., from $34$ up to $60$ inputs),
our strong lower bounds have a high impact.
Note that, as stated in \cref{thm-opt-alg},
for each number $m$ of inputs,
for scenario $1$, we have $E \approx 2^m$,
and that the running time increases by a factor of roughly $\closesqrt{6}$
when $m$ increases by $1$.
For scenario 2, we have checked that -- as proven in \cref{opt-aop-depth-runtime} --
the precise number of entries for $m$ inputs
is exactly the Fibonacci number $F_{m+1}$.
Note that from $m$ to $m+1$,
the running time roughly doubles,
matching the running time guarantee
of $\mathcal O(2.02^m)$ shown in \cref{thm-depth-table-padovan}.

% \FloatBarrier

\subsection{Optimum Depths of Adder Circuits} \label{sec-opt-adder-depths}

\begin{table}[h]
\centering
 \begin{tabular}{p{0.5cm}rrrrrrrrr}
 \toprule
  $d$ & & \multicolumn{4}{c}{Using \citet{Hegerfeld} results}
      & \multicolumn{4}{c}{Using \cref{alg::opt_dp} results} \\  \midrule
%    0  & & 1    & $\leq m \leq $  & 1    &  & 1     &  $\leq m \leq$ &   1   &     \\
   1  & & 2    & $\leq m \leq $  & 2    &  & 2     &  $\leq m \leq$ &   2   &     \\
   2  & & 3    & $\leq m \leq $  & 3    &  & 3     &  $\leq m \leq$ &   3   &     \\
   3  & & 4    & $\leq m \leq $  & 6    &  & 4     &  $\leq m \leq$ &   6   &     \\
   4  & & 7   & $\leq m \leq $   & 10   &  &  7    &  $\leq m \leq$ &   10  &     \\
   5  & & 11   & $\leq m \leq $  & 19   &  & 11    &  $\leq m \leq$ &   19  &     \\
   6  & & 20    & $\leq m \leq $ & 33   &  & 20    &  $\leq m \leq$ & 33    &    \\
   7  & & 39    & $\leq m \leq $ & 60   &  & 34    &  $\leq m \leq$ & 60    &    \\
   8  & & 77    & $\leq m \leq $ & 109  &  & 61    &  $\leq m \leq$ & 109   &    \\
   9  & & 153   & $\leq m \leq $ & 202  &  & 121   &  $\leq m \leq$ & 202   &    \\
   10 & & 305   & $\leq m \leq $ & 375  &  & 241   &  $\leq m \leq$ & 375   &    \\
   11 & & 609   & $\leq m \leq $ & 698  &  & 481   &  $\leq m \leq$ & 698   &    \\
   12 & & 1217  & $\leq m \leq $ & 1311 &  & 961   &  $\leq m \leq$ & 1311  &    \\
   13 & & 2433  & $\leq m \leq $ & 2466 &  & 1921  &  $\leq m \leq$ & 2466  &    \\
   14 & &       &                &      &  & 3841  &  $\leq m \leq$ & 4645  &    \\
   15 & &       &                &      &  & 7681  &  $\leq m \leq$ & 8782  &    \\
   16 & &       &                &      &  & 15361 &  $\leq m \leq$ & 16627 &    \\
   17 & &       &                &      &  & 30721 &  $\leq m \leq$ & 31548 &    \\ \bottomrule
  \end{tabular}
 \caption{
 Numbers $m$ of inputs for which we can show that the optimum depth of an \aop{} circuit
 over the basis $\set{\AND{}2, \OR{}2}$ on $m$ inputs is $d$, for $d \leq 17$.
 In both columns, \citet{grinchuk2013low} yields the upper bounds on $m$;
 the lower bounds are derived
 from the results by \citet{Hegerfeld} and \cref{alg::opt_dp}, respectively,
 using our theoretical statement from \cref{max-m-for-d}.}
 \label{table-optimum-m}
\end{table}

In \cref{depth-opt-runtimes},
we see the optimum depth of \aop{} circuits for up to $64$ inputs.
From these, we will now deduce the optimum adder depths for
all $2^k$-bit adders with $k \leq 13$.

As a first step,
for fixed $d \in \N$, we want to determine ranges of inputs $m$
for which we can prove that the optimum depth of
any \aop{} circuit on $m$ inputs is $d$, see \cref{table-optimum-m}.
For this, we use the empirically good results from the heuristic by \citet{grinchuk2013low}.
There, a table in Section 5 displays the maximum $m$ such that the heuristic
computes a circuit with depth $d$ for all $d \in \set{0, \dotsc, 32}$.
From this table, we obtain the upper bounds on $m$ in \cref{table-optimum-m}.
Comparing with our results in \cref{depth-opt-runtimes},
we directly see that Grinchuk's circuits are depth-optimum for up to $109$ inputs.
But using \cref{max-m-for-d} and our result that an \aop{} on $61$ inputs
cannot be realized with depth $8$,
we deduce that an \aop{} on $121$ inputs has depth at least $9$.
Together with \cref{max-m-for-d}, this implies that
an \aop{} on $241$ inputs has depth at least $10$, and so on.
This yields the lower bounds on $m$ in the right part of \cref{table-optimum-m};
while the lower bounds in the left part can be obtained by applying \cref{max-m-for-d}
to the results computed by \citet{Hegerfeld}.
In Hegerfeld's case, we apply \cref{max-m-for-d} iteratively with the
basic result that an \aop{} on $20$ inputs cannot be realized with depth $5$
and thus obtain significantly smaller ranges of $m$
for which the optimum depth can be computed.

As a second step,
recall from \cref{carry-aop}
that the carry bits $c_1, \dotsc, c_n$ of an $n$-bit adder
are \aop{}s on $1, 3, \dotsc, 2n - 1$ inputs
and that -- when circuit size is not regarded --
a depth-optimum adder circuit on $n$ bits
can be computed via depth-optimum \aop{} circuits
computing each carry bit.
Hence, in particular, \cref{table-optimum-m} yields the
optimum depths of all adder circuits with $2^k$ inputs for $k \leq 13$ over the basis $\set{\AND{}2, \OR{}2}$.
We show these in \cref{table-optimum-adders}.
Note that when the optimum depths computed by Hegerfeld were used instead
of the results computed by our algorithm,
the optimum adder depths could only be computed up to $k = 4$.

\newcommand{\cw}{0.6cm}
\begin{table}
 \centering
 \begin{tabularx}{\textwidth}{p{0.8cm}RRRRRRRRRRRRRr}\\ \toprule
 $n$      & 1 & 2 & 4 & 8  & 16 & 32 & 64  & 128 & 256 & 512  & 1024 & 2048 & 4096 & 8192  \\ \midrule
 $2n - 1$ & 1 & 3 & 7 & 15 & 31 & 63 & 127 & 255 & 511 & 1023 & 2047 & 4095 & 8191 & 16383 \\
 $d$      & 0 & 2 & 4 & 5  & 6  & 8  & 9   & 10  & 11  & 12   & 13   & 14   & 15   & 16\\
 \bottomrule
  \end{tabularx}
 \caption
 {Optimum depths $d$ of $n$-bit adder circuits over the basis $\set{\AND{}2, \OR{}2}$, where $n$ is a power of $2$.
 The middle row shows the number $2n - 1$ of inputs of the \aop{} computing the most significant carry bit.}
 \label{table-optimum-adders}
\end{table}
               % 7
\FloatBarrier % floats before and after the barrier are not mixed up
% \newpage
\section*{Conclusions}

We presented a new exact algorithm
for constructing depth- and delay-optimum \aop{} and adder circuits
over the basis $\set{\AND{}2, \OR{}2}$.
Our algorithm is much faster than previous approaches
-- both empirically and regarding provable worst-case running time --
and hence can solve significantly larger instances.
For all \aop{} instances with up to 64 inputs,
the optimum depth was computed in reasonable time.

Using these empirical computations and new theoretical results,
we derived the optimum depths for binary carry-propagate adders on $2^k$ bits
for all $k \leq 13$,
previously known only for $k \leq 4$.
Thus,
for any practically relevant number of bits,
the problem of constructing depth-optimum adder circuits
over the basis $\set{\AND{}2, \OR{}2}$,
which for decades was a subject of research,
is now settled.
Further research may improve secondary objectives like fan-out and size
or consider a trade-off between different objective functions.

         % conclusion
\FloatBarrier % floats before and after the barrier are not mixed up

% We actually should use one of the styles suggested in template,
% but they all sort references by first appearence in the text,
% not in alphabetical order.
%\section*{References}
\bibliographystyle{plainnat}
\bibliography{all_citations}

\end{document}